\def\pmb#1{\setbox0=\hbox{#1}%
   \kern-.025em\copy0\kern-\wd0
   \kern.05em\copy0\kern-\wd0
   \kern-0.025em\raise.0433em\box0}
\def\gta{\mathrel{{\lower 3pt\hbox{$\mathchar"218$}}\hskip-8pt
   \raise 2pt\hbox{$\mathchar"13E$}}}
\def\lta{\mathrel{{\lower 3pt\hbox{$\mathchar"218$}}\hskip-8pt
   \raise 2pt\hbox{$\mathchar"13C$}}}
\def\half{{\scriptstyle{1\over2}}}
\def\quarter{{\scriptstyle{1\over4}}}
\def\dagg{\phantom{\dagger}}            

\documentclass[twocolumn,showkeys,preprintnumbers,amsmath,amssymb,floatfix]{revtex4}

\usepackage{graphicx}
\usepackage{dcolumn}
\usepackage{bm}

\begin{document}

\title{A Theory for the High-$T_c$ Cuprates: Anomalous Normal-State and \\ 
Spectroscopic Properties, Phase Diagram, and Pairing }

\author{J. Ashkenazi}
 \email{jashkenazi@miami.edu}
\affiliation{%
Physics Department, University of Miami, P.O. Box 248046, Coral
Gables, FL 33124, U.S.A.\\
}%

\date{\today}

\begin{abstract}
A theory of highly correlated layered superconducting materials is
applied for the cuprates. Differently from an independent-electron
approximation, their low-energy excitations are approached in terms of
auxiliary particles representing combinations of atomic-like electron
configurations, where the introduction of a Lagrange Bose field enables
treating them as bosons or fermions. The energy spectrum of this field
accounts for the tendency of hole-doped cuprates to form stripe-like
inhomogeneities. Consequently, it induces a different analytical
behavior for auxiliary particles corresponding to ``antinodal'' and
``nodal'' electrons, enabling the existence of different pairing
temperatures at $T^*$ and $T_c$. This theory correctly describes the
observed phase diagram of the cuprates, including the non-Fermi-liquid
to FL crossover in the normal state, the existence of Fermi arcs below
$T^*$ and of a ``marginal-FL'' critical behavior above it. The
qualitative anomalous behavior of numerous physical quantities is
accounted for, including kink- and waterfall-like spectral features, the
drop in the scattering rates below $T^*$ and more radically below $T_c$,
and an effective increase in the density of carriers with $T$ and
$\omega$, reflected in transport, optical and other properties. Also is
explained the correspondence between $T_c$, the resonance-mode energy,
and the ``nodal gap''. 
\end{abstract}


\maketitle


\section{Introduction}

High-$T_c$ superconductivity (SC) has been in the forefront of
condensed-matter physics research since its discovery in the cuprates
over 23 years ago. This system is characterized by anomalous behavior of
many of its physical properties \cite{Takagi, Kubo, Hwang, Fisher,
Tanaka, Puchkov, Bonn, Zasadzinski, Luo, Tanner1, Boeb} which led to the
suggestion of non-Fermi-liquid (non-FL) models \cite{Anderson, Varma,
Lee, Rice} for its electronic structure. The recent discovery of
high-$T_c$ SC in iron-based compounds (referred to below as FeSCs),
provides a new test case for high-$T_c$ theories, especially in view
\cite{AshkHam} of the striking similarity of their anomalous properties
to those of the cuprates. 

Recently, a unified theory for the cuprates and the FeSCs has been
derived \cite{AshkHam} on the basis of common features in their
electronic structures, including quasi-two-dimensionality, and the
large-$U$ nature of the electron orbitals close to the Fermi level
($E_{_{\rm F}}$). Electrons of such orbitals {\it cannot} be treated
through a mean-field independent-particle approach; therefore, their
behavior is studied in terms of auxiliary particles, representing
combinations of atomic-like electron configurations. 

Within this auxiliary-particles method \cite{Barnes}, configurations in
sites $i$ can be, approximately, treated as bosons or fermions, if each
site is mathematically constrained to be occupied by one (and only one)
configuration. There is a freedom in choosing configurations of an odd
number of electrons as fermions, and of an even number of electrons as
bosons, or {\it vice versa} \cite{AshkHam}. The choice here is that
configurations corresponding to undoped cuprates or FeSCs (thus the 
parent compounds) are bosons.

These auxiliary particles have been treated {\it beyond} mean-field
theory \cite{Lee}. A grand-canonical Hamiltonian ${\cal H}$ has been
written down \cite{AshkHam}, where (in addition to the chemical
potential $\mu$) a field of Lagrange multipliers $\lambda_i$ was
introduced ($\lambda \equiv \langle \lambda_i \rangle$) to maintain the
above auxiliary-particles' constraint. This Lagrange field represents an
effective fluctuating potential which enables the treatment of the above
configurations as bosons or fermions; its effect on them is analogous to
the effect of vibrating atoms on electrons, and similarly to lattice
dynamics, this is a Bose field \cite{AshkHam} (see discussion about it
below). 

These auxiliary particles are assigned the following {\it ad hoc} names
\cite{AshkHam}: ({\it i}) combinations of boson atomic-like
configurations are referred to as ``svivons''; their Bose condensation
results in static or dynamical inhomogeneities which could be manifested
in the existence of a commensurate or an incommensurate resonance mode
\cite{AshkHam}; ({\it ii}) combinations of fermion atomic-like
configurations are referred to as ``quasi-electrons'' (QEs); they carry
charge $-{\rm e}$ and are introduced through electron or hole doping (in
the second case the configurations correspond to holes of QEs); ({\it
iii}) the Lagrange-field bosons are referred to as ``lagrons'', and
their coupling to the svivons and QEs provides a dynamical enforcement
of the auxiliary-particles' constraint. 

Stripe-like inhomogeneities have been observed \cite{Tran1, Yamada,
Kapitul, Davis1, Hudson} within the phase diagram of the cuprates. Even
though such an inhomogeneous behavior is often dynamical within the SC
regime, it is not treated here as a perturbation to a homogeneous state,
but as a correction to a static inhomogeneous structure. The application
of the constraint, approached here through the Lagrange field, must take
into account the existence of these inhomogeneities. This results in a
multi-component scenario. 

In small-$U$ systems, where low-energy electrons are appropriate
quasiparticles (QPs), the FLEX approximation \cite{Scalapino1} could be
applied \cite{manske} to derive an effective pairing interaction between
them, due to their own spin and charge fluctuations. The low-energy
large-$U$ electrons, discussed here, {\it cannot} be considered as
appropriate QPs; but their description in terms of independent auxiliary
Fermi and Bose fields provides an intrinsic pairing mechanism through
QE-lagron coupling; it can be treated \cite{AshkHam} within the
Migdal--Eliashberg theory \cite{Migdal, Eliashberg, Scalapino},
similarly to electron-phonon coupling in the case of phonon-induced
pairing, but with coupling constants which are so strong that lattice
instabilities would have been driven in the electron-phonon case. 

In this paper auxiliary particles which are specific for the cuprates
are introduced, and their Green's functions are applied to derive those
of the electrons. The detailed QE and electron spectral functions are
calculated, with further emphasis on the anomalous low-energy features.
The QE, svivon and electron scattering rates are derived in the non-FL
regime, and the resulting physical anomalies are discussed.  ${\cal H}$
is then applied to derive QE and electron pairing which is demonstrated
to result in the distinct pseudogap (PG) and SC phases; the associated
anomalous physical behavior is discussed. Further calculations on part
of these aspects (specifically in the paired states) will appear in
separate papers. 

\section{Hamiltonian}

Electronic-structure calculations, and a variety of experimental data,
indicate that the ``universal'' (thus {\it not} characteristic of
just specific compounds) low-energy properties of the cuprates are
primarily determined by Cu($d$) and O($p$) orbitals of the CuO$_2$
planes. Within the LDA, these orbitals generate a multiple-band
structure extending over a range of \cite{Andersen1} about $9\;$eV.
However, the electronic states at the vicinity of $E_{_{\rm F}}$
dominantly reside \cite{Andersen1, Andersen, Bansil} in {\it one} of
these bands, of Wannier functions \cite{AshMer} which are centered at
the planar Cu atoms, and correspond mainly to $\sigma$-antibonding
states between Cu($d_{x^2-y^2}$) and O($p_x,p_y$) orbitals (though a
minor contribution of other orbitals exists as well). Due to this
orbital nature of the band, its treatment requires effective transfer
(hopping) integrals up to the third-nearest neighbor (see below). 

The effective intra-site Coulomb parameter $U$, for electrons in this
band, corresponds to its Wannier functions which are, primarily,
combinations of Cu($d$) and O($p$) functions. Estimates of $U$ must take
into account the fact that the $d$ intra-atomic integral $U_d$ is
considerably larger than the $p$ integral $U_p$. In an early derivation
of the effective one-band Hamiltonian \cite{Zhang} in hole-doped
cuprates, the large-$U_d$ and small-$U_p$ limits have been assumed,
resulting in approximately ionic Cu($d$) and itinerant O($p$) states.
The resulting \cite{Zhang} upper Hubbard band corresponds to Cu$^{+1}$
ions, and the lower Hubbard band to Cu$^{+2}$ ions, part of which are
forming singlet states with the doped O($p$) holes (known as
``Zhang-Rice singlets''). 

Consequently, the value of $U$, derived through this analysis, consists
of $U_d$ plus corrections, due to the singlet energy and the difference
between the one-electron $p$ and $d$ energies which are considered to be
substantially smaller. A more realistic evaluation of $U$ would result
in a somewhat smaller value than $U_d$; however, since the width of the
effective band, within a two-dimensional (2D) Brillouin zone (BZ), is
\cite{Andersen1, Andersen, Bansil} about $3\;$eV, it is appropriate to
treat it using a large-$U$ approach. 

Within this one-band approach, a 2D square lattice of $N$ sites $i$ is
considered (assuming translational symmetry between them); the Cu atoms
are located at the lattice points ${\bf R}_i$, generated by the unit
vectors $a{\hat x}$ and $a{\hat y}$. An electron of spin $\sigma$ in
site $i$ is created by $d_{i\sigma}^{\dagger}$ (the two spin states are
denoted as $\sigma = \uparrow, \downarrow$, or $\sigma = \pm 1$). The
Hamiltonian is expressed as: 
\begin{eqnarray}
{\cal H}^d &\cong& \sum_{i \sigma}\big[ (\epsilon^d - \mu)
d_{i\sigma}^{\dagger} d_{i\sigma}^{\dagg} + \sum_{j \ne i } t({\bf R_i}
- {\bf R_j}) d_{i\sigma}^{\dagger} d_{j\sigma}^{\dagg} \nonumber \\ &\ &
+ \half U d_{i,-\sigma}^{\dagger} d_{i\sigma}^{\dagger}
d_{i\sigma}^{\dagg} d_{i,-\sigma}^{\dagg} \big], \label{eq0a} 
\end{eqnarray}
where the transfer integrals $t({\bf R})$ are considered up to the
third-nearest neighbor, and expressed in terms of the parameters $t$,
$t^{\prime}$ and $t^{\prime\prime}$: 
\begin{eqnarray}
t(\pm a {\hat x}) = t(\pm a {\hat y}) = -t, &\ & \label{eq0b} \\ t( \pm
a{\hat x} \pm a{\hat y}) = -t^{\prime}, &\ & t(\pm 2a {\hat x}) = t(\pm
2a {\hat y}) = -t^{\prime\prime}. \nonumber 
\end{eqnarray}

The present auxiliary-particles approach \cite{AshkHam} then becomes the
``slave-fermion'' method, applied in previous works by the author
\cite{Ashk94, Ashk01, ashk}. This approach is {\it different} from
homogeneous RVB models \cite{Anderson, Lee, Rice} which are also based
on auxiliary particles, but generally within the ``slave-boson'' method,
where the fermions are spin-carrying ``spinons'' and the bosons are
charge-carrying ``holons''. Within the present approach, the roles of
fermions and bosons are switched, and consequently inhomogeneities are
inherently introduced through Bose condensation. An electron creation
operator is then expressed as \cite{AshkHam, Barnes}: 
\begin{equation}
d_{i\sigma}^{\dagger} = s_{i\sigma}^{\dagger} h_{i}^{\dagg} + 
\sigma e_{i}^{\dagger} s_{i,-\sigma}^{\dagg},
\label{eq0c0} 
\end{equation}
where $s_{i\sigma}^{\dagger}$, $h_{i}^{\dagger}$ and $e_{i}^{\dagger}$
create a boson svivon state, and fermion holon and ``doublon'' states,
respectively. The auxiliary-particles' constraint is expressed as
\cite{AshkHam, Barnes}: 
\begin{equation}
\sum_{\sigma} s_{i\sigma}^{\dagger} s_{i\sigma}^{\dagg} + h_i^{\dagger}
h_i^{\dagg} + e_i^{\dagger} e_i^{\dagg} = 1. \label{eq0d0} 
\end{equation}

For hole-doped cuprates, the effect of the upper-Hubbard-band doublons
is ignored in the lowest order, and the creation operator of a fermion
QE state is defined as: $q_{i}^{\dagger} \equiv h_{i}^{\dagg}$.
Eqs.~(\ref{eq0c0},\ref{eq0d0}) are then approximated as: 
\begin{eqnarray}
d_{i\sigma}^{\dagger} &\cong& s_{i\sigma}^{\dagger} q_{i}^{\dagger},
\label{eq0c} \\
\sum_{\sigma} s_{i\sigma}^{\dagger} s_{i\sigma}^{\dagg} &\cong&
q_i^{\dagger} q_i^{\dagg}. \label{eq0d} 
\end{eqnarray}
By expressing the electron operators in the rhs of Eq.~(\ref{eq0a}) in
terms of the auxiliary-particle operators, through Eq.~(\ref{eq0c}), one
gets that the $U$ term in ${\cal H}^d$ drops, as is expected far below
the upper Hubbard band. 

Including in ${\cal H}^d$ the effect of doublons, through
Eq.~(\ref{eq0c0}), within a second-order perturbation expansion,
introduces \cite{AshkHam} corrections to the transfer integrals, and
antiferromagnetic (AF) exchange integrals: 
\begin{equation}
\Delta t({\bf R},{\bf R}^{\prime}) \cong - {t({\bf R}) t({\bf
R}^{\prime}) \over U}, \ \ J({\bf R}) \cong {t({\bf R}) t(-{\bf R})
\over U}, \label{eq4b} 
\end{equation}
for ${\bf R}\ne 0$ and ${\bf R}^{\prime}\ne 0$. Considering such
integrals only for nearest-neighbor ${\bf R}$ and ${\bf R}^{\prime}$
results in their expression in terms of the parameter $J \cong t^2/U$: 
\begin{eqnarray}
\Delta t(\pm a {\hat x}, \pm a {\hat x}) = \Delta t(\pm a {\hat y}, \pm
a {\hat y}) = \Delta t( \pm a {\hat x}, \pm a {\hat y}) &\cong& -J,
\nonumber \\ J(\pm a{\hat x}) = J(\pm a{\hat y}) = J. \ \ &\ &
\label{eq4e} 
\end{eqnarray}

The parameters in Eqs.~(\ref{eq0b},\ref{eq4e}) have been determined on
the basis of first-principles calculations \cite{Andersen1, Andersen,
Bansil, Anisimov1, Wan}. The set of their values chosen here is: $t =
0.43\;$eV, $t^{\prime} = -0.07\;$eV, $t^{\prime\prime} = 0.03\;$eV, and
$J = 0.1\;$eV (other sets of values, within the range of the calculated
parameters, yield similar results to those below). 

Applying the above approximations, and the constraint in
Eq.~(\ref{eq0d}) which is approached through a grand-canonical scheme,
results in the expression of ${\cal H}^d$ in Eq.~(\ref{eq0a}) in terms
of ${\cal H}^a$, given by \cite{AshkHam}: 
\begin{eqnarray}
{\cal H}^a &=& \sum_{ij\sigma} \Big\{\big[ \half \delta_{ij} (\epsilon^d
+ \lambda - \mu) + (1 - \delta_{ij}) [t({\bf R}_i - {\bf R}_j) \nonumber
\\ &\ & + \sum_{k \ne i,j} \Delta t({\bf R}_i - {\bf R}_k , {\bf R}_k -
{\bf R}_j) s_{k,-\sigma}^{\dagger} s_{k,-\sigma}^{\dagg}] \nonumber \\
&\ & \times s_{i\sigma}^{\dagger} s_{j\sigma}^{\dagg} \big]
q_i^{\dagger} q_j^{\dagg} - \big[ \delta_{ij} \lambda + \half (1 -
\delta_{ij}) J({\bf R}_{i} - {\bf R}_{j}) \nonumber \\ &\ & \times
s_{j,-\sigma}^{\dagger} s_{j,-\sigma}^{\dagg} \big]
s_{i\sigma}^{\dagger} s_{i\sigma}^{\dagg} \Big\} + \Delta{\cal H},
\label{eq4f}
\end{eqnarray}
where the mean value $\lambda$ of the constraint Lagrange multipliers
$\lambda_i$ is treated similarly to the chemical potential $\mu$, while
$\lambda_i - \lambda$ fluctuations are treated through: 
\begin{equation}
\Delta{\cal H} = \sum_{i} (\lambda_i - \lambda) \big[ q_i^{\dagger}
q_i^{\dagg} - \sum_{\sigma} s_{i\sigma}^{\dagger} s_{i\sigma}^{\dagg}
\big]. \label{eq4g} 
\end{equation}

${\cal H}^a$ describes an auxiliary system of fermions and bosons,
behaving according to the laws of physics, but it is relevant for the
real physical system only under the set of values of time-dependent
$\lambda_i$ for which the constraint is maintained. Assuming lattice
translational symmetry, let $\lambda({\bf k})$, $q^{\dagger}({\bf k})$
and $s_{\sigma}^{\dagger}({\bf k})$ be, respectively, the Fourier
transforms of $\lambda_i$, $q_i^{\dagger}$ and $s_{i\sigma}^{\dagger}$
to the wave vector (${\bf k}$) representation, determined within the
lattice BZ. 

By Eq.~(\ref{eq4g}), nonzero $\lambda_i - \lambda$ introduce
hybridization between QE and between svivon states, and the time
derivatives of $\lambda_i$ induce transitions between such states.
Consequently, the Lagrangian of the auxiliary system depends both on
$\lambda({\bf k})$, playing the role of generalized coordinates, and on
their time derivatives $\dot{\lambda}({\bf k})$ which must {\it not}
vanish (for ${\bf k} \ne 0$) in order to prevent constraint-violating
fluctuations. This dependence could be applied \cite{Goldstein} to
derive the conjugate momenta $p^{\lambda}({\bf k})$ of $\lambda({\bf
k})$, and the dependence of ${\cal H}^a$ on them, reflecting the effect
of $\dot{\lambda}({\bf k})$. 

Since $\lambda({\bf k})$ and $p^{\lambda}({\bf k})$ are coordinates and
momenta in a physical-like quantum system, there exist coefficients
$\gamma({\bf k})$ for which one can express (for ${\bf k} \ne 0$), in a
similar manner as in lattice dynamics \cite{Vliet} (units where $\hbar = 
1$ are used): 
\begin{eqnarray}
\lambda({\bf k}) &=& \gamma({\bf k}) [l({\bf k}) + l^{\dagger}(-{\bf k})], 
\nonumber \\
p^{\lambda}({\bf k}) &=& {1 \over 2 i \gamma({\bf k})} [l(-{\bf k}) -
l^{\dagger}({\bf k})], \label{eq4h}
\end{eqnarray}
where $l^{\dagger}({\bf k})$ are the creation operators of boson lagron
states within the BZ (omitting the ${\bf k} = 0$ point which corresponds
to $\lambda$ in Eq.~(\ref{eq4f})). Thus, one could express $\Delta{\cal
H}$ in Eq.~(\ref{eq4g}) as: 
\begin{eqnarray}
\Delta{\cal H} &=& \sum_{{\bf k},{\bf q} \ne 0} \gamma({\bf q}) [l({\bf
q}) + l^{\dagger}(-{\bf q})] \label{eq4i} \\ &\ & \times \big[
q^{\dagger}({\bf k}) q({\bf k} + {\bf q}) -
\sum_{\sigma}s_{\sigma}^{\dagger}({\bf k}) s_{\sigma}^{\dagg}({\bf k} +
{\bf q}) \big]. \nonumber 
\end{eqnarray}

The coupling of QEs and svivons to lagrons, through $\Delta{\cal H}$ in
Eq.~(\ref{eq4i}), introduces an implicit dependence of the auxiliary
system on $\lambda({\bf k})$ and $p^{\lambda}({\bf k})$, due to the
modifications introduced to the QE and svivon spectra. The effect of
$\lambda({\bf k})$ is manifested in the real parts of their induced
self-energies \cite{Vliet, Mahan}, and that of $p^{\lambda}({\bf k})$ in
the corresponding imaginary parts. These effects on the QE and svivon
spectra introduce modifications in the Helmholtz free energy ${\cal
F}^a$ of the auxiliary system which could be expressed in terms of an
effective lagron Hamiltonian ${\cal H}^{\lambda}$, depending on
$\lambda({\bf k})$ and $p^{\lambda}({\bf k})$. 

The effect of the magnitudes and rates of the $\lambda_i - \lambda$
fluctuations on ${\cal F}^a$ could be generally treated within a
second-order perturbation expansion. Since a constrained minimum of
${\cal F}^a$ cannot be lower than the unconstrained one, the
contribution of each of these time-dependent Lagrange multipliers to
${\cal H}^{\lambda}$ must not be negative. Thus, in analogy to the
harmonic approximation in lattice dynamics \cite{AshMer,Vliet}, the
dependence of ${\cal H}^{\lambda}$ on $\lambda({\bf k})$ and
$p^{\lambda}({\bf k})$ has a positive definite quadratic form.
Consequently, by applying the canonical transformation: 
\begin{eqnarray}
{\tilde \lambda}({\bf k}) &=& \cos{(\xi_{\bf k}^{\lambda})} \lambda({\bf
k}) + \sin{(\xi_{\bf k}^{\lambda})} p^{\lambda}({\bf k}), \nonumber \\ 
{\tilde p}^{\lambda}({\bf k}) &=& \cos{(\xi_{\bf k}^{\lambda})}
p^{\lambda}({\bf k}) - \sin{(\xi_{\bf k}^{\lambda})} \lambda({\bf k}),
\label{eq4i1} 
\end{eqnarray}
an appropriate choice of $\xi_{\bf k}^{\lambda}$ results in an
expression of the form: 
\begin{equation} 
{\cal H}^{\lambda} \cong {1 \over 2} \sum_{{\bf k} \ne 0}
[A^{\lambda}({\bf k}) {\tilde \lambda}^{\dagger}({\bf k}) {\tilde
\lambda}({\bf k}) + B^{\lambda}({\bf k}) {\tilde
p}^{\lambda\dagger}({\bf k}) {\tilde p}^{\lambda}({\bf k})], 
\label{eq4j0} 
\end{equation} 
where the coefficients $A^{\lambda}({\bf k})$ and $B^{\lambda}({\bf k})$
are positive. Expressing ${\tilde \lambda}({\bf k})$ and ${\tilde
p}^{\lambda}({\bf k})$ through Eqs.~(\ref{eq4h},\ref{eq4i1}) yields: 
\begin{eqnarray}
{\cal H}^{\lambda} &\cong& \sum_{{\bf k} \ne 0} \omega^{\lambda}({\bf
k}) [l^{\dagger}({\bf k}) l({\bf k}) + \half], \ \ \ \text{where}
\label{eq4j} \\ 
\omega^{\lambda}({\bf k}) &=& \big\{ [\sin{}^2(\xi_{\bf k}^{\lambda})
A^{\lambda}({\bf k}) + \cos{}^2(\xi_{\bf k}^{\lambda}) B^{\lambda}({\bf
k})] \nonumber \\ &\ & \times [\cos{}^2(\xi_{\bf k}^{\lambda})
A^{\lambda}({\bf k}) + \sin{}^2(\xi_{\bf k}^{\lambda}) B^{\lambda}({\bf
k})] \big\}^{\half}, \ \ \ \ \ \label{eq4j1} \\ 
|\gamma({\bf k})|^2 &=& {1 \over 2} \bigg\{ {\sin{}^2(\xi_{\bf
k}^{\lambda}) A^{\lambda}({\bf k}) + \cos{}^2(\xi_{\bf k}^{\lambda})
B^{\lambda}({\bf k}) \over \cos{}^2(\xi_{\bf k}^{\lambda})
A^{\lambda}({\bf k}) + \sin{}^2(\xi_{\bf k}^{\lambda}) B^{\lambda}({\bf
k})} \bigg\}^{\half}. \nonumber 
\end{eqnarray}

The derivation of ${\cal H}^{\lambda}$ from ${\cal H}^a$ is conceptually
analogous to the derivation of a phonon Hamiltonian from the Hamiltonian
of electrons and nuclei in a crystal. In the same manner that the
auxiliary system represents the real physical system only under the
correct lagron spectrum and coupling constants, a system of electrons
coupled to phonons in a crystal precisely represents the physical system
only under the correct phonon spectrum and coupling constants. 

The evaluation of the lagron energies $\omega^{\lambda}({\bf k})$ in
Eq.~(\ref{eq4j}) could be carried out self-consistently, on the basis of
${\cal H}^a$ in Eq.~(\ref{eq4f}), applying diagrammatic techniques
\cite{Vliet, Mahan} (this introduces lagron linewidths, ignored in
Eq.~(\ref{eq4j})). The coefficients $\gamma({\bf k})$, appearing in
Eq.~(\ref{eq4h}), should be chosen such that Eq.~(\ref{eq4j1}) is
satisfied, and the svivon and QE spectra, obtained on the basis of
Eq.~(\ref{eq4f}), maintain the constraint in Eq.~(\ref{eq0d}). 
 
These conditions are approached by checking trial $\omega^{\lambda}({\bf
k})$ spectra, determined on the basis of physical considerations (see
below), and looking for the coefficients $\gamma({\bf k})$ for which
these spectra are obtained self-consistently. Convergence is reached
when the lagron spectrum and $\gamma({\bf k})$ satisfy
Eq.~(\ref{eq4j1}), and determine QE and svivon spectra which maintain
the constraint; it is checked through specific relations between these
spectra, expressed through the ``constraint susceptibility''
\cite{AshkHam} (see below). It turns out that the derivation of the
major features of the lagron, QE and svivon spectra could be performed
without an elaborate point by point self-consistent calculation. 

\section{Green's functions}

The elements of the ``normal'' svivon and QE Green's-function
\cite{Vliet, Mahan} matrices $\underline{\cal G}^s$ and $\underline{
\cal G}^q$ are based, in the site representation, on expectation values
of the form $\langle s_{i\sigma}^{\dagg} s_{j\sigma}^{\dagger} \rangle$
and $\langle q_{i}^{\dagg} q_{j}^{\dagger} \rangle$, respectively. The
Bose condensation of svivons, and the pairing of QEs introduce
``anomalous'' \cite{Mahan} svivon and QE Green's-function matrices
$\underline{\cal F}^s$ and $\underline{ \cal F}^q$ based, in the site
representation, on expectation values of the form $\langle
s_{i\uparrow}^{\dagg} s_{j\downarrow}^{\dagg} \rangle$ and $\langle
q_{i}^{\dagg} q_{j}^{\dagg} \rangle$, respectively. 

\begin{figure}[t] 
\begin{center}
\includegraphics[width=3.25in]{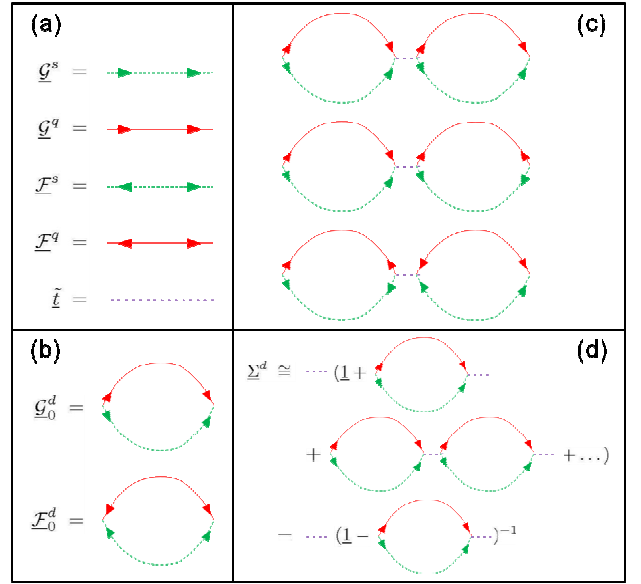}
\end{center}
\caption{(a) Propagator diagrams of the normal and anomalous svivon and
QE Green's-function matrices $\underline{\cal G}^s$, $\underline{ \cal
G}^q$, $\underline{\cal F}^s$, $\underline{ \cal G}^q$, and of the
renormalized transfer matrix $\underline{\tilde t}$ (see discussion in
the text); (b) bubble diagrams of the zeroth-order normal and anomalous
electron Green's-function matrices $\underline{\cal G}^d_0$ and
$\underline{\cal F}^d_0$; (c) diagrams of typical first-order
corrections to $\underline{\cal G}^d_0$ and $\underline{\cal F}^d_0$;
(d) diagrammatical presentation of the geometrical-series sum
determining the self-energy correction $\underline{\Sigma}^d$ to
$\underline{\cal G}^d_0$ for \cite{pairing} $T > T^q_{\rm p}$.} 
\label{fig1}
\end{figure}

The above Green's-function matrices are presented diagrammatically as
the propagators shown in Fig.~\ref{fig1}(a); in the ${\bf k}$ and
Matsubara representations \cite{Vliet, Mahan} they are diagonal and
expressed, at temperature $T$, as functions of $\omega_{\nu} = \nu\pi
k_{_{\rm B}} T$, where $k_{_{\rm B}}$ is the Boltzmann constant, and
$\nu$ is an integer which is even for bosons and odd for fermions. The
dependence of the Green's functions on $i\omega_{\nu}$ is analytically
continued \cite{Vliet, Mahan} to the complex $z$ plane, including the
$\pm i0^{+}$ vicinity of the real $\omega$-axis (where $0^{+}$ is an
infinitesimally small positive number). 

The normal and anomalous electron Green's-function matrices
$\underline{\cal G}^d$ and $\underline{\cal F}^d$ are also diagonal in
the ${\bf k}$ representation. By Eq.~(\ref{eq0c}), the elements of
$\underline{\cal G}^d$ (far below the upper Hubbard band) are based, in
the site representation, on expectation values of the form $\langle
q_{i}^{\dagg} s_{i\sigma}^{\dagg} s_{j\sigma}^{\dagger} q_{j}^{\dagger}
\rangle$, and of $\underline{\cal F}^d$ on expectation values of the
form $\langle q_{i}^{\dagg} s_{i\uparrow}^{\dagg} q_{j}^{\dagg}
s_{j\downarrow}^{\dagg} \rangle$. Thus the zeroth-order (in $t/U$)
normal electron Green's-function matrices $\underline{\cal G}^d_0$
correspond, in the site representation, to bubble diagrams formed by
$\underline{\cal G}^s$ and $\underline{\cal G}^q$, and the anomalous
matrices $\underline{\cal F}^d_0$ to bubble diagrams formed by
$\underline{\cal F}^s$ and $\underline{\cal F}^q$, as is presented
diagrammatically in Fig.~\ref{fig1}(b). In the ${\bf k}$ and Matsubara
representations, $\underline{\cal G}^d_0$ and $\underline{\cal F}^d_0$
become convolutions of such bubble diagrams (see below). 

The relevant term for single-electron excitations in ${\cal H}^a$ in
Eq.~(\ref{eq4f}) could be expressed (using Eqs.~(\ref{eq0c},\ref{eq0d}))
as: 
\begin{eqnarray}
{\cal H}^a_e &=& \sum_{ij\sigma} {\tilde t}_{ij} s_{i\sigma}^{\dagger}
q_i^{\dagger} s_{j\sigma}^{\dagg} q_j^{\dagg}, \ \ \ \text{where}
\nonumber \\ {\tilde t}_{ij} &\cong& \delta_{ij} (\epsilon^d + \lambda -
\mu) + (1 - \delta_{ij})[t({\bf R}_i - {\bf R}_j) \nonumber \\ &\ & +
\sum_{k \ne i,j} n^s_{ijk} \Delta t({\bf R}_i - {\bf R}_k , {\bf R}_k -
{\bf R}_j)], \label{eq5c} 
\end{eqnarray} 
where $n^s_{ijk}$ is the conditioned expectation value $\langle
s_{k,-\sigma}^{\dagger} s_{k,-\sigma}^{\dagg} \rangle$ when there is a
$\sigma$ electron in site $i$ and no electron in site $j$, or {\it vice
versa}. 

The elements ${\tilde t}_{ij}$ in Eq.~(\ref{eq5c}) introduce a matrix
$\underline{\tilde t}$ which is diagonal in the ${\bf k}$
representation, and its eigenvalues correspond (considerably below the
upper Hubbard band) to an LDA-like band structure, within the
one-orbital model. This matrix is also presented diagrammatically in
Fig.~\ref{fig1}(a). The dominant corrections to $\underline{\cal G}^d$
and $\underline{\cal F}^d$, beyond $\underline{\cal G}^d_0$ and
$\underline{\cal F}^d_0$, are introduced through ${\cal H}^a_e$ in
Eq.~(\ref{eq5c}); first-order corrections of this type are presented,
diagrammatically, in Fig.~\ref{fig1}(c). 

At $T$ above the QE pairing temperature \cite{pairing} $T^q_{\rm p}$ one
has $\underline{\cal F}^q = \underline{\cal F}^d =\underline{0}$ (though
$\underline{\cal F}^s \ne \underline{0}$). Similarly to the summation of
bubble diagrams within the RPA \cite{Vliet, Mahan}, the self-energy
correction to $\underline{\cal G}^d_0$ (due to multiple QE-svivon
scattering) could then be evaluated, through Eq.~(\ref{eq5c}), as a sum
of a geometrical series, presented diagrammatically in
Fig.~\ref{fig1}(d), yielding: 
\begin{equation} 
\underline{\Sigma}^d \cong \underline{\tilde t} \sum_{n=0}^{\infty}
(\underline{\cal G}^d_0 \underline{\tilde t})^n = \underline{\tilde t}
(\underline{1} - \underline{\cal G}^d_0 \underline{\tilde t})^{-1}.
\label{eq5d} 
\end{equation}

Using Dyson's equation \cite{Vliet, Mahan}, $(\underline{\cal G}^d)^{-1}
= (\underline{\cal G}^d_0)^{-1} - \underline{\Sigma}^d$, one can express
$\underline{\cal G}^d$ as: 
\begin{equation} 
\underline{\cal G}^d \cong \big( \underline{1} - \underline{\cal G}^d_0
\underline{\tilde t} \big) \big( \underline{1} - 2\underline{\cal G}^d_0
\underline{\tilde t} \big)^{-1} \underline{\cal G}^d_0. \label{eq6} 
\end{equation}
The consequence of Eq.~(\ref{eq6}) is that there are two types of poles
in the $z$-dependent $\underline{\cal G}^d$. One type, contributed by
the $\underline{\cal G}^d_0$ term, consists of a non-FL distribution of
convoluted QE-svivon poles; the other type is of poles contributed by
the multiple-scattering $(\underline{1} - 2\underline{\cal G}^d_0
\underline{\tilde t})^{-1}$ term, and it is consistent with electron
poles within FL theory. 

Since the evaluation of physical properties can be expressed in terms of
the electron Green's functions, and their poles, these properties are
characterized by FL-like and/or non-FL features due to the different
types of poles in $\underline{\cal G}^d$. The phase diagram of the
cuprates \cite{Honma} indicates a non-FL to FL crossover in the
normal-state behavior of the cuprates between the underdoped and the
overdoped regimes \cite{Takagi, Boeb}. In heavily overdoped cuprates, a
crossover is expected to a regime where the large-$U$ approach, applied
here, stops being valid, and there is no SC state. 

\section{Spectral functions}

\subsection{Lagrons}

Low-energy (soft) lagron modes play a significant role in the lagron
spectrum. In the phonon case, the emergence of a soft mode requires
(when it is not at the close vicinity of ${\bf k} = 0$) a considerable
self-energy correction, through electron-phonon coupling, in order to
lower the ``bare'' phonon energy; by the Kramers--Kronig relation
\cite{Vliet, Mahan} between the imaginary and real parts of the phonon
self-energy, such an energy lowering means that the soft phonon mode has
a considerable linewidth. 

On the other hand, in the lagron case, the entire mode energy
$\omega^{\lambda}({\bf q})$ consists of a self-energy correction, due to
its coupling to QEs and svivons, through $\Delta{\cal H}$ in
Eq.~(\ref{eq4i}). Thus, the Kramers--Kronig relation of the lagron
self-energy implies that both the energy and the linewidth of a lagron
mode could be extremely close to zero. Applying this relation, and
Eq.~(\ref{eq4j0}), indicates that $A^{\lambda}({\bf q})$ and
$B^{\lambda}({\bf q})$ almost vanish at this point, ${\bf q} = {\bf
Q}_m$, of the lagron energy minimum, and have the same type of
dependence on ${\bf q} - {\bf Q}_m$ within a BZ range of ${\bf q} \simeq
{\bf Q}_m$. 

Consequently, by Eq.~(\ref{eq4j1}), $\omega^{\lambda}({\bf q})$ also has
this type of dependence on ${\bf q} - {\bf Q}_m$, within this BZ range,
and $|\gamma({\bf q})|^2$ is fairly independent of ${\bf q}$ there. This
behavior of $\gamma({\bf q})$ is very different from the ${\bf k}$
dependence of the coupling constants between electrons and acoustic
phonons in metals which vanish \cite{AshMer,Vliet} for ${\bf k} \to 0$.
As will be discussed further below, the ${\bf q}$ dependence of
$\omega^{\lambda}({\bf q})$ which is consistent with the physics of the
cuprates is the one yielding a Bose-condensation-type behavior. For
their 2D BZ, this corresponds to $\omega^{\lambda}({\bf q}) \propto
|{\bf q} - {\bf Q}_m|$, for ${\bf q} \simeq {\bf Q}_m$; the apparent
singularity of $\omega^{\lambda}({\bf q})$ at ${\bf q} = {\bf Q}_m$ is a
self-consistent consequence of the macroscopic value of $\langle
l^{\dagger}({\bf Q}_m) l({\bf Q}_m) \rangle$.

By Eq.~(\ref{eq4h}), the ${\bf Q}_m$ points of the lagron spectrum
correspond to particularly large values of $\lambda({\bf q})$ and
$p^{\lambda}({\bf q})$, for ${\bf q} \cong {\bf Q}_m$. This is typical
of wave vectors corresponding to major spin and charge fluctuations,
where an inhomogeneous enforcement of the auxiliary-particles'
constraint is essential. Theoretical studies \cite{Zaanen1, Machida,
Emery1}, based on Hamiltonians like ${\cal H}^d$ in Eq.~(\ref{eq0a}),
predict that the interplay between the effects of electron hopping and
AF exchange is likely to drive the formation of stripe-like
inhomogeneities (studied in previous works by the author \cite{Ashk01,
ashk}). Thus the physics of the cuprates is consistent with ${\bf Q}_m$
points which correspond to the wave vectors of the striped structures. 

\begin{figure}[t] 
\begin{center}
\includegraphics[width=3.25in]{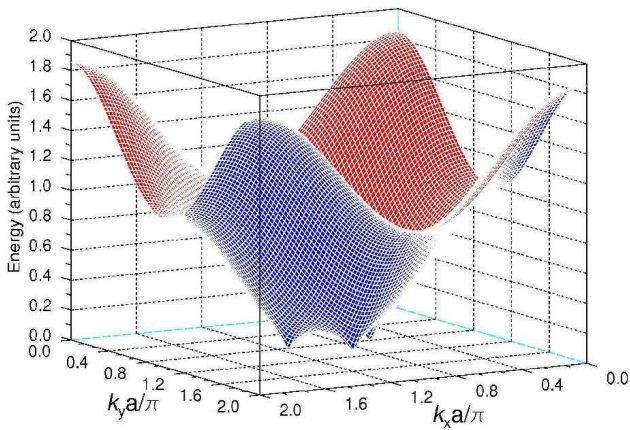}
\end{center}
\caption{A typical lagron spectrum in hole-doped cuprates; the minima
correspond to degenerate striped structures \cite{Tran1, Yamada,
Kapitul, Davis1, Hudson}.} 
\label{fig2}
\end{figure}

In agreement with the above theoretical predictions, the hole-doped
cuprates are characterized by dynamical or static stripe-like
inhomogeneities \cite{Tran1, Yamada, Kapitul, Davis1, Hudson}; in SC
stoichiometries, they correspond to stripes directed along the $a$- or
the $b$-axis; however, an intrinsic mechanism (of a nature discussed
below) introduces long-range symmetry between stripe segments directed
along these axes, resulting in a checkerboard-like structure
\cite{Davis1,Hudson}. Consequently, the lagron spectrum corresponding to
such a structure is of the type presented in Fig.~\ref{fig2}. 

Thus, the lagron energy band, $\omega^{\lambda}({\bf q})$, is
characterized by V-shape minima: 
\begin{eqnarray}
&\ &\omega^{\lambda}({\bf Q}_m) = k_{_{\rm B}} T /\mathcal{O}(N),\ \ \ 
\text{at} \ \ \ \nonumber \\ &\ & 
{\bf Q}_m = {\bf Q} + \delta {\bf q}_m,\ \text{for} \ m = 1, 2, 3\
\text{or} \ 4, \ \ \ \label{eq5} 
\end{eqnarray}
where ${\bf Q} = (\pi/a)({\hat x}+{\hat y})$ is the wave vector of the
AF order in the parent compounds; $\delta {\bf q}_m = \pm \delta q {\hat
x} \ \text{or} \ \pm \delta q {\hat y}$ are modulations around ${\bf
Q}$, indicative of the striped structures \cite{Tran1, Yamada, Kapitul,
Davis1, Hudson} (typically, $\delta q \cong \pi/4a$). The lagron
spectrum presented in Fig.~\ref{fig2} corresponds to degenerate states
of stripes directed along the $a$ and $b$ axes, and of different phases
of the spin periodicity (see below). The four lagron energy minima, and the
symmetry of the system, determine the main features of the spectrum, and
as was discussed above, the linewidths are generally larger at higher
energies. The high-energy extent of this spectrum is somewhat higher
than that of the svivon spectrum \cite{AshkHam}, and is thus expected to
be $\sim 0.3$--$0.4\;$eV. 

The existence of four equivalent lagron minima in points ${\bf Q}_m$ in
Eq.~(\ref{eq5}) results in a combination state of four Bose condensates 
of the lagron field \cite{condensate}, each consisting of a striped
structure of the symmetry corresponding to one of the ${\bf Q}_m$ wave
vectors. In SC stoichiometries, such a combination state is
energetically favorable, for $T \to 0$, on one of the symmetry-broken
condensates (corresponding to a long-range striped structure) since it
is the one which yields pairing (see below). Similarly to phonons which
are Goldstone bosons \cite{Goldstone} associated with spontaneous
breakdown of lattice translational symmetry, the lagrons are Goldstone
bosons associated with the breakdown of the symmetries of the long-range
striped structures, corresponding to each of the combined condensates. 

The svivon and QE spectra are evaluated through a self-consistent
second-order diagrammatic expansion, where a mean-field treatment of
${\cal H}^a$ in Eq.~(\ref{eq4f}) is applied at the zeroth order. The
expansion is carried out on two Hamiltonian terms; one of them is the
svivon-lagron and QE-lagron coupling term $\Delta{\cal H}$ in
Eq.~(\ref{eq4i}). The treatment of the analogous electron-phonon
coupling term in phonon-mediated SCs is based on the fact that the ratio
between phonon and electron energies is very small; consequently, by
Migdal's theorem \cite{Migdal}, ``vertex corrections'' beyond a
second-order diagrammatic expansion are negligible due to a small
integration phase space. 

Concerning the treatment of $\Delta{\cal H}$, as can be viewed in
Fig.~\ref{fig2}, the lagron spectrum consists of two regimes; a
low-energy one around point ${\bf Q}$ (including the points ${\bf Q}_m$
defined in Eq.~(\ref{eq5})), and a high-energy regime in the rest of the
BZ. As is discussed below, the peculiar behavior (including pairing)
introduced by $\Delta{\cal H}$ is primarily due to lagrons of the
low-energy regime, for which Migdal's theorem applies. Vertex
corrections due to lagrons of the high-energy regime may be not
negligible, but they are regular and could be approximated as
renormalization effects within the second-order diagrammatic expansion. 

\subsection{Svivons}

In the parent compounds, the lagron spectrum includes one V-shape
minimum at ${\bf q} = {\bf Q}$, corresponding to a condensate (see
above), and the Bose condensation of the svivon field  \cite{condensate}
is manifested in long-range AF order \cite{Ashk94}. Within such a svivon
condensate which corresponds to a combination of N\'eel states, shifted
from each other by $a{\hat x}$, the expectation value $n^s \equiv
\langle s_{i\sigma}^{\dagger} s_{i\sigma}^{\dagg} \rangle$ is site- and
spin- independent (and equals 0.5 for zero doping). 

The svivon band is then characterized \cite{Ashk94} by V-shape minima of
energy $k_{_{\rm B}} T /\mathcal{O}(N)$ at the points $\pm {\bf k}_0
\equiv \pm {\bf Q}/2$. There are four inequivalent values of ${\bf
k}_0$, introducing a broken symmetry: 
\begin{equation}
{\bf k}_0 \equiv {{\bf Q} \over 2}= \pm{\pi \over 2a} \Big( {\hat x} +
{\hat y} \Big) \ \text{or} \ \pm{\pi \over 2a} \Big( {\hat x} - {\hat y}
\Big). \label{eq6b0} 
\end{equation}
They correspond to two degenerate condensates of rectangular symmetry,
with axes along the ${\hat x} + {\hat y}$ and ${\hat x} - {\hat y}$
directions. Within the grand-canonical scheme, the svivon minima at $\pm
{\bf k}_0$ in the AF condensates are self-consistently formed due to
\cite{Ashk94} large values of both 
\begin{eqnarray}
&\ &n^s({\bf k}) \equiv \langle s_{\sigma}^{\dagger}({\bf k})
s_{\sigma}^{\dagg}({\bf k}) \rangle, \  \text{and} \ m^s({\bf k}) \equiv
\langle s_{\uparrow}^{\dagg}({\bf k}) s_{\downarrow}^{\dagg}(-{\bf k})
\rangle \ \ \ \nonumber \\ &\ &(\text{note that} \ n^s = {1 \over N}
\sum_{\bf k} n^s({\bf k})), \label{eq6b} 
\end{eqnarray}
around the minima (becoming $\mathcal{O}(N)$ at ${\bf k} = \pm {\bf
k}_0$). 

In SC stoichiometries, the minimum of the lagron spectrum at point ${\bf
Q}$ splits into the four ${\bf Q}_m$ points, introduced in
Eq.~(\ref{eq5}), and shown in Fig.~\ref{fig2}. As was mentioned above,
this spectrum corresponds to a combination of four condensates, in which
AF order is replaced by striped structures, corresponding to svivon
condensates \cite{condensate} of energy minima which are shifted
\cite{AshkHam} from $\pm {\bf k}_0$ (see discussion below) . 

The svivon minimal energies, within their condensates, are then
\cite{AshkHam} {\it not} as close to zero as $k_{_{\rm B}} T
/\mathcal{O}(N)$ (and the shape of the minima becomes parabolic close to
the bottom), since they are evaluated on the basis of a
lower-free-energy state (due to coupling to the QEs) which is a
combination of the condensates (and thus it consists of fluctuating
striped structures); but one still has (within each condensate)
$m^s({\bf k}) \ne 0$ which are large, together with $n^s({\bf k})$,
around the energy minima (though not as large as $\mathcal{O}(N)$---see
below). 

Note that $\lambda({\bf q})$ and $p^{\lambda}({\bf q})$ in
Eq.~(\ref{eq4h}) scale with $l(\pm{\bf q})$ and $l^{\dagger}(\pm{\bf
q})$ which behave like $\langle l^{\dagger}(\pm{\bf q}) l(\pm{\bf q})
\rangle^{\half}$ for ${\bf q} \cong {\bf Q}_m$, while $n^s({\bf k})$ and
$m^s({\bf k})$ are determined through Eq.~(\ref{eq6b})). Thus, the fact
that the svivon energy minima do not get too close to zero enables such
a lagron spectrum (thus with the V-shape energy minima specified in
Eq.~(\ref{eq5})) to yield values of $\lambda({\bf q})$ and
$p^{\lambda}({\bf q})$ which have appropriate values to maintain the
constraint \cite{condensate} within the ${\bf q} \simeq {\bf Q}_m$
ranges. 

Bose-condensed systems are characterized by nonzero expectation values
of the field operators \cite{Vliet, Mahan}. Treating the (combined)
svivon condensates adiabatically, one can express, through
Eq.~(\ref{eq6b}), $m^s({\bf k}) = \langle s_{\uparrow}^{\dagg}({\bf k})
\rangle \langle s_{\downarrow}^{\dagg}(-{\bf k}) \rangle$; these
$\langle s_{\uparrow}^{\dagg}({\bf k}) \rangle$ and $\langle
s_{\downarrow}^{\dagg}(-{\bf k}) \rangle$ have the same magnitude, but
different phases due to the different sites where the contribution to
$\sigma = \uparrow$ and $\sigma = \downarrow$ svivons is maximal, and
their phases fluctuate in a correlated manner. Terms including such
expectation values in the final result of a calculation, do not vanish
due to phase fluctuations only if they appear as products of the form
$\langle s_{\uparrow}^{\dagg}({\bf k}) \rangle \langle
s_{\downarrow}^{\dagg}(-{\bf k}) \rangle$ (which are expressed as
$m^s({\bf k})$). 

The second Hamiltonian term, treated through a self-consistent
second-order diagrammatic expansion, accounts for the coupling between
QEs and svivon fluctuations which are expressed through terms of the
form $s_{\sigma}^{\dagg}({\bf k}) - \langle s_{\sigma}^{\dagg}({\bf k})
\rangle$. It is obtained by expanding ${\cal H}^a$, in the rhs of
Eq.~(\ref{eq4f}), in such fluctuations (when the svivon operators there
are expressed in the ${\bf k}$ representation), and treating them as a
perturbation. This Hamiltonian term could be approximated as (see
Eq.~(\ref{eq5c})): 
\begin{eqnarray}
{\cal H}^{\prime} &\cong& {1 \over N} \sum_{{\bf k} {\bf k}^{\prime}
{\bf k}^{\prime\prime}\sigma} \big\{ {\tilde t}({\bf k}^{\prime} + {\bf
k}^{\prime\prime}) \langle s_{\sigma}^{\dagger}({\bf k}^{\prime} + {\bf
k}^{\prime\prime} - {\bf k}) \rangle q^{\dagger}({\bf k}) \nonumber \\
&\ & \times q({\bf k}^{\prime}) [ s_{\sigma}^{\dagger}({\bf
k}^{\prime\prime}) - \langle s_{\sigma}^{\dagg}({\bf k}^{\prime\prime})
\rangle] + h.c. \big\}, \label{eq6c} 
\end{eqnarray}
where ${\tilde t}({\bf k})$ is the Fourier transform of ${\tilde t}({\bf
R})$ which is, approximately, defined similarly to $t({\bf R})$ in
Eq.~(\ref{eq0b}), modifying $t^{\prime}$ and $t^{\prime\prime}$ (due to
the $\Delta t$ terms in Eq.~(\ref{eq4e})) to ${\tilde t}^{\prime} \cong
t^{\prime} + 2n^sJ$ and ${\tilde t}^{\prime\prime} \cong
t^{\prime\prime} + n^sJ$, respectively (while $t$ remains unchanged). 

Following the standard treatment of Bose-condensed systems \cite{Vliet,
Mahan}, the (Bogoliubov) boson creation operator of an excited svivon
state of wave vector ${\bf k}$ and spin $\sigma$ is derived as
\cite{Ashk94} a combination of $s_{\sigma}^{\dagger}({\bf k})$ and
$s_{-\sigma}^{\dagg}(-{\bf k})$. A $2\times 2$ energy plus self-energy
matrix is derived within the basis created by the spinor operator
($s_{\uparrow}^{\dagger}({\bf k})$, $s_{\downarrow}^{\dagg}(-{\bf k})$).
It is expressed as 
\begin{eqnarray}
\underline{\epsilon}^s_0({\bf k}) + \underline{\Sigma}^s({\bf k},z) &=&
[\epsilon^s_0({\bf k}) + \Sigma^s({\bf k},z)] \underline{1} + \Re
\Phi^s({\bf k},z) \underline{\tau}_1 \nonumber \\ &\ & + \Im \Phi^s({\bf
k},z) \underline{\tau}_2, \label{eq6d} 
\end{eqnarray}
where $\underline{1}$, $\underline{\tau}_1$ and $\underline{\tau}_2$ are
the unity and Pauli matrices, and $z$ are ``complex energies''
\cite{Vliet, Mahan}. The diagonal terms in the rhs of Eq.~(\ref{eq6d})
include $\epsilon^s_0({\bf k})$, derived within the mean-field
approximation, and $\Sigma^s({\bf k},z)$, derived through the
diagrammatic expansion. The non-diagonal terms, $\Phi^s({\bf k},z)$ and
$\Phi^s({\bf k},z)^*$, are derived through {\it both} the mean-field and
the diagrammatic calculations; they exist due to the nonzero values of
$m^s({\bf k}^{\prime})$. 

Two degenerate Bogoliubov svivon states, corresponding to $({\bf
k}\uparrow)$ and $(-{\bf k}\downarrow)$, are derived as a function of
$z$, and their boson annihilation operators are denoted by ${\tilde
s}_{\uparrow}^{\dagg}({\bf k},z)$ and ${\tilde
s}_{\downarrow}^{\dagg}(-{\bf k},z)$, respectively;
$s_{\uparrow}^{\dagg}({\bf k})$ and $s_{\downarrow}^{\dagg}(-{\bf k})$
are then expressed as \cite{Ashk94}: 
\begin{eqnarray}
s_{\sigma}^{\dagg}(\sigma{\bf k}) &=& \exp{(i\phi_{\bf k}^s(z)/2)}
[\cosh{(\xi_{\bf k}^s(z))} {\tilde s}_{\sigma}^{\dagg}(\sigma{\bf k},z)
\nonumber \\ &\ & + \sinh{(\xi_{\bf k}^s(z))} {\tilde
s}_{-\sigma}^{\dagger}(-\sigma{\bf k},z)], \label{eq6a}
\end{eqnarray}
where the coefficients $\xi_{\bf k}^s(z)$ and $\phi_{\bf k}^s(z)$ are
determined by the requirement that the matrix
$\underline{\epsilon}^s_0({\bf k}) + \underline{\Sigma}^s({\bf k},z)$ in
Eq.~(\ref{eq6d}) is diagonalized under the transformation to these
Bogoliubov states. This transformation diagonalizes Dyson's equation
\cite{Vliet, Mahan}, yielding the poles of the diagonalized svivon
Green's function ${\tilde {\cal G}}^s({\bf k},z)$ at: 
\begin{widetext}
\begin{eqnarray}
z &=& {\tilde \epsilon}^s_0({\bf k}) + {\tilde \Sigma}^s({\bf k},z), \
\text{where} \ \ 
\Im {\tilde \Sigma}^s({\bf k},z) = \cosh{(2\xi_{\bf k}^s(z))} \Im
\Sigma^s({\bf k},z), \nonumber \\ 
{\tilde \epsilon}^s_0({\bf k}) + \Re {\tilde \Sigma}^s({\bf k},z)
&\equiv& E^s({\bf k},z) = \text{sign}[\epsilon^s_0({\bf k}) + \Re
\Sigma^s({\bf k},z)] \sqrt{ [\epsilon^s_0({\bf k})
+ \Re \Sigma^s({\bf k},z)]^2 - |\cos{(\psi_{\bf
k}^s(z) - \phi_{\bf k}^s(z))} \Phi^s({\bf k},z)|^2 }, \nonumber \\ 
\Phi^s({\bf k},z) &=& |\Phi^s({\bf k},z)| \exp{(i\psi_{\bf k}^s(z))}, \
\text{and} \ \ |\cos{(\psi_{\bf k}^s(z) - \phi_{\bf k}^s(z))}| =
\min{\Big[1, \Big| {\epsilon^s_0({\bf k}) + \Re \Sigma^s({\bf k},z)
\over \Phi^s({\bf k},z)} \Big| \Big]}. \label{eq6e} 
\end{eqnarray}
\end{widetext}

The coefficients $\xi_{\bf k}^s(z)$ are obtained through:
\begin{eqnarray}
\cosh{(2\xi_{\bf k}^s(z))} &=& {\epsilon^s_0({\bf k}) + \Re
\Sigma^s({\bf k},z) \over E^s({\bf k},z)}, \label{eq6f} \\
\sinh{(2\xi_{\bf k}^s(z))} &=& - {\cos{(\psi_{\bf k}^s(z) - \phi_{\bf
k}^s(z))} |\Phi^s({\bf k},z)| \over E^s({\bf k},z)}. \nonumber
\end{eqnarray}
The svivon spectral functions are obtained through:
\begin{eqnarray}
A^s({\bf k}, \omega) &\equiv& \Im {\tilde {\cal G}}^s({\bf k},
\omega-i0^+) / \pi \nonumber \\ &=& {\Gamma^s({\bf k}, \omega)/2\pi
\over [\omega - E^s({\bf k},\omega)]^2 + [\half\Gamma^s({\bf k},
\omega)]^2}, \label{eq6g}
\end{eqnarray}
where $\Gamma^s({\bf k},\omega) \equiv 2 \Im {\tilde \Sigma}^s({\bf k},
\omega-i0^+)$. ${\tilde {\cal G}}^s({\bf k},z)$ has positive- and
negative-energy poles \cite{AshkHam} (see below); $\Gamma^s({\bf
k},\omega)$ and $A^s({\bf k}, \omega)$ are positive for $\omega > 0$ and
negative for  $\omega < 0$, and the $\omega$-integral of $A^s$
(including the contributions of the different poles) is normalized to
one. 

The values of $n^s({\bf k})$ and $m^s({\bf k})$ in Eq.~(\ref{eq6b}) are
obtained through \cite{Ashk94}: 
\begin{eqnarray}
n^s({\bf k}) &=& \int d\omega \cosh{(2\xi_{{\bf k}}^s(\omega))} 
A^s({\bf k}, \omega) [ b_{_T}(\omega) + \half ] - \half, \nonumber \\ 
m^s({\bf k}) &=& \int d\omega \sinh{(2\xi_{{\bf k}}^s(\omega))} A^s({\bf
k}, \omega) [ b_{_T}(\omega) + \half ] \nonumber \\ &\ & \times
\exp{(i\phi_{\bf k}^s(\omega))}, \label{eq6h}
\end{eqnarray}
where $b_{_T}(\omega)\equiv 1/[\exp{(\omega/k_{_{\rm B}}T)}-1]$ is the
Bose distribution function; one has $b_{_T}(\omega) + \half \cong \half
\text{sign}(\omega)$ in the high-$|\omega| / k_{_{\rm B}}T$ limit. 

A lagron spectrum of the type presented in Fig.~\ref{fig2} determines
eight degenerate svivon condensates with energy minima at
the points:
\begin{equation}
\pm {{\bf Q}_m \over 2} = \pm \Big( {\bf k}_0 + {\delta {\bf q}_m \over
2} \Big), \label{eq6i0} 
\end{equation}
for each of the four values of $m$ in Eq.~(\ref{eq5}), and the two
possibilities for $\pm {\bf k}_0$ in Eq.~(\ref{eq6b0}). Lagrons at the
$\pm {\bf Q}_m$ points (of the energies in Eq.~(\ref{eq5})) induce
symmetric positive- and (lower weight) negative-energy svivon spectral
branches \cite{AshkHam}, corresponding to the splitting of the Green's
functions poles, due to the inhomogeneities. The phases $\psi_{\bf
k}^s(\omega)$ and $\phi_{\bf k}^s(\omega)$ then satisfy: 
\begin{eqnarray}
\psi_{-{\bf k}}^s(\omega) &=& \psi_{\bf k}^s(-\omega) = \psi_{\bf
k}^s(\omega) + \pi, \nonumber \\
\phi_{-{\bf k}}^s(\omega) &=& \phi_{\bf k}^s(-\omega) = \phi_{\bf
k}^s(\omega). \label{eq6i} 
\end{eqnarray}
Thus, by Eqs.~(\ref{eq6e},\ref{eq6i}), the svivon condensation order
parameter $\Phi^s({\bf k},\omega)$ reverses its sign upon the sign
reversal of either ${\bf k}$ or $\omega$. 

The effects of entropy at high $T$, and of pairing at low $T$ (see
below), drive the system to a combination state of the eight
broken-symmetry condensates, specified in Eq.~(\ref{eq6i0}), at high
$T$, and of four of them, corresponding to one of the two possibilities
for $\pm {\bf k}_0$ in Eq.~(\ref{eq6b0}), at low $T$. Such combinations
reflect fluctuations between the condensates. Consequently, the planar
symmetry of the system remains square at high $T$, while at low $T$ it
is broken to a rectangular symmetry with axes along the ${\hat x} +
{\hat y}$ and ${\hat x} - {\hat y}$ directions, in agreement with
experiment. Svivon spectral functions in the SC state have been
presented elsewhere \cite{AshkHam}. 

\subsection{Unpaired QEs and electrons}

\subsubsection{General features}

The QE spectrum is evaluated treating fluctuations between the combined
svivon condensates adiabatically. Thus, in analogy to the effect of
lattice dynamics on single-electron states in a crystal, the
contribution of coupling to the different svivon condensates to the QE
states is averaged out, and $\underline{ \cal G}^q$ remains diagonal in
the ${\bf k}$ representation (corresponding to the lattice plus a
superstructure determined by ${\bf Q}$---see below), though scattering
is introduced. This conclusion remains approximately correct also under
the random long-range distribution of static inhomogeneities, observed
in a certain range of the phase diagram \cite{Davis1,Hudson}, since the
variation that they introduce to the charge distribution, and thus to
the effective potential of the QEs, is rather small and insufficient to
introduce Anderson localization. 

Since, by Eqs.~(\ref{eq6b0},\ref{eq6i0}), the points $\pm {\bf Q}_m/2$,
of the svivon energy minima within their condensates \cite{AshkHam}, are
symmetrical around $\pm {\bf k}_0 \equiv \pm {\bf Q}/2$, the averaged
svivon spectral contribution to the QE states is symmetrical around
these points as well; consequently, it is invariant \cite{AshkHam} under
a ${\bf k}$-shift of ${\bf Q}$ (note that $2{\bf Q}$ is
reciprocal-lattice vector), and thus it corresponds to an effective AF
order. The symmetry of the electron spectrum is the same as that of the
convoluted QE-svivon spectrum (see Eqs.~(\ref{eq0c},\ref{eq6})). Thus,
since the electron spectrum is invariant under a ${\bf k}$-shift of
$2{\bf Q}$, the QE spectrum must also be invariant under a ${\bf
k}$-shift of ${\bf Q}$. 

By introducing appropriate phase factors to the QE and svivon states,
their ${\bf k}$ values can be shifted by ${\bf k}_0$ or $-{\bf k}_0$, so
that corresponding QE and electron bands appear in the same BZ areas.
Thus, by Eq.~(\ref{eq6i0}), a shift of $-{\bf k}_0$ in the svivon ${\bf
k}$ values results in the shift of the energy minima in their
condensates from $\pm {\bf Q}_m/2$ to $\delta {\bf q}_m/2$ and ${\bf Q}
- \delta {\bf q}_m/2$. 

\begin{figure*}[t] 
\begin{center}
\includegraphics[width=3.25in]{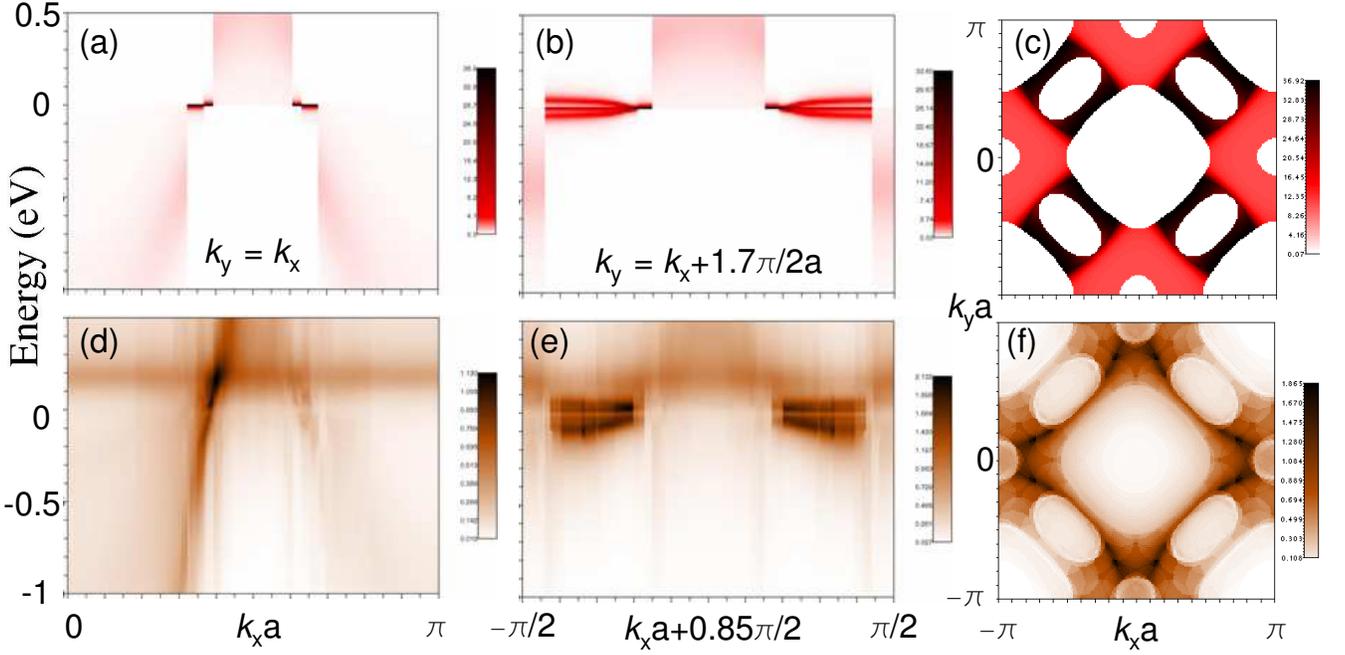}
\end{center}
\caption{The QE (a-c) and electron (d-f) spectral functions in optimally
hole-doped cuprates along a ``line of nodes'' (a,d), along a parallel
line (b,e) shifted by $0.425\pi({\hat y}-{\hat x})$, and on the ``Fermi
surface'' (c,f), determined at $\omega = -0.1 k_{_{\rm B}}T$.} 
\label{fig3}
\end{figure*}

The evaluated QE spectral functions $A^q({\bf k}, \omega)$ for
hole-doped cuprates, corresponding to $n^s = 0.42$ (thus close to
``optimal stoichiometry''), at $k_{_{\rm B}}T = 0.01\;$eV, within the
``hump phase'' \cite{Honma}, are presented in Fig.~\ref{fig3}(a-c). The
calculation is based on a typical svivon spectrum, where the averaged
svivon-condensate spectra \cite{AshkHam} have broad energy minima which
are approximately linear in $T$ in this phase (see below). 

The determination of the svivon and QE linewidths is discussed below,
and they are, approximately, consistent (within the hump phase) with the
marginal-Fermi-liquid (MFL) phenomenology \cite{Varma}. The fact that
the QE spectrum is invariant under a ${\bf k}$-shift of ${\bf Q}$, is
reflected in the existence of {\it equivalent} ``main'' and ``shadow''
QE bands obtained from each other by a ${\bf Q}$ shift.

As was discussed above, the QE spectrum has a mean-field ``bare-band''
part, $\epsilon^q_0({\bf k})$, and a ``dressed'' part, introduced by
self-energy corrections $\Sigma^q = \Sigma^q_s + \Sigma^q_{\lambda}$,
due to coupling to svivon fluctuations (through ${\cal H}^{\prime}$) and
to lagrons (through $\Delta{\cal H}$), respectively. The QE spectral 
functions are obtained through: 
\begin{eqnarray}
A^q({\bf k}, \omega) &\equiv& \Im {\cal G}^q({\bf k}, \omega-i0^+)/\pi
\label{eq8} \\ &=& {\Gamma^q({\bf k}, \omega)/2\pi \over [\omega -
\epsilon^q_0({\bf k}) - \Re \Sigma^q({\bf k},\omega)]^2 + [\Gamma^q({\bf
k}, \omega)/2]^2}, \nonumber 
\end{eqnarray}
where $\Gamma^q({\bf k}, \omega) \equiv 2  \Im \Sigma^q({\bf k}, 
\omega-i0^+)$. The effect of inhomogeneity is that the QE Green's 
function ${\cal G}^q({\bf k},z)$ has more than one pole (see below). The 
$\omega$-integral of $A^q({\bf k}, \omega)$, including the contributions 
of the different poles. is normalized to one. 

The effect of the nearest-neighbor transfer integral $t$ on
$\epsilon^q_0({\bf k})$ drops out (due to opposite-sign contributions of
svivon states shifted by ${\bf Q}$). The effects of $t^{\prime}$ and
$t^{\prime\prime}$ (see Eq.~(\ref{eq0b})) on $\epsilon^q_0({\bf k})$ are
renormalized, due to svivon and $\Delta t$ terms in
Eqs.~(\ref{eq4e},\ref{eq4f}), and they could be, approximately, replaced
by ${\bar t}^{\prime} = 2n^s(t^{\prime} + 4n^sJ)$ and ${\bar
t}^{\prime\prime} = 2n^s(t^{\prime\prime} + 2n^sJ)$, respectively. The
$n^s$ factors here are derived from the summation of expectation values
of products of svivon operators which could be expressed both in terms
of the $n^s({\bf k})$ and of the $m^s({\bf k})$ factors (which have a
major contribution at the vicinity of the svivon energy minima), and
applying Eqs.~(\ref{eq6b}), (\ref{eq6f}) and (\ref{eq6h}). 

The independence of $\epsilon^q_0({\bf k})$ on $t$ reflects the fact
that a QE represents \cite{AshkHam} an approximate electron for which
the creation operator of its svivon component, through Eq.~(\ref{eq0c}),
is replaced by its expectation value. Since averaging the svivon
spectrum over its degenerate condensates corresponds to an effective AF
order (see above), the dependence of the QE spectrum on nearest-neighbor
hopping processes (which disturb this order) must involve svivon
fluctuations, through ${\cal H}^{\prime}$ in Eq.~(\ref{eq6c}). Thus, the
effect of $t$ on it is introduced by the self-energy term: 
\begin{eqnarray}
\Sigma^q_s({\bf k},z) &\cong& {2 \over N^2} \sum_{{\bf k}^{\prime} {\bf
k}^{\prime\prime}} \int d\omega^q A^q({\bf k}^{\prime} , \omega^q)
\Bigg\{ |{\tilde t}( {\bf k}^{\prime} + {\bf k}^{\prime\prime})|^2
\nonumber \\ &\times& |m^s({\bf k}^{\prime} + {\bf k}^{\prime\prime} -
{\bf k})| \bigg\{\int d\omega^s A^s({\bf k}^{\prime\prime} , \omega^s)
\nonumber \\ &\times& \Big\{ \cosh{^2(\xi_{\bf k}^s(\omega^s))} \Big[
{b_{_T}(\omega^s) + f_{_T}(-\omega^q) \over z - \omega^s - \omega^q}
\Big] \nonumber \\ &+& \sinh{^2(\xi_{\bf k}^s(\omega^s))} \Big[
{b_{_T}(\omega^s) + f_{_T}(\omega^q) \over z + \omega^s - \omega^q}
\Big] \Big\} \nonumber \\ &-& {|m^s({\bf k}^{\prime\prime})| \over z -
\omega^q} \bigg\} + |{\tilde t}( {\bf k} + {\bf k}^{\prime\prime})|^2
|m^s({\bf k} + {\bf k}^{\prime\prime} - {\bf k}^{\prime})| \nonumber \\
&\times& \bigg\{\int d\omega^s A^s({\bf k}^{\prime\prime} , \omega^s)
\Big\{ \cosh{^2(\xi_{\bf k}^s(\omega^s))} \label{eq8a} \\ &\times& \Big[
{b_{_T}(\omega^s) + f_{_T}(\omega^q) \over z + \omega^s - \omega^q}
\Big] + \sinh{^2(\xi_{\bf k}^s(\omega^s))} \nonumber \\ &\times& \Big[
{b_{_T}(\omega^s) + f_{_T}(-\omega^q) \over z - \omega^s - \omega^q}
\Big] \Big\} - {|m^s({\bf k}^{\prime\prime})| \over z - \omega^q}
\bigg\} \Bigg\}, \nonumber 
\end{eqnarray}
where $f_{_T}(\omega)\equiv 1/[\exp{(\omega/k_{_{\rm B}}T)}+1]$ is the
Fermi distribution function. 

By Eq.~(\ref{eq6h}), the magnitudes of the $|m^s({\bf
k}^{\prime\prime})|$ terms in the rhs of Eq.~(\ref{eq8a}), and of the
integrated terms that they are subtracted from, differ (in the vicinity
of the svivon energy minima, where the major contribution to the ${\bf
k}^{\prime\prime}$ summation comes from) mainly because of their
different energy denominators. When the signs of the denominators of
these terms are opposite, they {\it both} contribute to ``pushing'' the
QE energies towards zero. Consequently, the $\Sigma^q_s$-induced
self-consistent renormalization of the QE energies is a shift towards
zero which becomes larger closer to zero. 

This results in BZ areas of ``flat'' QE bands which would have been at
energies of the $k_{_{\rm B}}T$ scale, if it were not for the effect of
QE-lagron coupling (introducing to them the structure discussed below).
The transition between these low-energy areas and the higher-energy BZ
areas is through an almost discontinuous wide-energy zone, as is viewed
in Fig.~\ref{fig3}(a-b). This peculiar spectral structure reflects the
distinction between strongly renormalized low-energy QEs which are
subject to hopping processes maintaining the stripe-like
inhomogeneities, and high-energy QEs which are not; it is primarily
determined by the contribution of the hopping parameter $t$ to ${\cal
H}^{\prime}$ in Eq.~(\ref{eq6c}). 

The corresponding electron spectral functions $A^d({\bf k}, \omega)
\equiv \Im {\cal G}^d({\bf k}, \omega-i0^+)/\pi$, obtained on the basis
of Eq.~(\ref{eq6}), are presented in Fig.~\ref{fig3}(d-f). The high
weight of the svivon poles around their minimal energies \cite{AshkHam}
results in band-like features in the electron spectrum which follow the
QE band (see below). 

The effect of the multiple-scattering FL-like electron poles in
Eq.~(\ref{eq6}) is the removal of the equivalence between the main and
the shadow electron bands, as is observed in experiment \cite{Campuzano,
Mesot}. However, non-FL behavior associated with the QE Fermi surface
(FS) in Fig.~\ref{fig3}(c) persists, contributing the hole and electron
pocket-like features appearing in Fig.~\ref{fig3}(f), and detected in
experimental data \cite{Chang} mainly in the underdoped regime. The
existence of such FS pockets has also been confirmed in the observation
of quantum oscillations \cite{Taillefer, Yelland}. On the other hand,
the FL-like contribution to the normal-state electron spectrum is
dominant in the overdoped regime, where quantum oscillations typical of
a large FS have been observed \cite{Hussey}. 

The low-energy flat QE bands, and the almost discontinuous transition
from them to the high-QE-energy BZ areas are modified in the electron
spectrum (see Fig.~\ref{fig3}(d-e)) to low-energy kinks followed by
``waterfalls'', in agreement with experiment \cite{Xie, Shen1}. As was
discussed above, this spectral anomaly results from the renormalization
of the QEs, such that their low-energy excitations maintain the
stripe-like inhomogeneities. 

\subsubsection{Low-energy range}

The self-energy term, introduced by QE-lagron coupling through
$\Delta{\cal H}$ in Eq.~(\ref{eq4i}) is expressed (for un-paired QEs)
as: 
\begin{eqnarray}
\Sigma^q_{\lambda}({\bf k},z) &\cong& {1 \over N} \sum_{\bf q} \int
d\omega^q A^q({\bf k}-{\bf q}, \omega^q) \nonumber \\ &\ &\times
F_{_T}({\bf q}, \omega^q, z), \ \ \text{where} \label{eq7} \\ 
F_{_T}({\bf q}, \omega^q, z) &=& |\gamma({\bf q})|^2 \bigg[
{b_{_T}(\omega^{\lambda}({\bf q})) + f_{_T}(-\omega^q) \over z -
\omega^{\lambda}({\bf q}) - \omega^q} \nonumber \\ &\ &\ \ \ \ \ \ \ \ \ +
{b_{_T}(\omega^{\lambda}({\bf q})) + f_{_T}(\omega^q) \over z +
\omega^{\lambda}({\bf q}) - \omega^q} \bigg]. \nonumber 
\end{eqnarray}

Within the low-QE-energy BZ areas, the $\omega$-dependence of
$\Sigma^q_s$ is weak, and its imaginary part is relatively small.
Consequently $\epsilon^q_0$ could be renormalized there to include the
effect of $\Sigma^q_s$, and the anomalous behavior of $A^q({\bf k},
\omega))$ is obtained approximating $\Sigma^q$ in Eq.~(\ref{eq8}) by the
QE-lagron term $\Sigma^q_{\lambda}$ in Eq.~(\ref{eq7}).

As is viewed in Fig.~\ref{fig2}, one could specify the lagrons according
to five ranges in the BZ: (1) those at the four energy minima points
${\bf Q}_m$ in Eq.~(\ref{eq5}), referred to as ``${\bf Q}_m$ lagrons'';
(2) those at the lagron ``extended saddle point'' (ESP) around point
${\bf Q}$, referred to as ``${\bf Q}$-ESP lagrons'' which introduce an
energy scale $\sim$$\omega^{\lambda}({\bf Q})$; (3) those of energies
$\lta \omega^{\lambda}({\bf Q})$, around the ${\bf Q}_m$ points,
referred to as ``${\bf Q}_m$-vicinity lagrons''; (4) those around the
high-energy saddle points at $(\pi/a){\hat x}$ and $(\pi/a){\hat y}$,
referred to as ``high-energy SP lagrons''; (5) those within the rest of
the lagron spectrum, referred to as ``continuum lagrons'' which occupy
most of their phase space. 

Analogous expressions to Eq.~(\ref{eq7}), where $A^q$, $\omega^q$ and
$f_{_T}(\pm\omega^q)$ are replaced by $A^s$, $\omega^s$ and
$-b_{_T}(\pm\omega^s)$ (and $\cosh{^2}$ and $\sinh{^2}$ factors are
introduced---see below), are obtained for the contribution of
$\Delta{\cal H}$ in Eq.~(\ref{eq4i}) to the svivon self-energy terms,
appearing in Eq.~(\ref{eq6d}). The ${\bf Q}$-ESP lagrons then help
stabilize high-spectral-weight svivon states somewhat below
$\omega^{\lambda}({\bf Q})$ in the SC state. Consequently
\cite{AshkHam}, the resonance-mode energy $E_{_{\rm res}}$ is somewhat
above $\omega^{\lambda}({\bf Q})$ \cite{Eres}. 
 
The spectral functions of low-energy QEs, coupled to each other through
${\bf Q}_m$ lagrons, split (due to the large values of the Bose
functions in Eq.~(\ref{eq7})---see Eq.~(\ref{eq5}) and the discussion
above) into positive- and negative-energy peaks, referred to as
``humpons''; such a splitting is expected due to the spin-density waves
(SDW) associated with the stripe-like inhomogeneities which are
generated by the lagron spectrum in Fig.~\ref{fig2}. 

\begin{figure}[t] 
\begin{center}
\includegraphics[width=3.25in]{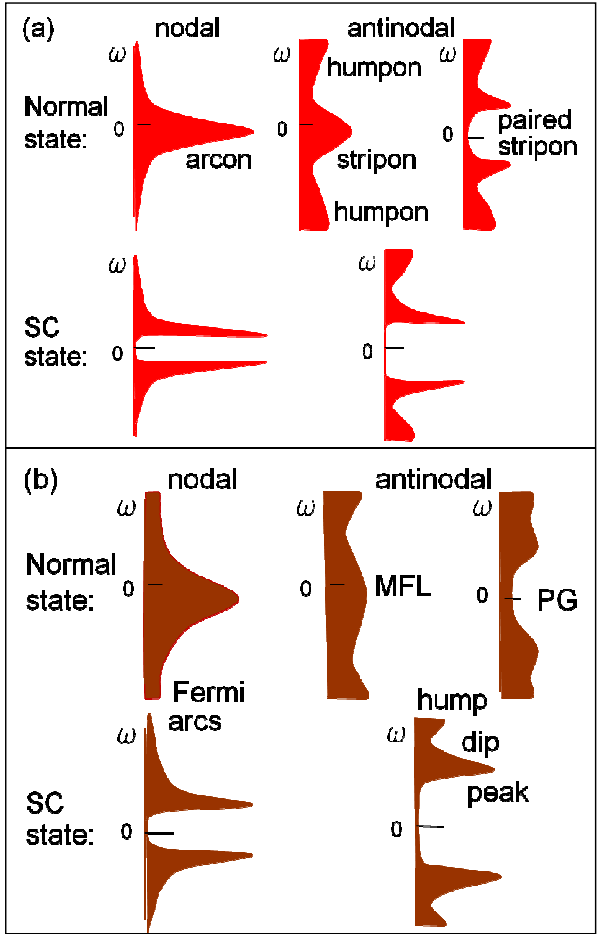}
\end{center}
\caption{Typical low-energy QE (a) and electron (b) spectral functions
in the normal and SC states, in nodal (arcon, Fermi arcs) and antinodal
(humpon, stripon, peak, dip, hump) points in hole-doped cuprates. Their
structures are presented in the MFL, PG, and SC states (see discussion
in the text).} 
\label{fig4}
\end{figure}

As was discussed above, within some range of temperatures, free energy
(where the effects of both energy and entropy are accounted for) is
minimized for a phase of fluctuating inhomogeneities. For unpaired QEs,
a third QE spectral peak appears in this phase, between the positive-
and negative-energy humpons (as is sketched in Fig.~\ref{fig4}(a)), and
it is referred to as a ``stripon''; it represents low-energy charge
carriers due to the fluctuating charged stripes. 

A stripon at point ${\bf k}$ is associated with the pole of ${\cal
G}^q({\bf k},z)$ which corresponds to lattice periodicity, while the
humpons result from poles corresponding to the inhomogeneities. The
energies $\sim$$\omega_{_{\rm H}}$ of the humpon peaks correspond to the
energy associated with the inhomogeneities, and could be as high as
$\sim$$J$. As is discussed below, the stripon peak is sharper than the
humpon peaks for $k_{_{\rm B}}T \ll \omega_{_{\rm H}}$, and its width
increases with $T$. 

The $\gamma({\bf q})$ coefficients in Eq.~(\ref{eq4i}) remain finite for
${\bf q} \to {\bf Q}_m$ (see Eq.~(\ref{eq4j1}) and the above discussion)
and their values there, as well as the values of $\omega^{\lambda}({\bf
Q}_m)$ in Eq.~(\ref{eq5}), are self-consistently determined to yield
such humpon energies due to coupling to lagrons, and specifically those
around the ${\bf Q}_m$ points, through Eq.~(\ref{eq7}). The contribution
of the ${\bf Q}$-ESP lagrons results in $\omega_{_{\rm H}} >
\sim$$\omega^{\lambda}({\bf Q})$. 

Within the low-QE-energy BZ areas there are ``nodal areas'', close to
the lines of nodes $y = \pm x$, and ``antinodal areas'', close to
circles of radii of about $(\pi/4a)$--$(\pi/3a)$ around the
$(\pi/a){\hat x}$ and $(\pi/a){\hat y}$ points. As can be seen in
Fig.~\ref{fig3}(c), a difference between the QEs in the antinodal and
nodal areas is that the first, but {\it not} the latter (if they are
close enough to the lines of nodes), can be coupled to other low-energy
QEs through (up to four) ${\bf Q}_m$ lagrons (see Eq.~(\ref{eq5})).

Consequently, the spectral functions of unpaired QEs within the nodal
areas (referred to as ``arcons'') are characterized by a simple peak
(due to one-${\cal G}^q$ pole per ${\bf k}$ point), of the $k_{_{\rm
B}}T$ energy scale, as is sketched in Fig.~\ref{fig4}(a). On the other
hand, the spectral functions of QEs within the antinodal areas are
characterized by the above humpon--stripon--humpon structure. As will be
discussed below, these spectral features of the QEs (specified as
arcons, stripons and humpons) are associated with different physical
features in the cuprates. 

Thus a two-component scenario is described here, concerning the
symmetries of the low-energy QEs in the nodal and antinodal areas, due
to the different number of poles in ${\cal G}^q({\bf k},z)$ they
correspond to. An increase in $T$ results in an increase in the spectral
weight within the stripons peaks in the antinodal areas (on the expense
of those within the humpon peaks), and in the shift of the borderline
between the nodal and antinodal areas (concerning the symmetry of the
QEs) towards the antinodal ones. When $k_{_{\rm B}}T$ crosses
$\omega_{_{\rm H}}$, a crossover occurs from the above
inhomogeneity-derived two-component scenario to a homogeneous one, where
all the low-energy QEs correspond to one-${\cal G}^q$ pole per ${\bf k}$
point. 

For $k_{_{\rm B}}T \ll \omega_{_{\rm H}}$, the arcon and stripon peaks
at different ${\bf k}$ points can be treated as an almost flat band. The
energy center of this band is close to the centers of most of the peaks,
determined through (see Eq.~(\ref{eq8})) the relation: $\omega =
\epsilon^q_0({\bf k}) + \Re \Sigma^q({\bf k},\omega)$. The $T$
dependence of the band center approximately corresponds to that of the
effective QE chemical potential $\mu - \lambda$ (see Eq.~(\ref{eq4f}));
thus \cite{Ashk01}, this $T$ dependence scales with $k_{_{\rm B}}T$, if
the band is not half filled and its width in not larger than
$\sim$$k_{_{\rm B}}T$, and it is weaker, otherwise. 

The above low-energy spectral features of the QEs are consistent with
the nodal--antinodal dichotomy occurring in the cuprates. Furthermore,
they provide convenient QPs to study, discuss, and analyze (see below)
their peculiar low-energy spectrum and its consequences on the
derivation of their anomalous physical properties. 

\subsubsection{QE scattering rates}

The scattering rates $\Gamma^q({\bf k}, \omega) \equiv 2  \Im
\Sigma^q({\bf k}, \omega-i0^+)$ of unpaired low-energy QEs can be
expressed, on the basis of Eq.~(\ref{eq7}), as: 
\begin{eqnarray}
\Gamma^q({\bf k},\omega) &\cong& {2\pi \over N} \sum_{\bf q}
|\gamma({\bf q})|^2 \big\{ A^q({\bf k}-{\bf q}, \omega -
\omega^{\lambda}({\bf q})) \nonumber \\ &\ &\times
[b_{_T}(\omega^{\lambda}({\bf q})) + f_{_T}(\omega^{\lambda}({\bf q}) -
\omega)] \nonumber \\ &\ & + A^q({\bf k}-{\bf q}, \omega +
\omega^{\lambda}({\bf q})) \nonumber \\ &\ &\times
[b_{_T}(\omega^{\lambda}({\bf q})) + f_{_T}(\omega^{\lambda}({\bf q}) +
\omega)] \big\}, \label{eq9}
\end{eqnarray}
where an approximately ${\bf q}$-independent $|\gamma({\bf q})|^2$ could
be assumed (see Eq.~(\ref{eq4j1}) and the above discussion). 

Eq.~(\ref{eq9}) is applied to study the behavior of $\Gamma^q$,
specifically in the low- and high-$|\omega| / k_{_{\rm B}}T$ limits. As
can be viewed in Fig.~\ref{fig3}(a-b), the ${\bf k}-{\bf q}$ points
where $A^q$ contributes to Eq.~(\ref{eq9}), for $|\omega| \lta 0.3\;$eV,
consist mainly of those within the low-QE-energy BZ areas, shown in
Fig.~\ref{fig3}(c). The $\omega$ dependencies of $A^q({\bf k},\omega)$
within the low-energy arcon and stripon peaks, the mid-energy humpon
peaks, and the high-energy areas, are specified in terms of canonical
functions $A^q_{\rm le}(\omega)$, $A^q_{\rm me}(\omega)$ and $A^q_{\rm
he}(\omega)$, respectively. 

The energies at the maxima of $A^q_{\rm he}(\omega)$ are generally too
high to have a significant contribution to $\Gamma^q$, through
Eqs.~(\ref{eq8},\ref{eq9}), within the ranges of $T$ and $\omega$
studied here. The humpon peaks (characterized by $A^q_{\rm me}$) exist
within the low-energy antinodal areas in the $k_{_{\rm B}}T \ll
\omega_{_{\rm H}}$ regime, and if also $|\omega| \ll \omega_{_{\rm H}}$,
their role is negligible, similarly to that of $A^q_{\rm he}$; for
$|\omega| \gta \omega_{_{\rm H}}$, they are approached as a case of
$A^q_{\rm le}$. In the $k_{_{\rm B}}T \gta \omega_{_{\rm H}}$ regime the
humpon and stripon peaks are merged into an arcon peak which is again
approached as a case of $A^q_{\rm le}$. 

\begin{figure}[t] 
\begin{center}
\includegraphics[width=3.25in]{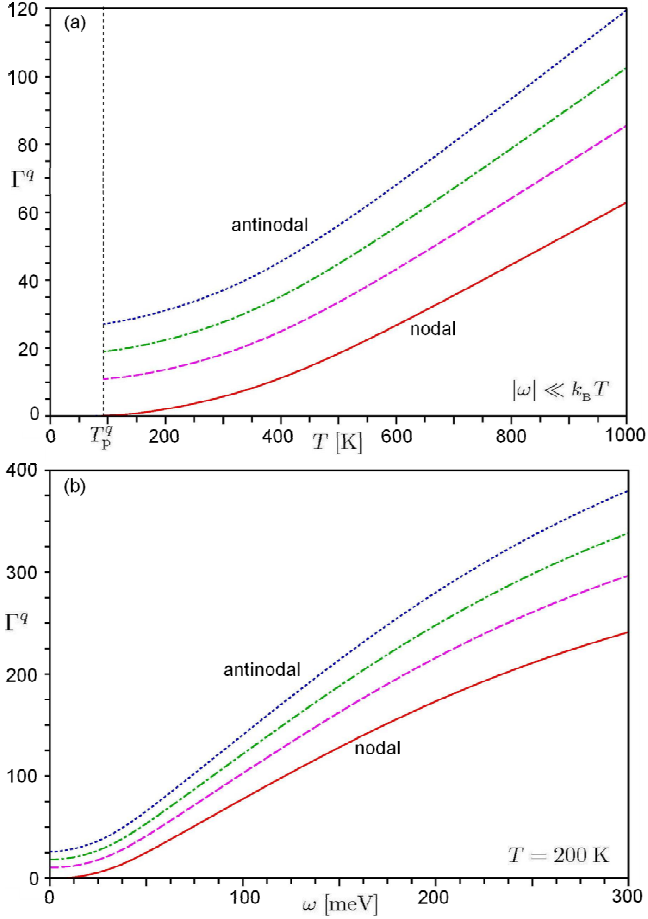}
\end{center}
\caption{Typical results (in arbitrary units) for the QE scattering
rates $\Gamma^q$, at \cite{pairing} $T > T^q_{\rm p}$, in ${\bf k}$
points ranging between the lines of nodes and the antinodal BZ areas, as
(a) a function of $T$ for $|\omega| < k_{_{\rm B}} T^q_{\rm p}$, and (b)
a function of $\omega$, for $k_{_{\rm B}}T < \Gamma^q_{0{\rm a}}$ (see
discussion in the text).} 
\label{fig5}
\end{figure}

The evaluation of $\Gamma^q({\bf k}, \omega)$ is detailed in Appendix A;
typical results are presented in Figs.~\ref{fig5}(a-b), as functions of
$T$ and $\omega$, in the low $|\omega| / k_{_{\rm B}}T$ limit, and at
$T$ above \cite{pairing} $T^q_{\rm p}$ and below $\Gamma^q_{0{\rm a}} /
k_{_{\rm B}}$ (see Appendix A) and $\omega^{\lambda}({\bf Q}) / k_{_{\rm
B}}$, respectively; these results include ${\bf k}$ points ranging
between the nodal and antinodal BZ areas. Approximate linear
dependencies of $\Gamma^q({\bf k}, \omega)$ on $T$ and on $\omega$ are
found in the low- and high-$|\omega| / k_{_{\rm B}}T$ limits,
respectively, above a ${\bf k}$-dependent low-energy scale
$\omega_{_{\rm L}}({\bf k})$. The high-$|\omega|$ extent of this linear
dependence is limited by the high-energy extent of the lagron spectrum
in Fig.~\ref{fig2} (which could exceed $\sim$$0.3\;$eV), and the
high-$T$ extent is limited by the phase stability. The contribution of
humpons to $\Gamma^q$ results in the increase of the slopes of the
$\Gamma^q$ {\it vs} $k_{_{\rm B}}T$ and $|\omega|$ curves when their
increasing values approach $\omega_{_{\rm H}}$. 

For $k_{_{\rm B}}T\; \&\; |\omega| \ll \omega_{_{\rm L}}({\bf k})$,
$\Gamma^q({\bf k}, \omega)$ can be approximated as $\Gamma^q_0({\bf k})$
(determined in Eq.~(\ref{eq10a})) which is close zero on the lines of
nodes, and to $\Gamma^q_{0{\rm a}}$ in the antinodal areas. As is
discussed in Appendix A, the value of $\omega_{_{\rm L}}({\bf k})$ is
close to $\Gamma^q_{0{\rm a}}$ in the antinodal areas, and is somewhat
smaller than \cite{Eres} $\omega^{\lambda}({\bf Q}) \simeq {3 \over 4}
E_{_{\rm res}}$ on the lines of nodes; $\omega^{\lambda}({\bf Q})$ is
smaller than $\Gamma^q_{0{\rm a}}$ in the heavily underdoped regime, and
their values cross each other when the doping level is increased. 

\subsubsection{Svivon scattering rates}

The svivon scattering rates $\Gamma^s$ include a term
$\Gamma^s_{\lambda}$, due to their coupling to lagrons through
$\Delta{\cal H}$ in Eq.~(\ref{eq4i}), yielding an analogous expression
to that for $\Gamma^q$ in Eq.~(\ref{eq9}). However, also a term
$\Gamma^s_q$, due to svivon-QE coupling, through ${\cal H}^{\prime}$ in
Eq.~(\ref{eq6c}), has a low-energy contribution to $\Gamma^s$ which
becomes significant above \cite{pairing} $T^q_{\rm p}$. Since $A^s({\bf
k}, \omega)$ includes symmetric positive- and negative-energy branches
(of different weights), and the sign of $\Gamma^s({\bf k}, \omega)$ is
that of $\omega$, one has (differently from QEs---see
Eq.~(\ref{eq10a})): 
\begin{equation}
\Gamma^s({\bf k}, \omega \to 0) = 0. \label{eq13}
\end{equation}
Using Eqs.~(\ref{eq6b}), (\ref{eq6c}) and (\ref{eq6e}), one can express
$\Gamma^s$ (for unpaired QEs) as: 
\begin{widetext}
\begin{eqnarray}
\Gamma^s({\bf k},\omega) &=& \Gamma^s_q({\bf k},\omega) +
\Gamma^s_{\lambda}({\bf k},\omega), \ \ \ \text{where} \nonumber \\
\Gamma^s_q({\bf k},\omega) &\cong& {2\pi \cosh{(2\xi_{\bf k}^s(\omega))}
\over N^2} \sum_{{\bf k}^{\prime} {\bf k}^{\prime\prime}} |{\tilde
t}({\bf k} + {\bf k}^{\prime\prime})|^2 |m^s({\bf k} + {\bf
k}^{\prime\prime} - {\bf k}^{\prime})| \int d\omega^{\prime} A^q({\bf
k}^{\prime}, \omega^{\prime}) A^q({\bf k}^{\prime\prime},
\omega^{\prime} - \omega) [f_{_T}(\omega^{\prime} - \omega) -
f_{_T}(\omega^{\prime})], \nonumber \\ 
\Gamma^s_{\lambda}({\bf k},\omega) &\cong& {2\pi \cosh{(2\xi_{\bf
k}^s(\omega))} \over N} \sum_{\bf q} |\gamma({\bf q})|^2 \big\{ A^s({\bf
k}-{\bf q}, \omega - \omega^{\lambda}({\bf q})) \cosh{^2(\xi_{{\bf k} -
{\bf q}}^s(\omega - \omega^{\lambda}({\bf q})))}
[b_{_T}(\omega^{\lambda}({\bf q})) - b_{_T}(\omega^{\lambda}({\bf q}) -
\omega)] \nonumber \\ &\ & + A^s({\bf k}-{\bf q}, \omega +
\omega^{\lambda}({\bf q})) \cosh{^2(\xi_{{\bf k} - {\bf q}}^s(\omega +
\omega^{\lambda}({\bf q})))} [b_{_T}(\omega^{\lambda}({\bf q})) -
b_{_T}(\omega^{\lambda}({\bf q}) + \omega)] \big\} \label{eq12}  \\ &\ &
+ A^s({\bf k}-{\bf q}, \omega^{\lambda}({\bf q}) - \omega)
\sinh{^2(\xi_{{\bf k} - {\bf q}}^s(\omega^{\lambda}({\bf q}) - \omega))}
[b_{_T}(\omega^{\lambda}({\bf q}) - \omega) -
b_{_T}(\omega^{\lambda}({\bf q}))] \nonumber \\ &\ & + A^s({\bf k}-{\bf
q}, - \omega - \omega^{\lambda}({\bf q})) \sinh{^2(\xi_{{\bf k} - {\bf
q}}^s(- \omega - \omega^{\lambda}({\bf q})))}
[b_{_T}(\omega^{\lambda}({\bf q}) + \omega) -
b_{_T}(\omega^{\lambda}({\bf q}))] \big\}. \nonumber 
\end{eqnarray}
\end{widetext}

Let ${\bar \epsilon}^s({\bf k})$ be the positive svivon energies
\cite{AshkHam} (within their condensates) at the maxima of $A^s({\bf k},
\omega)$ ($-{\bar \epsilon}^s({\bf k})$ are the lower-weight negative
energies---see above); let ${\bar \epsilon}^s_{\rm min} = {\bar
\epsilon}^s({\bf k}_{\rm min})$ be the minimum of ${\bar
\epsilon}^s({\bf k})$ (within each condensate there are two ${\bf
k}_{\rm min}$ points, separated from each other by a ${\bf Q}_m$
vector). Above \cite{pairing} $T^q_{\rm p}$, the $T$ dependence of
${\bar \epsilon}^s({\bf k})$ is determined by the variation of $\lambda$
in Eq.~(\ref{eq4f}) to maintain a $T$-independent $n^s$ through
Eqs.~(\ref{eq6b}, \ref{eq6h}). By Eq.~(\ref{eq6e}), this results in the
approximate scaling: 
\begin{eqnarray}
{\bar \epsilon}^s({\bf k}) &\propto& \sqrt{(c^s |{\bf k} - {\bf k}_{\rm
min}|)^2 + (T-T_0)^2}, \nonumber \\ &\ & \text{for } |{\bf k} - {\bf
k}_{\rm min}| < k_{\rm max}, \label{eq13a} 
\end{eqnarray}
where $c^s$ and $T_0$ are constants (one has $T_0 < T^q_{\rm p}$, and it
could be in principle either positive or negative---see below), and
$k_{\rm max}$ represents the radius of an approximate circle around
${\bf k}_{\rm min}$, in the BZ, where there is a noticeable effect of
$\Phi^s$ on ${\bar \epsilon}^s({\bf k})$. 

By Eq.~(\ref{eq13a}), $\bar \epsilon^s_{\rm min}$ approximately scales
with $T-T_0$, such that it remains sufficiently larger than $k_{_{\rm
B}}T$ (as \cite{AshkHam} below $T^q_{\rm p}$) that $b_{_T}(|\omega|)$ is
relatively small (though not negligible), within the range where ${\bf k} 
\simeq {\bf k}_{\rm min}$, and $|A^s({\bf k}, \omega)|$ is significant. 
On the other hand, Eq.~(\ref{eq13a}) yields that the effect of $T$ on 
${\bar \epsilon}^s({\bf k}) \gg \bar \epsilon^s_{\rm min}$ is through a 
minor additional term scaling, approximately, with $(T-T_0)^2$, and in 
such ${\bf k}$ points one has $b_{_T}(|\omega|) \ll 1$ within the 
$\omega$ range where $|A^s({\bf k}, \omega)|$ has a considerable 
magnitude. 

When $T$ is increased, the minima of ${\bar \epsilon}^s({\bf k})$ around
${\bar \epsilon}^s_{\rm min}$ become broader and flatter, and since
$|T_0/T|$ is decreased, the ratio $\bar \epsilon^s_{\rm min} / T$
changes as is necessary in order to maintain a constant $n^s$, through
Eqs.~(\ref{eq6b}, \ref{eq6h}) (under the opposing effects of $T$ on it
through the minima and $\cosh{(2\xi_{\bf k}^s(\omega)}$---see below---as
well as the effect of the linewidth of ${\bar \epsilon}^s({\bf k})$, for
${\bf k} \simeq {\bf k}_{\rm min}$). 

By Eqs.~(\ref{eq6e}), (\ref{eq6f}) and (\ref{eq13a}), the
$\cosh{(2\xi_{\bf k}^s(\omega))}$ factors, appearing in
Eq.~(\ref{eq12}), can be approximated as: 
\begin{eqnarray}
\cosh{(2\xi_{\bf k}^s(\omega))} &\simeq& \sqrt{(c^s {\bf k}_{\rm max})^2
+ (T-T_0)^2 \over (c^s |{\bf k} - {\bf k}_{\rm min}|)^2 + (T-T_0)^2},
\label{eq13b} \\ &\ & \text{for } |{\bf k} - {\bf k}_{\rm min}| < k_{\rm
max}, \ \text{and } |\omega| \lta {\bar \epsilon}^s({\bf k}). \nonumber 
\end{eqnarray}
These factors are largest in ${\bf k}$ points at the low-svivon-energy
(LE) BZ areas, close to the ${\bf k}_{\rm min}$ points, where
$\Gamma^s({\bf k},\omega)$ is significant in the $|\omega| \lta {\bar
\epsilon}^s_{\rm min}$ range, and by Eq.~(\ref{eq13b}) they,
approximately, scale there with $1/(T-T_0)$, for $|\omega| \lta {\bar
\epsilon}^s({\bf k})$. On the other hand, in ${\bf k}$ points at the
mid-svivon-energy (ME) BZ areas, where $\Gamma^s({\bf k},\omega)$ is
significant in the $\sim$${\bar \epsilon}^s_{\rm min} < |\omega| <
\sim$$2J$ range, and Eq.~(\ref{eq13b}) is still valid, it yields a weak
$T$ dependence for the $\cosh{(2\xi_{\bf k}^s(\omega))}$ factors,
through a term $\propto$$(T-T_0)^2$. The $\cosh{(2\xi_{\bf
k}^s(\omega))} \cong 1$ limit specifies ${\bf k}$ points at
high-svivon-energy (HE) BZ areas, where the effect of condensation is
missing and Eqs.~(\ref{eq13a}, \ref{eq13b}) are invalid; $\Gamma^s({\bf
k},\omega)$ is significant there in the high-$\omega$ and -$T$ limits.
At the LE and ME areas one has $\cosh{(2\xi_{\bf k}^s(\omega))} \cong 1$
for $|\omega| \gg {\bar \epsilon}^s({\bf k})$. 

\begin{figure}[t] 
\begin{center}
\includegraphics[width=3.25in]{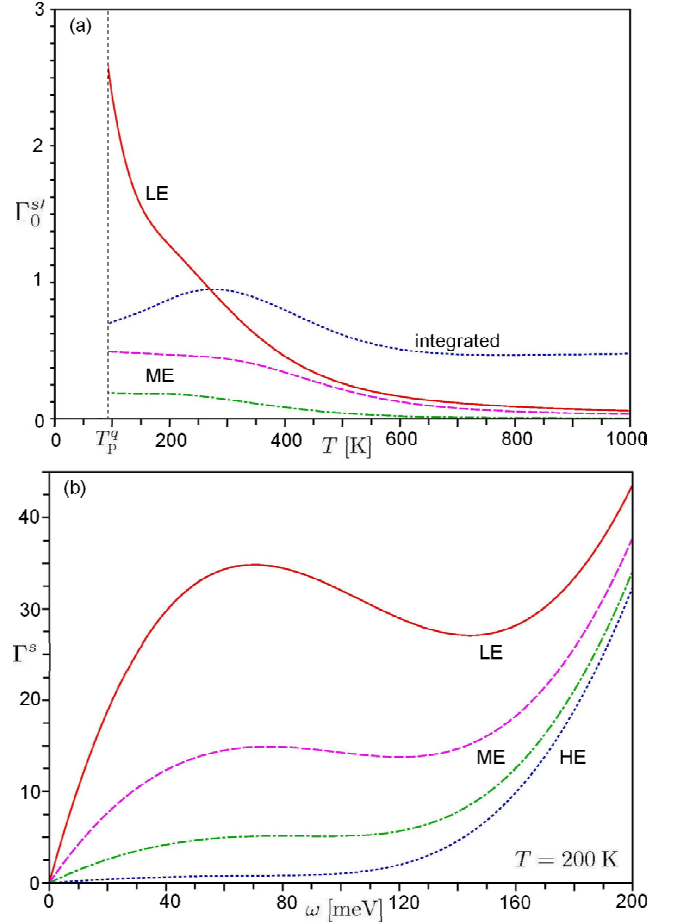}
\end{center}
\caption{Typical results (in arbitrary units), at \cite{pairing} $T >
T^q_{\rm p}$, for (a) the $T$ dependence of the $\omega$ derivative
$\Gamma^{s\prime}_0$ of the svivon scattering rates $\Gamma^s$ at
$\omega = 0$, in LE and ME svivon ${\bf k}$ points (see discussion in
the text) and of its ${\bf k}$-integrated value, and for (b) the
$\omega$ dependence of $\Gamma^s$, in LE, ME and HE svivon ${\bf k}$
points.} 
\label{fig6}
\end{figure}

Thus, one could classify the svivon spectrum within the BZ according to
LE, ME and HE areas, and crossover areas which could be, approximately,
split between them. Since the minima of ${\bar \epsilon}^s({\bf k})$
around ${\bar \epsilon}^s_{\rm min}$ become broader and flatter when $T$
is increased, the LE areas are growing with $T$ on the expense of the ME
areas. By Eq.~(\ref{eq13a}), the size of the LE areas scales with
$(T-T_0)^2$, and thus the size of the ME areas is decreased by a
relatively minor term (for $k_{_{\rm B}}T \ll \sim$$J$) which scales
with $(T-T_0)^2$. 

The evaluation of $\Gamma^s({\bf k},\omega)$ is detailed in Appendix B.
In the $|\omega| \ll {\bar \epsilon}^s_{\rm min}$ regime, one has
$\Gamma^s({\bf k},\omega) \propto \omega$ for \cite{pairing} $T >
T^q_{\rm p}$ (which is enabled by Eq.~(\ref{eq13})). This results,
through Eq.~(\ref{eq14}), in the existence of low-energy tails in
$A^s({\bf k},\omega)$, and thus in the disappearance of the spin gap and
the sharp resonance mode, existing \cite{AshkHam} at $T < T^q_{\rm p}$,
in agreement with experiment. 

Typical results for the $T$ dependence of $\Gamma^{s\prime}_0({\bf k})
\equiv \lim_{_{\omega \to 0}}[\Gamma^s({\bf k},\omega)/\omega]$ in the
LE and ME areas are presented in Fig.~\ref{fig6}(a). For $k_{_{\rm B}}T
\ll \Gamma^q_{0{\rm a}}$ (see Eq.~(\ref{eq10a})), $\Gamma^{s\prime}_0$
is characterized by scaling with $1/(T-T_0)$ in the LE areas, and by a
constant plus a minor term $\propto (T-T_0)^2$ in the ME areas (it
practically vanishes in the HE areas). For $k_{_{\rm B}}T \gg
\Gamma^q_{0{\rm a}}$, $\Gamma^{s\prime}_0$ is characterized by scaling
with $1/[(T-T_0)^2(1-T_0/T)]$ in the LE areas, with
$1/[(T-T_0)(1-T_0/T)]$ in the neighboring ME areas (which turn into LE
areas at higher $T$---see Fig.~\ref{fig6}(a)), and with $1/T^2$ in
higher-energy ME areas. A typical $T$ dependence of the ${\bf
k}$-integrated $\Gamma^{s\prime}_0({\bf k})$ is also presented in
Fig.~\ref{fig6}(a); it is characterized by some linear increase with $T$
at low $T > T^q_{\rm p}$, followed by a decrease around $k_{_{\rm B}}T
\simeq \Gamma^q_{0{\rm a}}$, and a weak dependence at higher $T$. 

Typical results for $\Gamma^s({\bf k},\omega)$, as a function of
$\omega$ (at constant $T > T^q_{\rm p}$), within the LE, ME and HE
areas, are presented in Fig.~\ref{fig6}(b). They are characterized (see
Appendix B) by scaling with $\omega$ for $|\omega| < \sim$${\bar
\epsilon}^s({\bf k})$ (which is somewhat larger than $k_{_{\rm B}}T$ in
the LE areas), with a slope which is approximately $\propto 1 / {\bar
\epsilon}^s({\bf k})$, for low ${\bar \epsilon}^s({\bf k})$, and is
smaller for high ${\bar \epsilon}^s({\bf k})$; in the case that ${\bar
\epsilon}^s({\bf k}) < \sim$$J$, this is followed by a decrease in the
$|\Gamma^s|$ {\it vs} $|\omega|$ slope at $|\omega| \simeq {\bar
\epsilon}^s({\bf k})$, and for small ${\bar \epsilon}^s({\bf k})$, also
by a wide maximum of $|\Gamma^s|$, centered somewhat above $|\omega|
\simeq {\bar \epsilon}^s({\bf k})$, and its decrease above it; the
$|\Gamma^s|$ {\it vs} $|\omega|$ slope rises at ${\bar \epsilon}^s({\bf
k}) \simeq J$, and an approximately linear increase of $|\Gamma^s|$ with
$|\omega|$ is maintained at higher $|\omega|$ values, up to about the
high-energy limits of the lagron spectrum (see Fig.~\ref{fig2}). 

\subsubsection{Non-FL electron scattering rates}

Let $\epsilon^q_{\rm p}({\bf k}) + \half i\Gamma^q_{\rm p}({\bf k})$ and
$\epsilon^s_{\rm p}({\bf k}) + \half i\Gamma^s_{\rm p}({\bf k})$ be the
poles of $\underline{\cal G}^q$ and $\underline{ \cal G}^s$,
respectively; they have weight factors $w^q_{\rm p}({\bf k})$ and
$w^s_{\rm p}({\bf k})$ normalized as: $\sum_{\rm p} w^q_{\rm p}({\bf k}) 
= \sum_{\rm p} w^s_{\rm p}({\bf k}) = 1$. $\Gamma^q_{\rm p}({\bf k})$ 
and $\Gamma^s_{\rm p}({\bf k})$ are related to the scattering rates 
$\Gamma^q({\bf k},\omega)$ and $\Gamma^s({\bf k},\omega)$, analyzed 
above, through the Kramers--Kronig relation \cite{Vliet, Mahan}:
\begin{equation}
\int {d\omega |\Gamma({\bf k},\omega)| / 2\pi \over (\omega -
\epsilon_{\rm p}({\bf k}))^2 + (\Gamma_{\rm p}({\bf k})/2)^2} =1.
\label{eq18a} 
\end{equation}
This indicates that the dependence of $\Gamma^q_{\rm p}({\bf k})$ on $T$
and $\epsilon^q_{\rm p}({\bf k})$ is close to that of $\Gamma^q({\bf
k},\omega)$ on $T$ and $\omega \simeq \epsilon^q_{\rm p}({\bf k})$. 

The elements of the approximate electron Green's-function matrix
$\underline{\cal G}^d_0$ (see Eq.~(\ref{eq6})), represented by the
bubble diagrams in Fig.~\ref{fig1}(b), could be expressed as
\cite{Vliet, Mahan}: 
\begin{widetext}
\begin{eqnarray}
{\cal G}^d_0({\bf k},z) &=& \sum_{{\rm p} {\rm p}^{\prime} {\bf
k}^{\prime}} w^s_{{\rm p}^{\prime}}({\bf k}^{\prime}) w^q_{\rm p}({\bf
k} - {\bf k}^{\prime}) \text{sign}(\epsilon^s_{{\rm p}^{\prime}}({\bf
k}^{\prime})) \nonumber \\ &\ & \times \bigg\{
\cosh{^2(\xi_{{\bf k}^{\prime}}^s(\epsilon^s_{{\rm p}^{\prime}}({\bf
k}^{\prime}) + \half i|\Gamma^s_{{\rm p}^{\prime}}({\bf k}^{\prime})|))}
{ [b_{_T}(\epsilon^s_{{\rm p}^{\prime}}({\bf k}^{\prime}) + \half
i|\Gamma^s_{{\rm p}^{\prime}}({\bf k}^{\prime})|) +
f_{_T}(-\epsilon^q_{\rm p}({\bf k} - {\bf k}^{\prime}) - \half
i\Gamma^q_{\rm p}({\bf k} - {\bf k}^{\prime}))] \over [z -
(\epsilon^q_{\rm p}({\bf k} - {\bf k}^{\prime}) + \epsilon^s_{{\rm
p}^{\prime}}({\bf k}^{\prime})) - \half i(\Gamma^q_{\rm p}({\bf k} -
{\bf k}^{\prime}) + |\Gamma^s_{{\rm p}^{\prime}}({\bf k}^{\prime})|) ]}
\nonumber \\ &\ & + 
\sinh{^2(\xi_{{\bf k}^{\prime}}^s(\epsilon^s_{{\rm p}^{\prime}}({\bf
k}^{\prime}) - \half i|\Gamma^s_{{\rm p}^{\prime}}({\bf k}^{\prime})|))}
{ [b_{_T}(\epsilon^s_{{\rm p}^{\prime}}({\bf k}^{\prime}) - \half
i|\Gamma^s_{{\rm p}^{\prime}}({\bf k}^{\prime})|) +
f_{_T}(\epsilon^q_{\rm p}({\bf k} - {\bf k}^{\prime}) + \half
i\Gamma^q_{\rm p}({\bf k} - {\bf k}^{\prime}))] \over [z -
(\epsilon^q_{\rm p}({\bf k} - {\bf k}^{\prime}) - \epsilon^s_{{\rm
p}^{\prime}}({\bf k}^{\prime})) - \half i(\Gamma^q_{\rm p}({\bf k} -
{\bf k}^{\prime}) + |\Gamma^s_{{\rm p}^{\prime}}({\bf k}^{\prime})|) ]}
\bigg\}, \ \ \label{eq18b} \\
A^d_0({\bf k}, \omega) &\equiv& \Im {\cal G}^d_0({\bf k},
\omega-i0^+)/\pi \cong {1 \over N} \sum_{{\bf k}^{\prime}} \int
d\omega^{\prime} A^s({\bf k}^{\prime}, \omega^{\prime}) \big\{ A^q({\bf
k} - {\bf k}^{\prime}, \omega - \omega^{\prime}) \cosh{^2(\xi_{{\bf
k}^{\prime}}^s(\omega^{\prime}))} \nonumber \\ &\ & \times
[b_{_T}(\omega^{\prime}) + f_{_T}(\omega^{\prime} - \omega)] + A^q({\bf
k} - {\bf k}^{\prime}, \omega + \omega^{\prime}) \sinh{^2(\xi_{{\bf
k}^{\prime}}^s(\omega^{\prime}))} [b_{_T}(\omega^{\prime}) +
f_{_T}(\omega^{\prime} + \omega)] \big\}. 
\label{eq18} 
\end{eqnarray}
\end{widetext}
Since the svivons are in a combination state of degenerate condensates
(corresponding to the different ${\bf Q}_m$ points), the ${\bf
k}^{\prime}$ summation in Eq.~(\ref{eq18}) includes the averaging of the
$A^s({\bf k}^{\prime}, \omega^{\prime}) \cosh{^2(\xi_{{\bf
k}^{\prime}}^s(\omega^{\prime}))}$ and $A^s({\bf k}^{\prime},
\omega^{\prime}) \sinh{^2(\xi_{{\bf k}^{\prime}}^s(\omega^{\prime}))}$
terms over these condensates. 

In the non-FL regime, where major features of the physics of the
cuprates are described correctly replacing $\underline{\cal G}^d$ in
Eq.~(\ref{eq6}) by $\underline{\cal G}^d_0$, the evaluation of these
features is approached approximating the electron spectral functions
$A^d({\bf k}, \omega) \equiv \Im {\cal G}^d({\bf k} \omega-i0^+)/\pi$ by
$A^d_0({\bf k}, \omega)$ in Eq.~(\ref{eq18}). Note that the
$\omega$-integral of $A^d_0({\bf k}, \omega)$, given in
Eq.~(\ref{eq18}), is short of one {\it exactly} by the contribution of
the electron states of the upper Hubbard band, created by the second
term in the rhs of Eq.~(\ref{eq0c0}), ignored in Eq.~(\ref{eq0c}). 

The evaluation of $A^d_0({\bf k}, \omega)$ is detailed in Appendix C,
and it is found to be, approximately, expressed as a sum of two terms,
presented in Eq.~(\ref{eq19}). One term, $A^d_{0{\rm c}}({\bf k},
\omega)$, represents a convolution of QE and modified svivon spectral
functions, where the low-energy ($< k_{_{\rm B}}T$) tails of $A^s$ have
been truncated (see Eq.~(\ref{eq14}) and Fig.~\ref{fig6}(b)). The other
term, $A^d_{0{\rm b}}({\bf k}, \omega)$, represents anomalous effective
electron bands, generated by the QE spectral functions and the truncated
$A^s$ tails. 

As was discussed above, the QE spectral functions are specified as
low-energy (LE), mid-energy (ME) and high-energy (HE) types of
functions: $A^q_{\rm le}(\omega)$, $A^q_{\rm me}(\omega)$ and $A^q_{\rm
he}(\omega)$, respectively; consequently, the $A^d_{0{\rm b}}$ electron
bands corresponding to them are referred to as LE, ME and HE bands. The
LE $A^d_{0{\rm b}}$ band, due to the arcon and stripon peaks, is flat
and extends over the low-QE-energy BZ areas (see Fig.~\ref{fig3}(c,f)).
The ME $A^d_{0{\rm b}}$ band, due to the humpons, merges with the LE
band at high $T$ (see above), and has a similar role to that of the HE
$A^d_{0{\rm b}}$ band at low $T$. 

\begin{figure}[t] 
\begin{center}
\includegraphics[width=3.25in]{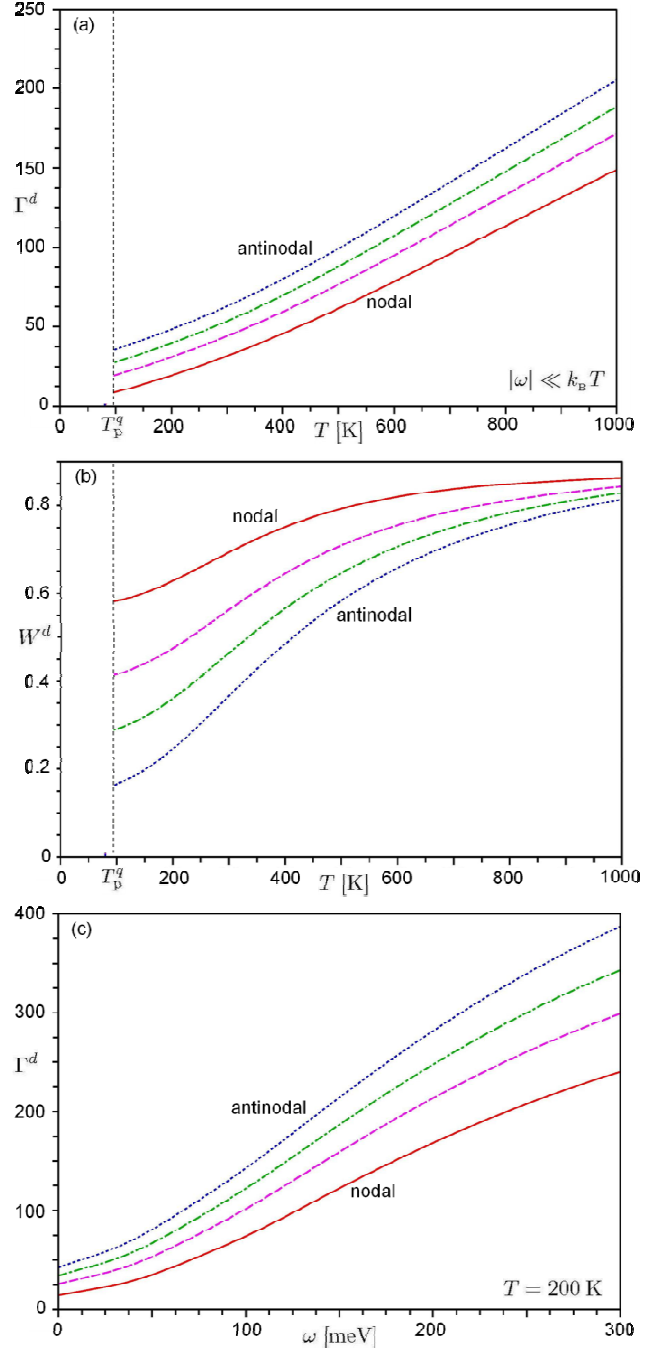}
\end{center}
\caption{Typical results (in arbitrary units), within the non-FL regime
(corresponding to the $\Gamma^q$ and $\Gamma^s$ results presented in
Figs.~\ref{fig5} and \ref{fig6}), for the $T$ dependence of (a) the
electron scattering rates $\Gamma^d$, (b) the spectral weights $W^d$
within effective LE electron band (see discussion in the text), and (c)
for the $\omega$ dependence of $\Gamma^d$.} 
\label{fig7}
\end{figure}

In the low $|\omega|/k_{_{\rm B}}T$ limit, $A^d_{0}$ is dominated
(through Eq.~(\ref{eq19})) by the LE $A^d_{0{\rm b}}$ band
\cite{smooth}, and the physical properties based on it reflect its
anomalous features; thus, as is explained in Appendix C, the linewidth
of this effective band increases linearly with $T$, and the spectral
weight $W^d({\bf k})$ within it depends on $T$ (especially in the
antinodal areas), increasing with it at low $T > T^q_{\rm p}$, and
saturating at $k_{_{\rm B}}T \gta \omega_{_{\rm H}}$; the band linewidth
plays the role of the electron scattering rates $\Gamma^d({\bf
k},\omega)$ in physical properties derived from it. Typical results for
these effective $\Gamma^d$ and $W^d$, as functions of $T$ (for $\omega
\to 0$), in ${\bf k}$ points ranging between the nodal and antinodal BZ
areas, are presented in Figs.~\ref{fig7}(a-b). 

On the other hand, in the high-$|\omega|/k_{_{\rm B}}T$ limit $A^d_{0}$
is dominated by $A^d_{0{\rm c}}$ (based on a convolution of QE and
modified svivon spectral functions) \cite{smooth} plus a minor
contribution of the ME and HE $A^d_{0{\rm b}}$ bands; when $T$ is
increased, the spectral weight within them is decreased by the same
amount that it is increased within the LE $A^d_{0{\rm b}}$ band (see
Fig.~\ref{fig7}(b)). Physical properties could then often be formulated
in terms of QE and svivon contributions, where one of them may be
dominant. 

Thus, the electron scattering rates $\Gamma^d({\bf k},\omega)$ measured
in the high-$|\omega|/k_{_{\rm B}}T$ limit, in various physical
properties, are obtained as convoluted combinations of those of the QEs
and svivons, as is implied from Eqs.~(\ref{eq18a}, \ref{eq18b}); due to
the $\cosh{^2}$ and $\sinh{^2}$ factors in Eq.~(\ref{eq18b}), the major
svivon contribution comes (see Eq.~(\ref{eq13b})) from LE and ME BZ
areas at energies ranging between $k_{_{\rm B}}T$ and somewhat above
${\bar \epsilon}^s({\bf k})$, where a maximum or a plateau appears in the
$\Gamma^s$ curves in Fig.~\ref{fig6}(b). Consequently (see
Figs.~\ref{fig5}(b) and \ref{fig6}(b)), when $T > T^q_{\rm p}$,
$\Gamma^d({\bf k},\omega)$ are approximately linear in $\omega$, for
$k_{_{\rm B}}T < |\omega| \lta 0.3\;$eV. 

Typical results for $\Gamma^d$ (at $T$ above \cite{pairing} $T^q_{\rm
p}$ and below $\Gamma^q_{0{\rm a}} / k_{_{\rm B}}$ and
$\omega^{\lambda}({\bf Q}) / k_{_{\rm B}}$), as a function of $\omega$,
in ${\bf k}$ points ranging between the nodal and antinodal BZ areas,
are presented in Fig.~\ref{fig7}(c). The evaluation of these results is based on
precise Fermi and Bose distribution functions, and not on their
asymptotic values assumed in Eq.~(\ref{eq19}); thus, they reflect a
smooth crossover between the high- and the low-$(|\omega|/k_{_{\rm
B}}T)$ behaviors, around $|\omega| \simeq k_{_{\rm B}}T$, manifested in
lower slopes of the $\Gamma^d$ {\it vs.} $|\omega|$ curves for $|\omega|
\to 0$. 

Similarly, due to large $\cosh{^2}$ and $\sinh{^2}$ factors, svivons of
the LE and ME areas also have a major contribution to $A^d_{0{\rm c}}$
in Eq.~(\ref{eq19}). Consequently, the convoluted QE-svivon states at
energies $k_{_{\rm B}}T < |\omega| \lta 0.3\;$eV have a band-like
behavior of linewidth which approximately increases linearly with
$|\omega|$, as can seen in Fig.~\ref{fig3}(d-e) \cite{smooth}. 

The linear dependence of the linewidths on $\omega$ is in agreement with
ARPES results \cite{Valla, Chang1}; furthermore, the measured
zero-energy-interpolated linewidths are closest to zero along the line
of nodes and largest in the antinodal direction, in agreement with the
results obtained here for $\Gamma^q_0({\bf k})$ in nodal and antinodal
points (see Eq.~(\ref{eq10a}) and the discussion in Appendix A). This
linear dependence is found here both for the main and the shadow bands,
as has been observed by ARPES \cite{Campuzano}. Note, however, that
$A^d_{0}$ misses the effect of the multiple-scattering FL-like electron
poles in Eq.~(\ref{eq6}), which contribute only to the main bands (see
Fig.~\ref{fig3}(d)). 

Thus, in the regime where the contribution of the non-FL $A^d_{0}({\bf
k}, \omega)$ to $A^d({\bf k}, \omega)$ provides a good approximation for
the evaluation of the effective $\Gamma^d({\bf k},\omega)$ in various
experiments, the measured $\Gamma^d$ are expected to be characterized by
a linear dependence on $T$ in the low-$|\omega|/k_{_{\rm B}}T$ limit,
and by a linear dependence on $\omega$ in the high-$|\omega|/k_{_{\rm
B}}T$ limit. Such a behavior corresponds to the MFL phenomenology
\cite{Varma}, and thus, the hump phase \cite{Honma} is referred to here
as the MFL phase. 

The linear dependence of the $\Gamma^d$ on $T$ or $\omega$, down to
their lowest value within the MFL phase, has been attributed to quantum
criticality \cite{Valla, Marel6}. In the present work a different,
though not contradictory, approach is applied, and the linear dependence
on $T$ or $\omega$ results from the role of the Bose svivon field which
has linear low-energy spectral tails, down to $\omega=0$, due to
Eq.~(\ref{eq13}). This linear dependence, and the switching in the roles
of $k_{_{\rm B}}T$ and $|\omega|$ between the low- and
high-$|\omega|/k_{_{\rm B}}T$ limits, result here from the behavior of
terms like $-|\omega| b_{_T}(-|\omega|)$ in these limits. 

The effect of convolution with svivons on the low-energy QE spectral
peaks, including the different contributions to $A^d_{0}$, is sketched
in Fig.~\ref{fig4}(b). The traces of the arcon and stripon peaks are
considerably broadened in the MFL phase \cite{smooth}; however, since
the evaluation of physical quantities in the low $|\omega|/k_{_{\rm
B}}T$ limit is dominated by the LE $A^d_{0{\rm b}}$ part of $A^d_{0}$,
for which the broadening of the QE peaks is $\sim$$2 k_{_{\rm B}} T$
(see Appendix C), these quantities are sensitive to the sharpness of
these peaks. 

\subsubsection{Anomalous physical properties}

The $T$ dependencies of the transport properties of the cuprates, whose
anomalous behavior has been associated with non-FL behavior from the
start, are the consequence of the unconventional $T$ behavior of the
effective scattering rates and spectral weight corresponding to the LE
$A^d_{0{\rm b}}$ band. The $T$ dependence of the electrical resistivity
in the cuprates \cite{Takagi} follows in the MFL phase that of the
evaluated effective $\Gamma^d$, presented in Fig.~\ref{fig7}(a). 

The effect of transfer of spectral weight from the $A^d_{0{\rm c}}$ and
ME $A^d_{0{\rm b}}$ bands to the LE $A^d_{0{\rm b}}$ band, when $T$
rises, is reflected in Hall-effect and thermoelectric power (TEP)
results which are dominantly determined by the latter. This transfer
occurs at both positive and negative energies of magnitude
$\sim$$k_{_{\rm B}} T$; thus, it should be reflected in such
measurements as an effective increase in the density of carriers,
subject to the question whether the occupation of the LE $A^d_{0{\rm
b}}$ band passes through half filling. 

In conventional bands both the Hall number ($n_{_{\rm H}}$), and the TEP
($S$), are negative when the band is almost empty, and positive when it
is almost full; the analytical periodic behavior of such bands over the
{\it entire} BZ generally results \cite{AshMer} in a coincidence between
their signs. As was discussed above, the LE $A^d_{0{\rm b}}$ band is not
conventional, and consists of the arcon--stripon contribution $A^q_{\rm
le}(\omega)$ to the QE spectral functions (see
Figs.~\ref{fig3},\ref{fig4}), and the low-energy tails of the svivon
spectral functions (see Fig.~\ref{fig6}). Thus, it occupies only a {\it
part} of the BZ, and a careful analysis should be made on the origin of
the signs of $n_{_{\rm H}}$ and $S$, derived from it. 

The sign of $S$ is \cite{Bar-Ad} that of the average energy (relative to
the chemical potential) within the range of the band close to the FS,
introduced through $\omega df_{_T}(\omega) /d\omega$. Since the LE
$A^d_{0{\rm b}}$ band lies almost entirely within this range, the TEP
derived from it could be either positive or negative, depending on the
band occupation, and thus on stoichiometry.

On the other hand, the sign of $n_{_{\rm H}}$ is \cite{Bar-Ad} opposite
from that of the averaged second derivatives of the band dispersion
curve over the FS (where $-df_{_T}(\omega) /d\omega$ contributes). Since
the LE $A^d_{0{\rm b}}$ band extends only over the arcon--stripon QE BZ
areas, shown in Fig.~\ref{fig3}(a-c), where the sign of the second
derivatives of the band dispersion curve is dominantly negative, the
sign of $n_{_{\rm H}}$ is expected to be positive, independently of the
occupation of the band (and thus of stoichiometry); the fact that this
low-energy band is flat results in small dispersion derivatives, but
also in a nonzero $-df_{_T}(\omega) /d\omega$ factor over a range of the
BZ, and the contradicting effects introduced to the expression
\cite{Bar-Ad} for $n_{_{\rm H}}$ compensate for each other. And indeed,
for hole-doped cuprates, studied here, $n_{_{\rm H}}$ is found to be
positive in the MFL phase, and its $T$ dependence follows \cite{Kubo,
Hwang} that of $W^d({\bf k})$, presented in Fig.~\ref{fig7}(b). 

The $T$ dependence of the TEP in YBCO has been analyzed \cite{Fisher} in
terms of a ``narrow-band model'', under which $|S|$ is larger than
typical metallic values, and increases with $T$ up to its saturation
value (when $k_{_{\rm B}} T$ exceeds the bandwidth), thus \cite{Fisher}:
$S_{\rm sat} = (k_{_{\rm B}} / {\rm e}) \ln{[x/(1-x)]}$; here $x$ is the
fractional band occupation (thus $x=0$ when the band is empty of
electrons, and $x=1$ when it is full) within the measured stoichiometry.

For hole-doped cuprates of an electronic structure similar to the one
presented in Fig.~\ref{fig3}, the TEP does not saturate with $T$, and
has an almost universal dependence on $T$ and the doping level
\cite{Tanaka}. Thus, in the underdoped and lightly overdoped regimes,
$S$ increases with $T$ at low $T$, reaching a positive maximum, and then
decreases with $T$, having an almost linear dependence on it at high
$T$; at the optimal stoichiometry, $S$ crosses zero at $T \cong 300\;$K
which has been applied \cite{Tanaka} to determine the doping level in
cuprates. In the heavily overdoped regime, $S$ is negative and decreases
monotonously with $T$, consistently with an FL state with a band which
is less than half filled (as is expected in this regime). 

The LE $A^d_{0{\rm b}}$ band is a narrow one (being based on the almost
dispersionless stripons and arcons); however, since its linewidth
increases linearly with $T$, the TEP saturation temperature
\cite{Fisher} cannot be reached. Furthermore, the asymmetry of the QE
and electronic spectra, presented in Fig.~\ref{fig3}, with respect to
the inversion of the sign of the energy, results in a
temperature-induced transfer of spectral weight to the LE $A^d_{0{\rm
b}}$ band (from the $A^d_{0{\rm c}}$ and ME $A^d_{0{\rm b}}$ bands)
which is larger for negative than for positive energies. Consequently,
the fractional occupation $x$ of this band decreases when $T$ is
increased. 

Thus, in the underdoped and lightly overdoped regimes, the LE
$A^d_{0{\rm b}}$ band is more than half filled at low $T$, resulting in
positive $S$ (within the MFL phase) which first increases with $T$ to
approach its saturation value, but the decrease in $x$ with $T$ results
also in a decrease in this saturation value \cite{Fisher}, and thus in
$S$ (after it reaches its maximal value). Consequently, a ``universal''
behavior of $S$ is obtained for cuprates of the same type of electronic
structure as in Fig.~\ref{fig3}, in agreement with experiment
\cite{Tanaka}. The value $S=0$, at $T \cong 300\;$K, at optimal
stoichiometry, implies that the LE $A^d_{0{\rm b}}$ band is then half
filled. 

The existence of a universal behavior in the cuprates has been observed
\cite{Luo} in various physical quantities, determined by the LE
$A^d_{0{\rm b}}$ band. This behavior is expressed through the scaling of
their $T$-dependencies in a typical energy corresponding to spectrum of
the low-energy QEs. These quantities include, in addition to the TEP,
the planar and $c$-axis resistivities, the Hall coefficient, the uniform
magnetic susceptibility, and the spin-lattice relaxation rate. 

Linearity of the electron scattering rates in $\omega$, for $|\omega|
\lta 0.3\;$eV (similarly to ones shown in Fig.~\ref{fig7}(c)), within
the MFL phase, has been observed in optical results
\cite{Marel6,Timusk2}. In different hole-doped cuprates, the optical
conductivity $\sigma^d(\omega)$ consists in this phase \cite{Tanner1} of
a low-energy Drude term, and a higher-energy ``mid-IR'' term, where the
effective density of carriers within the Drude term is 4--5 times
smaller than those integrated up to above the mid-IR energy (a behavior
referred to as ``Tanner's law''). 

Within the present approach, the electron spectral functions are
approximated by $A^d_0({\bf k}, \omega)$ in Eq.~(\ref{eq18}), including
both the $A^d_{0{\rm b}}$ and $A^d_{0{\rm c}}$ terms in
Eq.~(\ref{eq19}) \cite{smooth}. At low $\omega$, the QE contribution to
$A^d_{0}$ includes only the stripons and arcons peaks, resulting in the
Drude term in $\sigma^d(\omega)$. The mid-IR term in $\sigma^d(\omega)$
is introduced due to the growing role (in $A^d_{0}$) of humpons at
higher energies, and of the almost detached higher-energy QEs (see
Fig.~\ref{fig3}(a-b)) at further higher energies. Thus the increase in
the effective number of carriers with energy, expressed by Tanner's law
\cite{Tanner1}, reflects the difference between the occupied QE spectral
weight residing within their low-energy peaks, and that residing within
their entire spectrum. 

\subsection{Paired QEs and electrons}

\subsubsection{Formulation and general features}

QE pairing is approached adapting of the Migdal--Eliashberg theory
\cite{Scalapino} for the present case. Nambu spinors ($q^{\dagger}({\bf
k})$, $q^{\dagg}(-{\bf k})$) are considered as creation operators (the
${\bf k}$ values here are within a half of the BZ, but since there is a
freedom in choosing this half, the derived expressions are valid within
the entire BZ). Self-energy $2\times 2$ matrices are derived, and
expressed (in terms of the Pauli matrices $\underline{\tau}_1$,
$\underline{\tau}_2$ and $\underline{\tau}_3$) through: 
\begin{eqnarray}
\underline{\epsilon}^q_0({\bf k}) + \underline{\Sigma}^q({\bf k},z) &=&
[\epsilon^q_0({\bf k}) + \Sigma^q({\bf k},z)] \underline{\tau}_3 + \Re
\Phi^q({\bf k},z) \underline{\tau}_1 \nonumber \\ &\ & + \Im \Phi^q({\bf
k},z) \underline{\tau}_2. \label{eq20} 
\end{eqnarray}

The diagonalization of $\underline{\epsilon}^q_0({\bf k}) +
\underline{\Sigma}^q({\bf k},z)$ yields QE Bogoliubov states annihilated
by $q_{+}^{\dagg}({\bf k},z)$ and $q_{-}^{\dagg}({\bf k},z)$,
corresponding to positive and negative energies, respectively; the
transformation between them and the unpaired-state QE operators is
expressed as: 
\begin{eqnarray}
q({\bf k}) &=& \exp{(i\phi_{\bf k}^q(z)/2)} [\cos{(\xi_{\bf k}^q(z))}
q_{+}^{\dagg}({\bf k},z) \nonumber \\ &\ & - \sin{(\xi_{\bf k}^q(z))}
q_{-}^{\dagg}({\bf k},z)], \nonumber \\ 
q^{\dagger}(-{\bf k}) &=& \exp{(-i\phi_{\bf k}^q(z)/2)} [\sin{(\xi_{\bf
k}^q(z))} q_{+}^{\dagg}({\bf k},z) \nonumber \\ &\ & + \cos{(\xi_{\bf
k}^q(z))} q_{-}^{\dagg}({\bf k},z)]. \label{eq21} 
\end{eqnarray}
This transformation diagonalizes Dyson's equation \cite{Vliet, Mahan},
yielding the poles of the diagonalized QE Green's function ${\cal
G}^q_{\pm}({\bf k},z)$ at: 
\begin{eqnarray}
z &=& \pm\epsilon^q_0({\bf k}) + \Sigma^q_{\pm}({\bf k},z)  \nonumber \\
&\equiv& \pm E^q({\bf k},z) + i\Im \Sigma^q_{\pm}({\bf k},z), \
\text{where} \nonumber \\ 
\Im \Sigma^q_{\pm}({\bf k},z) &=& \pm\cos{(2\xi_{\bf k}^q(z))} \Im
\Sigma^q({\bf k},z), \nonumber \\ 
E^q({\bf k},z) &=& \sqrt{ [\epsilon^q_0({\bf k}) + \Re \Sigma^q({\bf
k},z)]^2 + [{\bar \Phi}^q({\bf k},z)]^2 }, \nonumber \\ \Phi^q({\bf
k},z) &=& |\Phi^q({\bf k},z)| \exp{(i\psi_{\bf k}^q(z))}, \label{eq22} \\ 
\phi_{\bf k}^q(z) &=& \psi_{\bf k}^q(z)\ \text{ or } \ \psi_{\bf
k}^q(z)+\pi, \ \text{ and} \nonumber \\
{\bar \Phi}^q({\bf k},z) &\equiv& \cos{(\psi_{\bf k}^q(z) - \phi_{\bf
k}^q(z))} |\Phi^q({\bf k},z)| = \pm |\Phi^q({\bf k},z)|. \nonumber 
\end{eqnarray}

The coefficients $\xi_{\bf k}^q(z)$ are obtained through: 
\begin{eqnarray}
\cos{(2\xi_{\bf k}^q(z))} &=& {\epsilon^q_0({\bf k}) + \Re \Sigma^q({\bf
k},z) \over E^q({\bf k},z)}, \nonumber \\ 
\sin{(2\xi_{\bf k}^q(z))} &=& {{\bar \Phi}^q({\bf k},z) \over E^q({\bf
k},z)}. \label{eq23} 
\end{eqnarray}
The spectral functions of the QE Bogoliubov states are obtained through: 
\begin{eqnarray}
A^q_{\pm}({\bf k}, \omega) &\equiv& \Im {\cal G}^q_{\pm}({\bf k},
\omega-i0^+) / \pi \nonumber \\ &=& {\Gamma^q_{\pm}({\bf k},
\omega)/2\pi \over [\omega \mp E^q({\bf k},\omega)]^2 +
[\half\Gamma^q_{\pm}({\bf k}, \omega)]^2}, \label{eq24} 
\end{eqnarray}
where $\Gamma^q_{\pm}({\bf k},\omega) \equiv 2 \Im \Sigma^q_{\pm}({\bf
k}, \omega-i0^+) \ge 0$, and thus $A^q_{\pm} \ge 0$ (note that
$\cos{(2\xi_{\bf k}^q(\omega))}$ is positive within the $\omega > 0$
range of $A^q_{+}$, and negative within the $\omega < 0$ range of
$A^q_{-}$---see Eqs.~(\ref{eq22},\ref{eq23})). 

The self-energy terms $\Sigma^q$ and $\Phi^q$ are primarily determined,
at low energies, by QE-lagron coupling, through $\Delta{\cal H}$ in
Eq.~(\ref{eq4i}). Their self-consistent expressions are derived
similarly to the un-paired case in Eq.~(\ref{eq7}), yielding: 
\begin{eqnarray}
\Sigma^q({\bf k},z) &\cong& {1 \over N} \sum_{\bf q} \int d\omega^q
[\cos{^2(\xi_{{\bf k}-{\bf q}}^q(\omega^q))} \nonumber \\ &\ & \times
A^q_{+}({\bf k}-{\bf q}, \omega^q) + \sin{^2(\xi_{{\bf k}-{\bf
q}}^q(\omega^q))} \nonumber \\ &\ & \times A^q_{-}({\bf k}-{\bf q},
\omega^q)] F_{_T}({\bf q}, \omega^q, z), \nonumber \\ 
\Phi^q({\bf k},z) &\cong& {1 \over 2N} \sum_{\bf q} \int d\omega^q
\sin{(2\xi_{{\bf k}-{\bf q}}^q(\omega^q))} \nonumber \\ &\ & \times
\exp{(i\phi_{{\bf k} - {\bf q}}^q(\omega^q))} [A^q_{+}({\bf k}-{\bf q},
\omega^q) \nonumber \\ &\ & - A^q_{-}({\bf k}-{\bf q}, \omega^q)]
F_{_T}({\bf q}, \omega^q, z). \label{eq25}
\end{eqnarray}

$\Im \Sigma^q({\bf k}, \omega-i0^+)$ is obtained by expressing
$(\omega-i0^+ \mp \omega^{\lambda}({\bf q}) - \omega^q)^{-1}$ in
$F_{_T}({\bf q}, \omega^q, \omega-i0^+)$, in Eqs.~(\ref{eq7},
\ref{eq25}), as \cite{Vliet, Mahan} $\pi i \delta((\omega \mp
\omega^{\lambda}({\bf q}) - \omega^q)$; such an expression could not be
applied to obtain $\Im \Phi^q({\bf k}, \omega-i0^+)$, because of the
$\exp{(i\phi_{{\bf k} - {\bf q}}^q(\omega^q))}$ phase factors in the rhs
of Eq.~(\ref{eq25}), but it could still be applied to obtain a term $i
{\hat \Phi}^q({\bf k}, \omega-i0^+)$ in $\Phi^q({\bf k}, \omega-i0^+)$.
Due to the $f_{_T}$ and $b_{_T}$ factors in Eq.~(\ref{eq7}), QEs and
svivons at energies $\omega^q$ and $\omega^{\lambda}({\bf q})$
contribute to $\Im \Sigma^q({\bf k}, \omega-i0^+)$ and ${\hat
\Phi}^q({\bf k}, \omega-i0^+)$ at $\omega = \omega^q \pm
\omega^{\lambda}({\bf q})$ when $\omega^{\lambda}({\bf q}) \ll k_{_{\rm
B}} T$, and at $\omega = \omega^q + \text{sign}(\omega^q)
\omega^{\lambda}({\bf q})$ when $\omega^{\lambda}({\bf q}) \gg k_{_{\rm
B}} T$. 

The most significant low-energy contribution to $\Sigma^q$ and $\Phi^q$
in Eq.~(\ref{eq25}) is from ${\bf q}$ points corresponding to ${\bf
Q}_m$ and ${\bf Q}_m$-vicinity lagrons (see above); an additional
significant contribution comes from ${\bf Q}$-ESP lagrons. Thus, by
Fig.~\ref{fig2}, a QE spectral peak around the energy $\omega^q$,
located at $\sim$${\bf k} - {\bf Q}_m$, contributes (due to ${\bf Q}_m$
and ${\bf Q}_m$-vicinity lagrons) peaks to $\Im \Sigma^q({\bf k},
\omega-i0^+)$ and ${\hat \Phi}^q({\bf k}, \omega-i0^+)$ at $\omega
\simeq \omega^q$, with asymmetry towards larger $|\omega^q|$; if such a
QE peak is located at $\sim$${\bf k} - {\bf Q}$, it contributes to them
(due to ${\bf Q}$-ESP lagrons) peaks at $\omega \simeq \omega^q +
\text{sign}(\omega^q) \omega^{\lambda}({\bf Q})$, if $k_{_{\rm B}} T \ll
\omega^{\lambda}({\bf Q})$, and at $\omega \simeq \omega^q \pm
\omega^{\lambda}({\bf Q})$, if $k_{_{\rm B}} T \gg \omega^{\lambda}({\bf
Q})$. 

$\Re \Sigma^q$ and $\Phi^q$ could be expressed through the
Kramers--Kronig relations \cite{Vliet, Mahan}: 
\begin{eqnarray}
\Re\Sigma^q({\bf k},\omega) &=& \wp \int {d\omega^{\prime} \Im
\Sigma^q({\bf k}, \omega^{\prime}-i0^+) \over \pi (\omega -
\omega^{\prime})}, \nonumber \\ 
\Phi^q({\bf k},\omega-i0^+) &=& \wp \int {d\omega^{\prime} {\hat
\Phi}^q({\bf k}, \omega^{\prime}-i0^+) \over \pi (\omega -
\omega^{\prime})} \nonumber \\ &\ & + i {\hat \Phi}^q({\bf k},
\omega-i0^+). \label{eq26} 
\end{eqnarray}

The effect of peaks in $\Im \Sigma^q({\bf k},\omega-i0^+)$ (which is
always positive) on $\Re\Sigma^q$, in Eq.~(\ref{eq26}), is to ``push''
the peaks in $A^q_{\pm}({\bf k}, \omega)$ away from them, through
Eqs.~(\ref{eq22}, \ref{eq24}). As was discussed above, the location of
peaks in ${\hat \Phi}^q({\bf k},\omega-i0^+)$ (and in $\Im \Sigma^q({\bf
k},\omega-i0^+)$) is close to ${\bf k} - {\bf Q}$, and symmetrically
around it. Consequently, in order to stabilize pairing, one should
choose (see Eqs.~(\ref{eq22}), (\ref{eq23}) and (\ref{eq25})): 
\begin{eqnarray}
&\ & \phi_{\bf k}^q(\omega) = \psi_{\bf k}^q(\omega), \ \text{getting}
\ \ \ \nonumber \\ 
&\ & \phi_{{\bf k} + {\bf Q}}^q(\omega) = \phi_{\bf k}^q(\omega) + \pi,
\ \text{and introduce} \ \ \ \nonumber \\ 
&\ & {\tilde \Phi}^q({\bf k},\omega) = |\Phi^q({\bf k},\omega)|
\cos{\phi_{\bf k}^q(\omega)}, \ \text{where} \ \ \ \nonumber \\ 
&\ & \text{sign}[{\tilde \Phi}^q({\bf k},\omega)] \ \text{is
independent of $\omega$}. \ \ \ \label{eq27} 
\end{eqnarray}

${\tilde \Phi}^q({\bf k},\omega)$ plays the role of the pairing order
parameter; by Eq.~(\ref{eq27}) it reverses its sign when ${\bf k}$ is
shifted by ${\bf Q}$, resulting to an approximate $d_{x^2-y^2}$ pairing
symmetry. For the choices in Eq.~(\ref{eq27}) one gets that the effect
of the peaks in ${\hat \Phi}^q({\bf k},\omega-i0^+)$ on $\Phi^q$ in
Eq.~(\ref{eq26}) is to ``push'' (through Eqs.~(\ref{eq22}, \ref{eq24}))
the peaks in $A^q_{\pm}({\bf k}, \omega)$ away from them, if they are at
energies of the opposite sign, and to ``pull'' these peaks towards them,
if they are at energies of the same sign. 

Such a scenario could result in a pairing gap, where in addition to the 
humpons, the stripon and arcon peaks (which are not on the lines of
nodes) are split into positive- and negative-energy peaks. As was
discussed above, the locations (at low $T$) of the centers of the peaks
in $\Im \Sigma^q({\bf k},\omega-i0^+)$ and ${\hat \Phi}^q({\bf
k},\omega-i0^+)$ are at higher values of $|\omega|$ than those of such
split peaks in $A^q_{\pm}({\bf k} - {\bf q}, \omega)$, where ${\bf q} =
{\bf Q} \text{ or } {\bf Q}_m$. 

By the symmetry of the QE spectrum under a shift in ${\bf Q}$, a split
peak in $A^q_{+}({\bf k}, \omega)$ or $A^q_{-}({\bf k}, \omega)$ is
pushed, on the average, away from zero by the opposite-energy-sign peaks
in $\Im \Sigma^q({\bf k},\omega-i0^+)$, and towards zero by the
same-energy-sign peaks in it, and this is insufficient to stabilize the
gap, unless the symmetry between the degenerate lagron and svivon
condensates (see Fig.~\ref{fig2}, Eq.~(\ref{eq5}), and the above
discussion) is broken, and a {\it static} striped structure sets in. 

The {\it additional} effect of peaks in ${\hat \Phi}^q({\bf
k},\omega-i0^+)$, at {\it both} positive and negative energies, is to
push this split peak, on the average, {\it away} from zero. Thus, a gap
is stabilized, under a certain temperature (in a state of combined
lagron and svivon condensates corresponding to fluctuating
inhomogeneities), by breaking symmetry through the introduction of a
nonzero ${\tilde \Phi}^q$. A quantitative calculation on this pairing
scenario will be published in a separate paper. 

Nonzero ${\hat \Phi}^q$ results in a nonzero anomalous QE Green's
function matrix $\underline{ \cal F}^q$ (see above), and consequently in
a nonzero anomalous electron Green's function matrix $\underline{ \cal
F}^d$. The expression for its zeroth-order term $\underline{\cal
F}^d_0$, in terms of $\underline{ \cal F}^s$ and $\underline{ \cal
F}^q$, is presented diagrammatically in Fig.~\ref{fig1}(b). Within the
${\bf k}$ representation, the expression for ${\cal F}^d_0({\bf k},z)$
is similar to the expression for ${\cal G}^d_0({\bf k},z)$ in
Eq.~(\ref{eq18b}); the $\cosh{^2(\xi_{{\bf k}^{\prime}}^s)}$ and
$\sinh{^2(\xi_{{\bf k}^{\prime}}^s)}$ factors there are replaced by
$\sinh{(2\xi_{{\bf k}^{\prime}}^s)}/2$, and additional factors are
included (for both ${\cal F}^d_0$ and ${\cal G}^d_0$) due to QE pairing;
thus, ${\cal G}^q_0({\bf k},z)$ and ${\cal F}^q_0({\bf k},z)$ are
expressed in terms of the diagonalized ${\cal G}^q_{\pm}({\bf k},z)$,
and have their poles specified in Eq.~(\ref{eq22}). 

As will be detailed elsewhere, the electron spectrum is determined
through the diagonalization of $2\times 2$ matrices whose diagonal terms
are determined by ${ \cal G}^d({\bf k},z)^{-1}$, and non-diagonal terms
by ${ \cal F}^d({\bf k},z)^{-1}$ and $[{ \cal F}^d({\bf k},z)^{-1}]^*$.
The evaluation of the approximate electron spectral functions
$A^d_0({\bf k}, \omega)$ in a paired state is analogous to their
evaluation, through Eqs.~(\ref{eq18}, \ref{eq19}), in the unpaired case;
but the $\cosh{^2(\xi_{{\bf k}^{\prime}}^s)}$ and $\sinh{^2(\xi_{{\bf
k}^{\prime}}^s)}$ factors there are modified, and expansion coefficients
due to the pairing diagonalization procedures are included. The obtained
electron spectrum is characterized by a pairing gap of the same symmetry
as that of ${\hat \Phi}^q$; the opposite signs of ${\hat \Phi}^q$ around
the lines of modes result in a zero electron pairing gap on these lines,
and in its continuous variation around them, due to QE-svivon
convolution. 

\subsubsection{Pseudogap phase}

As was discussed above, the number of poles in ${\cal G}^q({\bf k},z)$
is different for QEs in nodal and antinodal ${\bf k}$ points. Thus these
two types of low-energy QEs correspond to different symmetries, and it
is analytically possible to have ${\tilde \Phi}^q \ne 0$ for only one of
them (corresponding to a specific area within the QE BZ). Since (as was
discussed above with respect to humpon--humpon splitting) ${\bf Q}_m$
lagrons contribute significantly (through Eq.~(\ref{eq25})) to QE
pairing in antinodal points, and less (if at all) in nodal points, it is
likely that there exists a temperature range $T_c < T < T^*$, where
${\tilde \Phi}^q \ne 0$ in the antinodal, but {\it not} in the nodal QE
BZ areas. The symmetry of a state within this $T$ range is {\it
different} from the symmetries of the states where ${\tilde \Phi}^q \ne
0$ for either almost all the low-energy QEs, or almost none of them. 

Scattering between the ${\tilde \Phi}^q \ne 0$ antinodal, and the
${\tilde \Phi}^q = 0$ nodal QEs prevents the existence of supercurrent
in such a partial-pairing state, and it corresponds to the PG phase of
the cuprates \cite{Honma}. From a comparison between the expressions for
$\Sigma^q$ and $\Phi^q$ in Eq.~(\ref{eq25}), and the above discussion,
one concludes that $T^*$ could not exceed a maximal value which is
somewhat smaller than $\sim$$\omega_{_{\rm H}} / k_{_{\rm B}}$ (where
the effect of the inhomogeneities disappears). Consequently, $k_{_{\rm
B}} T^*$ could be, at the most, close to $\sim$$J$ (see discussion
above), in agreement with the values of $T^*$ observed in the phase
diagram \cite{Honma}. 

The stripon peaks split in this phase into positive- and negative-energy
peaks, as is sketched in Fig.~\ref{fig4}(a). The contribution of ${\bf
Q}$-ESP lagrons to $\Sigma^q$ and $\Phi^q$ in Eq.~(\ref{eq25}), and the
fact that $|\Re \Sigma^q| > |\Phi^q|$ within the higher-weight
Bogoliubov band, results in an energy separation \cite{Eres}
$>\omega^{\lambda}({\bf Q}) \simeq {3 \over 4} E_{_{\rm res}}$ between
same-energy-sign stripon and humpon peaks. 

There exists a low-$T$ regime within the PG phase where the
inhomogeneities become static (resulting in an energy gain), and they
are observed as a glassy structure \cite{Davis1, Hudson}; this regime is
characterized by a lower minimum ${\bar \epsilon}^s_{\rm min}$ of the
svivon dispersion curves, and a higher spectral weight in the humpon
peaks (which, as was discussed above, correspond to the
inhomogeneities), on the expense of that in the split stripon peaks. In
order for a static inhomogeneous structure to coexist with pairing, it
should not break the symmetry between the four ${\bf Q}_m$ points in
Eq.~(\ref{eq5}) and Fig.~\ref{fig2}, and ${\bar \epsilon}^s_{\rm min}$
should not be too close to zero; this requires the effect of coupling to
the lattice, manifested in the existence of local distortions. 

The static inhomogeneities are obtained through two types of
combinations of striped condensates, corresponding to different ${\bf
Q}_m$ points; one type is of the different phases of the SDWs introduced
by the stripes, resulting in static charge-density waves \cite{Hudson}
(CDWs) of wave vectors $2\delta {\bf q}_m$ (while the spins are
fluctuating); the other type is of the different directions (thus $a$
and $b$) of the stripes, resulting in a checkerboard-like structure
\cite{Davis1} of short stripe segments of $\sim$$4a$ periodicity, along
the two directions (similarly to a structure predicted earlier
\cite{ashk}). Longer unidirectional stripes can be obtained \cite{Tran1,
Yamada} when the resulting energy gain is larger than that due to
pairing which is suppressed altogether, as probably occurs in ``1/8
doping anomalies'' and below the minimal SC doping level. 

The above derivation of the low-$\omega$ QE and svivon scattering rates
in the MFL phase required the existence of low-$\omega$ QE (particularly
stripon) spectral weight, as occurs \cite{pairing} for $T>T^q_{\rm p}$.
Consequently, the opening of a partial QE gap in the PG phase results in
a substantial reduction in $\Gamma^q$ and $|\Gamma^s|$, for $|\omega|$
below the gap energy, but not above it, in agreement with experiment
\cite{Takagi, Puchkov}. 

However, since the gap is partial, the low-$\omega$ scattering features,
including the low-energy svivon spectral tails, existing for $T > T^*$,
partially persist for $T_c < T < T^*$. Consequently, the electron
spectral functions in the PG phase include, as in the MFL phase (see
Eq.~(\ref{eq19})), an effective LE band part $A^d_{0{\rm b}}({\bf k},
\omega)$ which consists here of the low-energy part of the Fermi arcs
around the lines of nodes, and a convoluted QE-svivon part $A^d_{0{\rm
c}}({\bf k}, \omega)$. The contribution of arcons to $A^d_{0{\rm c}}$
forms the extension of the Fermi arcs to higher energies, and the
contribution of stripons to it is modified by pairing (see above),
reflecting the existence of the pseudogap. The association of the latter
electron states with Bogoliubov states, due to pairing, is supported by
experiment \cite{Kanigel1, Valla1}. Also the existence of Fermi arcs
which are {\it distinct} from Fermi pockets (see above) has been
established by experiment \cite{Meng}. These two features in $A^d$ in
the PG phase are sketched in Fig.~\ref{fig4}(b). 

A decrease in $T$, within the PG phase, results in the transition of the
spectral functions of low-energy QEs from a structure of ${\tilde \Phi}^q
= 0$ arcons to that of ${\tilde \Phi}^q \ne 0$ stripons and humpons. It
is associated with the growing effect of the inhomogeneities at lower
$T$, within the PG phase, discussed above. Since all the arcons are
coupled to stripon--humpons through lagrons, the transition of unpaired
arcons into paired stripon--humpons would include at $T=0$ {\it all} the
arcons (except for those on the lines of nodes, where ${\tilde \Phi}^q =
0$ by symmetry) if the arcons remained unpaired. 

The signature of such a transition on the electron spectrum is the
observed reduction of the Fermi arcs with decreasing $T$, within the PG
phase, and indeed, the extrapolation of their reduction to $T=0$ results
in the points on the lines of nodes \cite{Kanigel}. Such a low-$T$
``$d$-wave nodal liquid'' has been observed \cite{Kanigel2} in the
regime where the PG state persists down to $T \to 0$. 

\subsubsection{Superconducting phase}

SC occurs when ${\tilde \Phi}^q \ne 0$ for the nodal arcons (except for
those on the lines of nodes), in addition to the antinodal stripons and
humpons. Thus the arcon peaks split below $T_c$ into positive- and
negative-energy peaks, as is sketched in Fig.~\ref{fig4}(a). Since their
coupling to other QEs through ${\bf Q}_m$ lagrons is weak, or absent,
the {\it additional} pairing necessary for SC to set in is induced
primarily through ${\bf Q}$-ESP lagrons. 

As was discussed above, the lagron energy involved in such pairing is
$\sim$$\omega^{\lambda}({\bf Q})$ which could be treated analogously to
the dominant boson energy \cite{Kresin} $k_{_{\rm B}} {\tilde \Omega}$
in electron-phonon systems. A coupling parameter ${\bar \lambda}$ is
obtained by approximating Eq.~(\ref{eq25}) similarly to the treatment of
such systems \cite{Kresin}, resulting in values as large as ${\bar
\lambda} \simeq 3$. The dependence of ${\bar \lambda}$ on the doping
level is {\it weak} since, below $T_c$, ${\tilde \Phi}^q \ne 0$ {\it
both} for the nodal and the antinodal QEs. 

Estimating $T_c$ on the basis of an expression (for electron-phonon
systems) which is valid within the {\it entire} coupling regime
\cite{Kresin}: $T_c \cong 0.25 {\tilde \Omega} / [\exp{(2/{\bar
\lambda})} - 1]^{1/2}$, results in an approximate scaling
$\omega^{\lambda}({\bf Q}) / k_{_{\rm B}} T_c \simeq 3.9$, and thus
\cite{Eres} $E_{_{\rm res}} / k_{_{\rm B}} T_c \simeq 5.2$, in agreement
with experiment \cite{Bourges}. This explains the enhancement of $T_c$
compared to conventional electron-phonon SCs, where such large ${\bar
\lambda}$ values would result in lattice instabilities. 

The crucial role played by lagrons at the ${\bf Q}_m$ points and their
vicinity (where $\omega^{\lambda}({\bf q}) \propto |{\bf q} - {\bf
Q}_m|$) in the pairing of stripons is not comparable to the minor role
played by low-energy acoustic phonons in conventional SCs; while the
coupling constants between electrons and acoustic phonons vanish
\cite{AshMer,Vliet} for ${\bf q} \to 0$, the coupling constants
$\gamma({\bf q})$ between QEs and lagrons remain constant for ${\bf q}
\to {\bf Q}_m$ (see Eq.~(\ref{eq4j1}) and the above discussion). 

High-energy lagrons also play some role in pairing, and specifically the
high-energy SP lagrons (see Fig.~\ref{fig2}). An optical analysis, in an
attempt to determine the energy spectrum of the bosons involved in
pairing in the cuprates, reveals \cite{Marel} significant contributions
of bosons at low energies, at energies \cite{Eres}
$\sim$$\omega^{\lambda}({\bf Q}) \simeq {3 \over 4} E_{_{\rm res}}$, and
at energies $\sim$0.2--0.3$\;$eV which are consistent with those of
high-energy SP lagrons. 

The opening of an SC gap prevents the above derivation of QE and svivon
scattering rates, depending on the condition $T>T^q_{\rm p}$ (which is
only partially fulfilled in the PG phase). This results in a drastic
reduction in $\Gamma^q$ and $|\Gamma^s|$, for $|\omega|$ below the gap
energy, but not above it. Consequently, the split stripon and arcon
peaks are sharp below $T_c$, while the humpon peaks remain wide (see
Fig.~\ref{fig4}(a)). 

Also the low-energy svivon spectral tails disappear below $T_c$, and
thus the electron spectral functions $A^d({\bf k}, \omega)$, can then be
approximated through a modification (see above) of the convoluted
QE-svivon term $A^d_{0{\rm c}}({\bf k}, \omega)$ in Eq.~(\ref{eq19})
{\it alone}. The resulting radical reduction (for $T<T_c$) in the
electron scattering rates, for $|\omega|$ below the gap energy, is
observed in optical \cite{Puchkov} and microwave \cite{Bonn} results. 

An aspect of Tanner's law \cite{Tanner1}, discussed above, is that the
optically determined SC-state superfluid density corresponds to the
density of carriers determined by the normal-state Drude term, and
misses the contribution of the mid-IR term. This observation is
consistent with the present prediction that both the Drude and the
superfluid density are determined by the contribution of stripons and
arcons to $A^d_{0}$. 

The traces of the arcon and stripon gaps on $A^d$, sketched in
Fig.~\ref{fig4}(b), are often referred to as nodal and antinodal gaps,
respectively. The antinodal gap-edge structure includes the
(stripon-derived) peak and (humpon-derived) hump (see
Fig.~\ref{fig4}(b)). The nodal gap has been found \cite{Tacon,
Sawatzky1} to scale with $\sim$$5 k_{_{\rm B}}T_c$, consistently
\cite{Kresin} with the ratio between the pairing gap and $T_c$ for the
large coupling parameters ${\bar \lambda}$ in this case (see above). 

Since Eq.~(\ref{eq25}) couples between the nodal and antinodal pairing
order parameters, the opening of a nodal gap below $T_c$ modifies the
antinodal gap (which opens below $T^*$) in a manner observed
\cite{Shen2, Khasanov} as an apparent onset (below $T_c$) of a ``second
energy gap'', superimposed on the pseudogap; however, it is clear that
the antinodal gap is ``smoothly connected'' to the nodal gap
\cite{Nakayama}. 

In the underdoped regime, most of the antinodal (but not nodal) Cooper
pairs are formed at $T_c < T < T^*$; this is consistent with the
observation \cite{Gallais2} that in this regime only a small density of
additional antinodal (but not nodal) pairs are formed at $T < T_c$.
Since the superfluid density consists of pairs formed both at $T < T_c$
and at $T_c < T < T^*$, it does not coincide with the density of the
additional pairs formed at $T < T_c$ \cite{Gallais2}. 

The svivon energy minima are lower below $T_c$ than their values (${\bar
\epsilon}^s_{\rm min}$) above it, and since the low-energy svivon
spectral tails existing above $T_c$ (see Fig.~\ref{fig6} and Appendix B)
have disappeared, a spin gap opens up. This results \cite{AshkHam} in
the existence of a {\it sharp} resonance mode below $T_c$. Since (see
above) the energy separation between the stripon and humpon peaks
\cite{Eres} $>\omega^{\lambda}({\bf Q}) \simeq {3 \over 4} E_{_{\rm
res}}$, the reduction in the width of the stripon peak results in the
appearance of a dip between the peak and the hump (see
Fig.~\ref{fig4}(b)), at $\sim$$E_{_{\rm res}}$ above the peak, in
agreement with experiment \cite{Zasadzinski, Fischer}. This structure is
manifested as a $T$-dependent ``antinodal kink'' in the electron
dispersion curves \cite{Sato}. 

The balance between energy gains due to pairing and inhomogeneity,
discussed above with respect to the PG phase, is relevant also for the
SC state, especially in the underdoped regime. The fact that the lattice
is involved in the establishment of the static checkerboard-like
\cite{Davis1} structure (see above) is supported by the observation of
an unconventional isotope effect within this regime \cite{Keller}. It is
viewed {\it not only} in $T_c$, but also in properties like the magnetic
penetration depth, in $T^*$, and in the onset temperatures of spin-glass
and AF ordering in adjacent inhomogeneous regimes of the phase diagram
\cite{Keller}. 

Since the local lattice distortions have a non-linear negative effect on
the pairing energy, a heterogeneous nanoscale structure sets in
\cite{Davis2}, including regions of varying strengths of the effects of
pairing and inhomogeneity. Regions of a stronger pairing effect are
characterized by a larger spectral weight within the peak (on the
expense of that within the hump), and a weaker checkerboard-like
structure. Since the static inhomogeneities have only a minor effect on
the nodal gap, this heterogeneous structure has almost no effect on it
(it has a significant effect on the antinodal gap), in agreement with
experiment \cite{Davis2}. 

\section{Constraint Susceptibility}

The spectrum $\omega^{\lambda}({\bf q})$ of the lagrons, and the
constants $\gamma({\bf q})$ of their coupling to QEs and svivons, have
to be determined through the condition that the resulting QE and svivon
spectra satisfy, in every site, the auxiliary-particles' constraint in
Eq.~(\ref{eq0d}). This condition can be expressed as a requirement
\cite{AshkHam} that two-site correlation functions of the svivon
operators in the lhs of Eq.~(\ref{eq0d}) is equal to that defined,
similarly,through the QE operators in the rhs of Eq.~(\ref{eq0d}); thus:
\begin{equation}
\sum_{\sigma\sigma^{\prime}} \langle s_{i\sigma}^{\dagger}
s_{i\sigma}^{\dagg} s_{j\sigma^{\prime}}^{\dagger}
s_{j\sigma^{\prime}}^{\dagg} \rangle \cong \langle q_{i}^{\dagger}
q_{i}^{\dagg} q_{j}^{\dagger} q_{j}^{\dagg} \rangle. \label{eq5a} 
\end{equation}

The terms in the lhs and the rhs of Eq.~(\ref{eq5a}) introduce a
susceptibility-like function, referred to as the
constraint-susceptibility \cite{AshkHam} ($\chi^s_{\rm c}$ or
$\chi^q_{\rm c}$); it should be the same (thus $\chi^s_{\rm c} =
\chi^q_{\rm c}$), whether it is evaluated on the basis of the svivon
spectrum (thus $\chi^s_{\rm c}$), or the QE spectrum (thus $\chi^q_{\rm
c}$), through the lhs or the rhs of Eq.~(\ref{eq5a}), respectively. 
Vertex corrections to $\chi^s_{\rm c}$ and $\chi^q_{\rm c}$, due to
$\Delta{\cal H}$ in Eq.~(\ref{eq4i}) and ${\cal H}^{\prime}$ in
Eq.~(\ref{eq6c}), do not vary them qualitatively.

A major feature of $\Im \chi^s_{\rm c}({\bf q}, \omega-i0^+)$, in the SC
state, is \cite{AshkHam} the existence, around ${\bf q}=0$, of a sharp
low-energy peak at $\omega \simeq E_{_{\rm res}}$, and a higher-energy
tail, extending up to $\sim$$J$; the $\chi^q_{\rm c} = \chi^s_{\rm c}$
equality implies that such a peak structure must be also a major feature
of $\Im \chi^q_{\rm c}({\bf q}, \omega-i0^+)$ in the SC state; it can
be, approximately, evaluated through \cite{Vliet, Mahan} (see
Eqs.~(\ref{eq21}), (\ref{eq22}), (\ref{eq24}) and (\ref{eq5a})): 
\begin{eqnarray}
\chi^q_{\rm c}({\bf q},z) &\cong& {1 \over N} \sum_{\bf k} \int
d\omega_1 \int d\omega_2 \Big\{ [\cos{^2(\xi_{{\bf k}}^q(\omega_1))}
A^q_{+}({\bf k}, \omega_1) \nonumber \\ &\ & + \sin{^2(\xi_{{\bf
k}}^q(\omega^q))} A^q_{-}({\bf k}, \omega_1)] [\cos{^2(\xi_{{\bf k}-{\bf
q}}^q(\omega_2))} \nonumber \\ &\ & \times A^q_{+}({\bf k}-{\bf q},
\omega_2) + \sin{^2(\xi_{{\bf k}-{\bf q}}^q(\omega_2))} \nonumber \\ &\
& \times A^q_{-}({\bf k}-{\bf q}, \omega_2)] + \quarter \sin{(2\xi_{{\bf
k}}^q(\omega_1))} \nonumber \\ &\ & \times \sin{(2\xi_{{\bf k}-{\bf
q}}^q(\omega_2))} \cos{[\phi_{{\bf k}}^q(\omega_1) - \phi_{{\bf k}-{\bf
q}}^q(\omega_2)]} \nonumber \\ &\ & \times [A^q_{+}({\bf k}, \omega_1) -
A^q_{-}({\bf k}, \omega_1)] [A^q_{+}({\bf k}-{\bf q}, \omega_2)
\nonumber \\ &\ & - A^q_{-}({\bf k}-{\bf q}, \omega_2)] \Big\}
{f_{_T}(\omega_2) - f_{_T}(\omega_1) \over z + \omega_2 - \omega_1}.
\label{eq28} 
\end{eqnarray}

The structure of a peak in $\Im \chi^q_{\rm c}({\bf q} = 0,
\omega-i0^+)$, obtained through Eq.~(\ref{eq28}), represents an average
on the same-${\bf k}$-point transitions between negative- and
positive-energy QE spectral features (thus on the two sides of the SC
gap, as is sketched in Fig.~\ref{fig4}(a)). These features consist of
single sharp arcon peaks in the nodal areas, and of separate stripon and
(broad) humpon peaks in the antinodal areas. 

Thus, an equality between $\Im \chi^q_{\rm c}$ and $\Im \chi^s_{\rm c}$,
at ${\bf q} = 0$ in the SC state, means that the sharp low-energy peak
corresponds to averaged same-${\bf k}$ transitions between the positive-
and negative-energy arcon peaks, while the higher-energy tail
corresponds to averaged same-${\bf k}$ transitions between the positive-
and negative-energy stripon--humpon spectral features. The proximity of
the low-energy peak to $E_{_{\rm res}}$, and of the extent of the
higher-energy tail to $\sim$$J$, is consistent with the above-mentioned
values of the SC nodal gap \cite{Kresin}, and the humpon energies,
respectively. This demonstrates the qualitative self-consistency of the
theoretical approach, the lagron spectrum, and the spectral functions
presented here. 

The constraint-susceptibility analysis demonstrates that the connection
between features (such as typical energies) of the QE and lagron
spectra, derived here, is closely related to the auxiliary-particles'
constraint. This results in similarities between spectroscopic features
derived here and in models where electrons are coupled to spin
fluctuations. However, since such models do not correspond to the
large-$U$ case (studied here), a quantitative analysis based on them
\cite{Carbotte} does not yield a realistic spectrum of the relevant spin
fluctuations. 

\section{Conclusions}

In conclusion, a theory for highly correlated layered SCs, where the
low-energy excitations are approached in terms of combinations of
atomic-like electron configurations, rather than approximately
independent electrons, has been shown to resolve qualitative mysteries
of the cuprates. A Lagrange Bose field which accounts for the tendency
of the doped system to form stripe-like inhomogeneities, enables
treating these configurations as bosons or fermions. 

The addressed anomalous properties of the hole-doped cuprates include
the observed phase diagram, non-FL to FL crossover, the existence of MFL
critical behavior and a PG phase with Fermi arcs, kink- and
waterfall-like spectral features, the drop in the scattering rates in
the PG phase, and further in the SC phase, an effective increase in the
density of carriers with $T$ and $\omega$, the correspondence between
$T_c$, $E_{_{\rm res}}$, and the SC nodal gap, {\it etc}. 

Electron-lattice coupling is not included in the Hamiltonian treated
here; however, such coupling is necessary to explain the establishment
of static inhomogeneous structures within the PG and SC states. This
coupling is expected to be strong for the derived low-energy spectral
peaks, introducing phonon anomalies at comparable energies, an anomalous
isotope effect \cite{Keller}, {\it etc.} 

The electronic structure of the FeSCs differs from that of the cuprates.
However, the similarity between their anomalous properties
\cite{AshkHam}, and the fact that a formally common many-body
Hamiltonian could be worked out \cite{AshkHam} for both systems, implies
that part of the conclusions drawn above about the physics of the
cuprates, apply also for the FeSCs, with some modification. Further
details of the theory, and a comparison between its consequences for the
cuprates and the FeSCs, will appear in forthcoming papers. 

\begin{acknowledgments} 
The author acknowledges the encouragement of Davor Pavuna, and
constructive discussions with him. He also benefited from stimulating
discussions with Stewart Barnes, Antonio Bianconi, Annette
Bussmann-Holder, Neil Johnson, Hugo Keller, Dirk Manske, Alex M\"uller,
David Tanner, Dirk van der Marel, Carolyne Van Vliet, and other
colleagues and members of the high-$T_c$ community. 
\end{acknowledgments}

\appendix

\section{Evaluation of the QE Scattering Rates}

The behavior of $\Gamma^q$ is determined at ${\bf k}$ points within the
low-energy nodal and antinodal areas, and the adjacent high-energy
areas; it could be expressed as a sum of two terms: 
\begin{equation}
\Gamma^q({\bf k},\omega) = \Gamma^q_1({\bf k},\omega) + \Gamma^q_2({\bf
k},\omega), \label{eq9a} 
\end{equation}
corresponding to the contributions of the ${\bf Q}_m$ lagrons
($\Gamma^q_1$), and of the other lagrons ($\Gamma^q_2$), to the ${\bf
q}$ summation in Eq.~(\ref{eq9}). 

By Eqs.~(\ref{eq5}, \ref{eq9}), $\Gamma^q_1({\bf k}, \omega)$ in
Eq.~(\ref{eq9a}) can be approximately expressed as: 
\begin{equation}
\Gamma^q_1({\bf k},\omega) \propto \sum_{m=1}^4 A^q({\bf k}-{\bf Q}_m,
\omega), \label{eq10} 
\end{equation}
where the combination in the rhs is of terms specified as $A^q_{\rm
le}(\omega)$, $A^q_{\rm me}(\omega)$ and $A^q_{\rm he}(\omega)$ (see
above), consisting of the arcon, stripon and humpon peaks, shown in
Fig.~\ref{fig4}(a), and the higher-energy spectrum, shown in
Fig.~\ref{fig3}(a-b); terms scaling with them contribute, through
Eq.~(\ref{eq10}), to low-$|\omega|$ $\Gamma^q$ within the antinodal
areas, and partially within the nodal areas (not too close to the lines
of nodes) and the adjacent high-energy areas. $A^q_{\rm le}(\omega \to
0)$ (and its contribution to $\Gamma^q_1(\omega \to 0)$) decreases when
$T$ increases (see below), and $\lim_{_{T \to 0}} A^q_{\rm le}(\omega
\to 0)$ does not vanish (for unpaired QEs). Consequently, in ${\bf k}$
points within the above areas there exists a nonzero limit: 
\begin{equation}
\Gamma^q_0({\bf k}) \equiv \lim_{_{T \to 0}} \Gamma^q_1({\bf k}, \omega
\to 0). \label{eq10a} 
\end{equation}

By Eq.~(\ref{eq8}), and the lack of contributions due to $\Gamma^q_2$
(see below), $\Gamma^q_0({\bf k})$ determine the zero-$T$ limit of the
widths of the stripon and arcon peaks which are, consequently, $\sim$$0$
close to the lines of nodes. Let $\Gamma^q_{0{\rm a}}$ be a typical
value of $\Gamma^q_0({\bf k})$ in the antinodal areas; since it is
determined by the coupling of QEs, through Eqs.~(\ref{eq9},\ref{eq10a}),
to ${\bf Q}_m$ lagrons, while $\omega_{_{\rm H}}$ is determined by their
coupling, through Eq.~(\ref{eq7}), to {\it both} ${\bf Q}_m$ and other
lagrons (see below), $\Gamma^q_{0{\rm a}}$ is few times smaller than
$\omega_{_{\rm H}}$. 

A ${\bf k}$-dependent low-energy scale $\omega_{_{\rm L}}({\bf k})$
exists which is close to $\Gamma^q_{0{\rm a}}$ in the antinodal areas;
for $k_{_{\rm B}}T\; \&\; |\omega| \ll \omega_{_{\rm L}}({\bf k})$,
$A^q({\bf k}, \omega)$ is dominated by the contribution of
$\Gamma^q_0({\bf k})$ to Eq.~(\ref{eq8}), and it is almost independent
of $T$ and $\omega$ within this range. For $k_{_{\rm B}}T \gg
\omega_{_{\rm L}}({\bf k})$, the $T$- and $\omega$-dependencies of the
stripon/arcon peak center (discussed above), and of $\Gamma^q$, largely
determine the behavior of $A^q_{\rm le}(\omega)$, through
Eq.~(\ref{eq8}). Its low-$\omega$ $T$-dependence is determined by the
broadening of the peak (through $\Gamma^q$) which, self-consistently,
scales with $T$ (see below), resulting in an approximate scaling of the
low-$\omega$ $A^q_{\rm le}$ with $1/T$. 

$\Gamma^q_2({\bf k}, \omega)$ in Eq.~(\ref{eq9a}) includes the part in
the rhs of Eq.~(\ref{eq9}) where only the four ${\bf Q}_m$ points in
Eq.~(\ref{eq5}) are omitted from the sum over ${\bf q}$. This summation
requires a 2D integration which could be carried out by dividing the BZ
into a mesh based on ${\bf q}_{\parallel}$ and ${\bf q}_{\perp}$ lines,
directed along lagron-energy gradients, and perpendicular to them,
respectively. 
 
As can be viewed in Figs.~\ref{fig2} and \ref{fig3}(c), for a
sufficiently fine mesh, each BZ section (specified as s) within it
could be approximated as a circular segment where $\omega^{\lambda}({\bf
q})$ is linear in ${\bf q}_{\parallel}$, and $A^q({\bf k} - {\bf q},
\omega \mp \omega^{\lambda}({\bf q}))$ in Eq.~(\ref{eq9}) could be
expressed as $A^q_{\rm s}(\omega \mp \omega^{\lambda}({\bf q}))$, where
$A^q_{\rm s}$ is of the form of either $A^q_{\rm le}$ or $A^q_{\rm me}$
or $A^q_{\rm he}$. Consequently, the ${\bf q}$ integral over the
circular segment s is, approximately, proportional to an integral on
$d\omega^{\lambda}$, where the integrand is multiplied by
($\omega^{\lambda} - \omega^{\lambda q}_{0{\rm s}}$), between positive
limits $\omega^{\lambda q}_{n{\rm s}}$ and $\omega^{\lambda q}_{x{\rm
s}}$ ($\omega^{\lambda q}_{0{\rm s}}$ is a constant). 

Fig.~\ref{fig2} indicates that $\omega^{\lambda q}_{0{\rm s}} =
\omega^{\lambda q}_{n{\rm s}} = 0$ within four circular BZ sections
around the V-shape lagron energy minima at the ${\bf Q}_m$ points, and
that $|\omega^{\lambda q}_{0{\rm s}}|$ are large in integration sections
corresponding to ${\bf Q}$-ESP lagrons (where $\omega^{\lambda} \simeq
\omega^{\lambda}({\bf Q})$) and to high-energy SP lagrons. 

The major features of $\Gamma^q_2({\bf k}, \omega)$, in the low- and
high-$|\omega| / k_{_{\rm B}}T$ limits, are obtained when the following
approximations are applied within each of the above $d\omega^{\lambda}$
integration ranges: (a) the asymptotic behavior of
$b_{_T}(\omega^{\lambda})$, for $\omega^{\lambda} \ll k_{_{\rm B}}T$ and
$\omega^{\lambda} \gg k_{_{\rm B}}T$, (thus $b_{_T}(\omega^{\lambda})
\cong k_{_{\rm B}}T / \omega^{\lambda}$, for $\omega^{\lambda} \ll
k_{_{\rm B}}T$, and $b_{_T}(\omega^{\lambda}) \cong 0$, for
$\omega^{\lambda} \gg k_{_{\rm B}}T$) is extended to $\omega^{\lambda} <
k_{_{\rm B}}T$ and $\omega^{\lambda} > k_{_{\rm B}}T$, respectively; (b)
the low-$T$ limit of $f_{_T}(\omega^{\prime})$ (thus
$f_{_T}(\omega^{\prime}) \cong 1$, for $\omega^{\prime} < 0$, and
$f_{_T}(\omega^{\prime}) \cong 0$, for $\omega^{\prime} > 0$) is applied
for $f_{_T}(\omega^{\lambda} \mp \omega)$. This yields, through
Eq.~(\ref{eq9}): 
\begin{eqnarray} 
\Gamma^q_2({\bf k},\omega) &=& \sum_{\rm s} [ \Gamma^q_{T{\rm s}}({\bf
k},\omega) + \Gamma^q_{\omega{\rm s}}({\bf k},\omega) ], \ \ \
\text{where} \nonumber \\ 
\Gamma^q_{T{\rm s}}({\bf k},\omega) &\propto& k_{_{\rm B}}T
\int_{\omega^{\lambda q}_{Tn{\rm s}}}^{\omega^{\lambda q}_{Tx{\rm s}}}
d\omega^{\lambda} (1 - \omega^{\lambda q}_{0{\rm s}} / \omega^{\lambda})
[A^q_{\rm s}(\omega - \omega^{\lambda}) \nonumber \\ &\ &+ A^q_{\rm
a}(\omega + \omega^{\lambda})], \label{eq11} \\ 
\Gamma^q_{\omega{\rm s}}({\bf k},\omega) &\propto& \int_{\omega^{\lambda
q}_{\omega n{\rm s}}}^{\omega^{\lambda q}_{\omega x{\rm s}}}
d\omega^{\lambda} (\omega^{\lambda} - \omega^{\lambda q}_{0{\rm s}})
A^q_{\rm s}(\omega - \omega^{\lambda} \text{sign}(\omega)), \nonumber \\
\omega^{\lambda q}_{Tn{\rm s}} &=& \min{(k_{_{\rm B}}T, \omega^{\lambda
q}_{n{\rm s}})}, \ \ \omega^{\lambda q}_{Tx{\rm s}} = \min{(k_{_{\rm
B}}T, \omega^{\lambda q}_{x{\rm s}})}, \nonumber \\ 
\omega^{\lambda q}_{\omega n{\rm s}} &=& \min{(|\omega|, \omega^{\lambda
q}_{n{\rm s}})}, \ \ \omega^{\lambda q}_{\omega x{\rm s}} =
\min{(|\omega|, \omega^{\lambda q}_{x{\rm s}})}. \nonumber 
\end{eqnarray} 
 
By the results presented in Figs.~\ref{fig2} and \ref{fig3}(a-c), one
could evaluate the contributions $\delta\Gamma^q_2$ to $\Gamma^q$,
within the different BZ areas, due to terms including different forms of
$A^q_{\rm s}$ in Eq.~(\ref{eq11}). The $T$- and $\omega$-dependencies of
these $\delta\Gamma^q_2$ terms depend on those of $A^q_{\rm s}$, on the
integration limits in Eq.~(\ref{eq11}), and on the number of nonzero
integrals there. Within the antinodal areas, significant values of
$\delta\Gamma^q_2$ are obtained through Eq.~(\ref{eq11}) for $\omega \to
0$, while close to the lines of nodes (where $\Gamma^q_0({\bf k})$ to
Eq.~(\ref{eq10a}) is negligible) they contribute significantly only when
$|\omega|$ exceeds an energy which is somewhat smaller than
$\omega^{\lambda}({\bf Q})$. Thus, this energy determines the low-energy
scale $\omega_{_{\rm L}}({\bf k})$ (see above) around the lines of
nodes. 

As was discussed above, for $k_{_{\rm B}}T\; \&\; |\omega| \ll
\omega_{_{\rm L}}({\bf k}^{\prime})$, $A^q_{\rm le}$ in this ${\bf
k}^{\prime}$ point is approximately independent of $T$ and $\omega$,
within the low-$\omega^{\lambda}$ integration sections in
Eq.~(\ref{eq11}), resulting in weak $T$ and $\omega$ dependencies of the
corresponding $\delta\Gamma^q_2$. For $k_{_{\rm B}}T \gg \omega_{_{\rm
L}}({\bf k}^{\prime})$ and $|\omega| \ll k_{_{\rm B}}T$, since $A^q_{\rm
le}$, self-consistently, scales as $1/T$ (see above), $\delta\Gamma^q_2$
approximately scales with $T$ by Eq.~(\ref{eq11}). For $|\omega| \gg
\omega_{_{\rm L}}({\bf k}^{\prime})$ and $k_{_{\rm B}}T \ll |\omega|$,
$A^q_{\rm le}$ can be approximated in Eq.~(\ref{eq11}) by a
$\delta$-function around the energy of the corresponding arcon or
stripon peak, resulting in $\delta\Gamma^q_2$ which is linear in
$\omega$. The high-$|\omega|$ extent of this term is limited by the
high-energy extent of the lagron spectrum in Fig.~\ref{fig2}, and the
high-$T$ extent is limited by the phase stability.

When $k_{_{\rm B}}T$ or $|\omega|$ exceeds $\omega_{_{\rm H}}$,
$A^q_{\rm me}$ (due to the humpons---see above) behaves similarly to
$A^q_{\rm le}$, and the $\delta\Gamma^q_2$ term it contributes also has,
in this range, a linear dependence on $T$ in the low-$|\omega|$ limit,
and a linear dependence on $\omega$, in the low-$T$ limit. This results
in an increase in the slope of the $\Gamma^q$ {\it vs} $T$ ($|\omega|$)
curve as the increasing $k_{_{\rm B}}T$ ($|\omega|$) approaches
$\omega_{_{\rm H}}$. A similar slope increase, due to $A^q_{\rm he}$, is
expected when their high-energy maxima are approached, {\it as long as}
they do not exceed the high-energy extent of the lagron spectrum.
However, since (see Fig.~\ref{fig3}(a-b)) there is an apparent
discontinuity between the QE low- and high-energy BZ areas, where the
energies at the maxima of $A^q_{\rm he}$ mostly exceed those of the
lagron spectrum, the derived contribution of $A^q_{\rm he}$ to
$\Gamma^q$ is minute. 

\section{Evaluation of the svivon Scattering Rates}

The $\cosh{^2(\xi_{{\bf k}^{\prime}}^s(\omega^{\prime}))}$ and
$\sinh{^2(\xi_{{\bf k}^{\prime}}^s(\omega^{\prime}))}$ factors,
appearing in the expression for $\Gamma^s_{\lambda}$ in
Eq.~(\ref{eq12}), are close to $\cosh{(2\xi_{{\bf
k}^{\prime}}^s(\omega^{\prime}))}$ within the LE and ME svivon BZ areas,
where the major contribution to the expression comes from; thus their
$T$ and $\omega$ dependencies there are close to those of
$\cosh{(2\xi_{{\bf k}^{\prime}}^s(\omega^{\prime}))}$, derived above
through Eq.~(\ref{eq13b}). The effect of their $\propto 1/(T-T_0)$
behavior within the LE areas is considered {\it together with} the size
$\propto (T-T_0)^2$ of these areas, resulting in an approximate scaling
with $(T-T_0)$. The associated decrease in the size of the ME areas
introduces in these areas a minor $\propto (T-T_0)^2$ term, as in
$\cosh{(2\xi_{{\bf k}^{\prime}}^s(\omega^{\prime}))}$. 

The values of $n^s({\bf k})$ are determined through Eq.~(\ref{eq6h}),
and their variation with $T$ takes place mainly in the LE BZ areas.
However, since $n^s$ is $T$ independent, one gets through
Eq.~(\ref{eq6b}) that the $T$ variation is minor for the averaged values
of $n^s({\bf k})$ over the LE areas and the crossover areas between them
and the ME areas. This applies also for the $T$ dependence of the
$|m^s({\bf k}^{\prime})|$ factors, appearing in the expression for
$\Gamma^s_q$ in Eq.~(\ref{eq12}), due to the proximity in the LE areas
between the expressions for $n^s({\bf k})$ and $m^s({\bf k})$ in
Eq.~(\ref{eq6h}). 
 
$\Gamma^s_q({\bf k}, \omega)$ represents the scattering of a svivon into
a particle-hole QE pair, while its spin is condensed; at low  $T$ and
$\omega$, it is determined by the stripon and arcon $A^q_{\rm le}$
spectral functions. In the $k_{_{\rm B}}T\; \&\; |\omega| \ll
\Gamma^q_{0{\rm a}}$ regime (see Eq.~(\ref{eq10a}) and the discussion in
Appendix A), $A^q_{\rm le}$ is, approximately, constant within the
integration range in Eq.~(\ref{eq12}); in the LE areas, this results in
$\Gamma^s_q$ which approximately scales with $\omega/(T-T_0)$, for
$|\omega| \lta {\bar \epsilon}^s_{\rm min}$, its magnitude passes
through a maximum at higher $|\omega|$ values, and crosses over to
scaling with $\omega$ in the $|\omega| \gg {\bar \epsilon}^s_{\rm min}$
limit, if it could be reached in this regime; as was discussed above, in
processes where $\Gamma^s_q$ is integrated over a part of the BZ, the
contribution of the LE areas to it scales with $\omega(T-T_0)$, for
$|\omega| \lta {\bar \epsilon}^s_{\rm min}$; $|\Gamma^s_q|$ is smaller,
in this regime, in the ME areas (and much smaller in the HE areas), and
approximately scales there with $\omega$, plus a minor term $\propto
\omega(T-T_0)^2$ (and it may pass through a maximum). 

When $k_{_{\rm B}}T \gg \Gamma^q_{0{\rm a}}\; \&\; |\omega|$, the
$T$-scaled broadening of $A^q_{\rm le}(\omega)$ (see Appendix A) results
in a $1/T^2$-scaled decrease of the above values of $\Gamma^s_q$ with
$T$. For $|\omega| \gg \Gamma^q_{0{\rm a}}\; \&\; k_{_{\rm B}}T$, the
contribution of particle-hole QE pairs, based on $A^q_{\rm le}$, to
$\Gamma^s_q$ is vanishing, resulting in its sharp decrease; a
high-$\omega$ contribution to $\Gamma^s_q$ due to humpons and
high-energy QEs is small compared to $\Gamma^s_{\lambda}$, for such
values of $\omega$. 

The major physical effect of $\Gamma^s_q$ is the introduction, {\it
right above} \cite{pairing} $T^q_{\rm p}$, of a width $|\Gamma^s|$ to
the svivon states around their energy minima which extends to
exceedingly low (nonzero) energies $|\omega|$. The corresponding
spectral functions, obtained through Eq.~(\ref{eq6g}), can be
approximately expressed as: 
\begin{eqnarray}
A^s({\bf k}, \omega) &\simeq& {\Gamma^s({\bf k}, \omega) / 2\pi \over
[\omega - {\bar \epsilon}^s({\bf k})]^2 + [\half \Gamma^s({\bf k},
\omega)]^2}  \nonumber \\ &\simeq& {\Gamma^s({\bf k}, \omega) \over 2\pi
[{\bar \epsilon}^s({\bf k})]^2}, \ \text{for}\ |\omega| \ll {\bar
\epsilon}^s({\bf k}). \label{eq14} 
\end{eqnarray}
The $\omega$ scaling of $\Gamma^s_q$, at low $\omega$ and $T$, results
in an $\omega$-independent product $A^s({\bf k}, \omega) b_{_T}(\omega)$
in the low-$\omega$ limit. 

Similarly to $\Gamma^q$ in Eq.~(\ref{eq9a}), $\Gamma^s_{\lambda}$ could
be expressed as a sum of two terms: 
\begin{equation}
\Gamma^s_{\lambda}({\bf k},\omega) = \Gamma^s_1({\bf k},\omega) +
\Gamma^s_2({\bf k},\omega), \label{eq15} 
\end{equation}
corresponding to the contributions of the ${\bf Q}_m$ lagrons
($\Gamma^s_1$), and of the other lagrons ($\Gamma^s_2$), to the ${\bf
q}$ summation in Eq.~(\ref{eq12}). As in the case of $\Gamma^q_1$ in
Eq.~(\ref{eq10}), $\Gamma^s_1$ can be approximately expressed as: 
\begin{eqnarray}
\Gamma^s_1({\bf k},\omega) &\propto&  \cosh{(2\xi_{\bf k}^s(\omega))}
\label{eq16} \\ &\times& \sum_{m=1}^4 [A^s({\bf k}-{\bf Q}_m, \omega)
\cosh{^2(\xi_{{\bf k}-{\bf Q}_m}^s(\omega))} \nonumber \\ &\ & -
A^s({\bf k}-{\bf Q}_m, -\omega) \sinh{^2(\xi_{{\bf k}-{\bf
Q}_m}^s(-\omega))}]. \nonumber 
\end{eqnarray}
Thus, $|\Gamma^s_1({\bf k},\omega)|$ has maxima combined of the maxima
of $A^s({\bf k}-{\bf Q}_m, |\omega|)$ at $|\omega| \cong {\bar
\epsilon}^s({\bf k}-{\bf Q}_m)$, for the four ${\bf Q}_m$ points, and
decreases to zero at higher energies. Within a svivon condensate, ${\bar
\epsilon}^s({\bf k}-{\bf Q}_m) \simeq {\bar \epsilon}^s({\bf k})$ for
one of the ${\bf Q}_m$ points, and $|\Gamma^s_1({\bf k},\omega)|$ is
largest when both ${\bf k}$ and ${\bf k}-{\bf Q}_m$ are at the LE areas.
In the low $\omega/k_{_{\rm B}}T$ limit, $\Gamma^s_1$ is negligible (see
Eqs.~(\ref{eq14}, \ref{eq16})) compared to $\Gamma^s_q$ and $\Gamma^s_2$
(see below). 

An expression for $\Gamma^s_2({\bf k},\omega)$ is derived analogously to
the one for $\Gamma^q_2({\bf k},\omega)$ in Eq.~(\ref{eq11}). The ${\bf
q}$ summation in Eq.~(\ref{eq12}) is, similarly, carried out by using a
BZ mesh based on circular segments s, where the $A^s({\bf k}-{\bf q},
\omega \mp \omega^{\lambda}({\bf q})) \cosh{^2(\xi_{{\bf k} - {\bf
q}}^s(\omega \mp \omega^{\lambda}({\bf q}))}$ and $A^s({\bf k}-{\bf q},
-\omega \pm \omega^{\lambda}({\bf q})) \sinh{^2(\xi_{{\bf k} - {\bf
q}}^s(-\omega \pm \omega^{\lambda}({\bf q}))}$ terms in Eq.~(\ref{eq12})
are expressed as $A^s_{\rm s}(\omega \mp \omega^{\lambda}({\bf q}))
\cosh{^2(\xi_{\rm s}^s(\omega \mp \omega^{\lambda}({\bf q}))}$ and
$A^s_{\rm s}(-\omega \pm \omega^{\lambda}({\bf q})) \sinh{^2(\xi_{\rm
s}^s(-\omega \pm \omega^{\lambda}({\bf q}))}$, respectively. 

As in the case of $\Gamma^q_2$, the ${\bf q}$ integral over section s
is, approximately, proportional to an integral on $d\omega^{\lambda}$,
where the integrand is multiplied by ($\omega^{\lambda} -
\omega^{\lambda s}_{0{\rm s}}$), between positive limits
$\omega^{\lambda s}_{n{\rm s}}$ and $\omega^{\lambda s}_{x{\rm s}}$;
again, one has $\omega^{\lambda s}_{0{\rm s}} = \omega^{\lambda
s}_{n{\rm s}} = 0$ within four circular BZ sections around the V-shape
lagron energy minima at the ${\bf Q}_m$ points, and $|\omega^{\lambda
s}_{0{\rm s}}|$ are large in integration sections corresponding to ${\bf
Q}$-ESP and high-energy SP lagrons. 

Similarly to the evaluation of $\Gamma^q_2$ in Appendix A, the
$b_{_T}(\omega^{\prime})$ functions in Eq.~(\ref{eq12}) (where
$\omega^{\prime}$ is either $\omega^{\lambda}$ or $\omega^{\lambda} \mp
\omega$) are approximated through their asymptotic behaviors: (a)
$b_{_T}(\omega^{\prime}) \cong k_{_{\rm B}}T / \omega$, for
$|\omega^{\prime}| \ll k_{_{\rm B}}T$; (b) $b_{_T}(\omega^{\prime})
\cong -1$, for $\omega^{\prime} \ll -k_{_{\rm B}}T$; (c)
$b_{_T}(\omega^{\prime}) \cong 0$, for $\omega^{\prime} \gg k_{_{\rm
B}}T$. In order to obtain the major features of $\Gamma^s_2({\bf k},
\omega)$ in the low- and high-$|\omega| / k_{_{\rm B}}T$ limits, it is
formulated in term of expressions where the asymptotic behavior of
$b_{_T}$ in (a) is extended to $|\omega^{\prime}| < k_{_{\rm B}}T$, and
its behaviors in (b) and (c) are extended to $\omega^{\prime} <
-k_{_{\rm B}}T$ and $\omega^{\prime} > k_{_{\rm B}}T$, respectively.
Furthermore, when both $0 < \omega^{\lambda} < k_{_{\rm B}}T$ and $0 <
\omega^{\lambda} \mp \omega < k_{_{\rm B}}T$, the differences
$b_{_T}(\omega^{\lambda}) - b_{_T}(\omega^{\lambda} \mp \omega)$ are
approximated (through derivation) as $\mp \omega k_{_{\rm B}}T /
[\omega^{\lambda} \mp \omega/2]^2$. Consequently, one gets through
Eq.~(\ref{eq12}): 
\begin{widetext}
\begin{eqnarray} 
\Gamma^s_2({\bf k},\omega) &=& \cosh{(2\xi_{\bf k}^s(\omega))} \sum_{\rm
s} [ \Gamma^s_{T{\rm s}}({\bf k},\omega) + \Gamma^s_{\tau{\rm s}}({\bf
k},\omega) + \Gamma^s_{\omega{\rm s}}({\bf k},\omega) ], \ \ \
\text{where} \nonumber \\ 
\Gamma^s_{T{\rm s}}({\bf k},\omega) &\propto& k_{_{\rm B}}T \bigg\{
\int_{\omega^{\lambda s}_{1n{\rm s}}}^{\omega^{\lambda s}_{1x{\rm s}}}
d\omega^{\lambda} (1 - \omega^{\lambda s}_{0{\rm s}} / \omega^{\lambda})
[A^s_{\rm s}(\omega - \omega^{\lambda}\text{sign}(\omega))
\cosh{^2(\xi_{\rm s}^s(\omega - \omega^{\lambda}\text{sign}(\omega)))} -
A^s_{\rm s}(\omega^{\lambda}\text{sign}(\omega) - \omega) \nonumber \\
&\ & \times \sinh{^2(\xi_{\rm s}^s(\omega^{\lambda}\text{sign}(\omega) -
\omega))}] 
- \bigg[ \int_{\omega^{\lambda s}_{2n{\rm s}}}^{\omega^{\lambda
s}_{2x{\rm s}}} + \int_{\omega^{\lambda s}_{3n{\rm s}}}^{\omega^{\lambda
s}_{3x{\rm s}}} \bigg] d\omega^{\lambda} \Big[{\omega^{\lambda} -
\omega^{\lambda s}_{0{\rm s}} \over \omega^{\lambda} - |\omega|} \Big]
[A^s_{\rm s}(\omega - \omega^{\lambda}\text{sign}(\omega)) \nonumber \\
&\ & \times \cosh{^2(\xi_{\rm s}^s(\omega -
\omega^{\lambda}\text{sign}(\omega)))} - A^s_{\rm
s}(\omega^{\lambda}\text{sign}(\omega) - \omega) \sinh{^2(\xi_{\rm
s}^s(\omega^{\lambda}\text{sign}(\omega) - \omega))}] \nonumber \\ &\ & 
+ \int_{\omega^{\lambda s}_{4n{\rm s}}}^{\omega^{\lambda s}_{4x{\rm s}}}
d\omega^{\lambda} (1 - \omega^{\lambda s}_{0{\rm s}} / \omega^{\lambda})
[A^s_{\rm s}(\omega + \omega^{\lambda}\text{sign}(\omega))
\cosh{^2(\xi_{\rm s}^s(\omega + \omega^{\lambda}\text{sign}(\omega)))}
\nonumber \\ &\ & - A^s_{\rm s}(-\omega^{\lambda}\text{sign}(\omega) -
\omega) \sinh{^2(\xi_{\rm s}^s(-\omega^{\lambda}\text{sign}(\omega)-
\omega))}] \bigg\} \nonumber \\ 
\Gamma^s_{\tau {\rm s}}({\bf k},\omega) &\propto& |\omega|k_{_{\rm B}}T
\bigg\{ -\int_{\omega^{\lambda s}_{5n{\rm s}}}^{\omega^{\lambda
s}_{5x{\rm s}}} d\omega^{\lambda} {(\omega^{\lambda} - \omega^{\lambda
s}_{0{\rm s}}) \over (\omega^{\lambda} - |\omega|/2)^2} [A^s_{\rm
s}(\omega - \omega^{\lambda}\text{sign}(\omega)) \cosh{^2(\xi_{\rm
s}^s(\omega - \omega^{\lambda}\text{sign}(\omega)))} \nonumber \\ &\ & -
A^s_{\rm s}(\omega^{\lambda}\text{sign}(\omega) - \omega)
\sinh{^2(\xi_{\rm s}^s(\omega^{\lambda}\text{sign}(\omega) - \omega))}] 
+ \int_{\omega^{\lambda s}_{6n{\rm s}}}^{\omega^{\lambda s}_{6x{\rm s}}}
d\omega^{\lambda} {(\omega^{\lambda} - \omega^{\lambda s}_{0{\rm s}})
\over (\omega^{\lambda} + |\omega|/2)^2} [A^s_{\rm s}(\omega +
\omega^{\lambda}\text{sign}(\omega)) \nonumber \\ &\ & \times
\cosh{^2(\xi_{\rm s}^s(\omega + \omega^{\lambda}\text{sign}(\omega)))} -
A^s_{\rm s}(-\omega^{\lambda}\text{sign}(\omega) - \omega)
\sinh{^2(\xi_{\rm s}^s(-\omega^{\lambda}\text{sign}(\omega) - \omega))}]
\bigg\}, \label{eq17} \\ 
\Gamma^s_{\omega{\rm s}}({\bf k},\omega) &\propto& \int_{\omega^{\lambda
s}_{7n{\rm s}}}^{\omega^{\lambda s}_{7x{\rm s}}} d\omega^{\lambda}
(\omega^{\lambda} - \omega^{\lambda s}_{0{\rm s}}) [A^s_{\rm s}(\omega -
\omega^{\lambda} \text{sign}(\omega)) \cosh{^2(\xi_{\rm s}^s(\omega -
\omega^{\lambda} \text{sign}(\omega)))} \nonumber \\ &\ & - A^s_{\rm
s}(\omega^{\lambda} \text{sign}(\omega) - \omega) \sinh{^2(\xi_{\rm
s}^s(\omega^{\lambda} \text{sign}(\omega) - \omega))}], \ \ \ \text{and}
\nonumber \\ 
\omega^{\dagg}_{T1} &=& \min{(|\omega|, k_{_{\rm B}}T)}, \ \ \ \
\omega^{\dagg}_{T2} = \max{(0, |\omega| - k_{_{\rm B}}T)}, \ \ \ \
\omega^{\lambda s}_{an{\rm s}} = \min{(\omega^{\dagg}_{T2},
\omega^{\lambda s}_{x{\rm s}})}, \ \ \ \ \omega^{\lambda s}_{bn{\rm s}}
= \max{(\omega^{\lambda s}_{an{\rm s}}, \omega^{\lambda s}_{n{\rm s}})},
\nonumber \\ 
\omega^{\dagg}_{T3} &=& \max{(|\omega|, k_{_{\rm B}}T)}, \ \ \ \
\omega^{\lambda s}_{cn{\rm s}} = \min{\omega^{\dagg}_{T3},
\omega^{\lambda s}_{x{\rm s}})}, \ \ \ \ \omega^{\lambda s}_{dn{\rm s}}
= \max{(\omega^{\lambda s}_{cn{\rm s}}, \omega^{\lambda s}_{n{\rm s}})},
\nonumber \\ 
\omega^{\dagg}_{T4} &=& |\omega| + k_{_{\rm B}}T, \ \ \ \
\omega^{\dagg}_{T5} = \max{(0, k_{_{\rm B}}T - |\omega|)}, \ \ \ \
\omega^{\lambda s}_{en{\rm s}} = \min{(\omega^{\dagg}_{T5},
\omega^{\lambda s}_{x{\rm s}})}, \ \ \ \ \omega^{\lambda s}_{fn{\rm s}}
= \max{(\omega^{\lambda s}_{en{\rm s}}, \omega^{\lambda s}_{n{\rm s}})},
\nonumber \\ 
\omega^{\lambda s}_{gn{\rm s}} &=& \min{(\omega^{\dagg}_{T1},
\omega^{\lambda s}_{x{\rm s}})}, \ \ \ \ \omega^{\lambda s}_{hn{\rm s}}
= \max{(\omega^{\lambda s}_{gn{\rm s}}, \omega^{\lambda s}_{n{\rm s}})},
\ \ \ \ \omega^{\lambda s}_{1n{\rm s}} = \min{(\omega^{\dagg}_{T1},
\omega^{\lambda s}_{n{\rm s}})}, \ \ \ \ \omega^{\lambda s}_{1x{\rm s}}
= \min{(\omega^{\dagg}_{T1}, \omega^{\lambda s}_{x{\rm s}})}, \nonumber \\ 
\omega^{\lambda s}_{2n{\rm s}} &=& \min{(|\omega|, \omega^{\lambda
s}_{bn{\rm s}})}, \ \ \ \ \omega^{\lambda s}_{2x{\rm s}} =
\min{(|\omega|, \omega^{\lambda s}_{x{\rm s}})}, \ \ \ \ \omega^{\lambda
s}_{3n{\rm s}} = \min{(\omega^{\dagg}_{T4}, \omega^{\lambda s}_{dn{\rm
s}})}, \ \ \ \ \omega^{\lambda s}_{3x{\rm s}} =
\min{(\omega^{\dagg}_{T4}, \omega^{\lambda s}_{x{\rm s}})}, \nonumber \\
\omega^{\lambda s}_{4n{\rm s}} &=& \min{(k_{_{\rm B}}T, \omega^{\lambda
s}_{fn{\rm s}})}, \ \ \ \ \omega^{\lambda s}_{4x{\rm s}} =
\min{(k_{_{\rm B}}T, \omega^{\lambda s}_{x{\rm s}})}, \ \ \ \
\omega^{\lambda s}_{5n{\rm s}} = \min{(k_{_{\rm B}}T, \omega^{\lambda
s}_{hn{\rm s}})}, \ \ \ \ \omega^{\lambda s}_{5x{\rm s}} =
\min{(k_{_{\rm B}}T, \omega^{\lambda s}_{x{\rm s}})}, \nonumber \\ 
\omega^{\lambda s}_{6n{\rm s}} &=& \min{(\omega^{\dagg}_{T5},
\omega^{\lambda s}_{n{\rm s}})}, \ \ \ \ \omega^{\lambda s}_{6x{\rm s}}
= \min{(\omega^{\dagg}_{T5}, \omega^{\lambda s}_{x{\rm s}})}, \ \ \ \
\omega^{\lambda s}_{7n{\rm s}} = \min{(\omega^{\dagg}_{T2},
\omega^{\lambda s}_{n{\rm s}})}, \ \ \ \ \omega^{\lambda s}_{7x{\rm s}}
= \min{(\omega^{\dagg}_{T2}, \omega^{\lambda s}_{x{\rm s}})}. \nonumber 
\end{eqnarray} 
\end{widetext}

The dominant contribution to $\Gamma^s_2$, in the low $|\omega|/k_{_{\rm
B}}T$ limit, is of the $\Gamma^s_{\tau {\rm s}}$ terms in
Eq.~(\ref{eq17}). By Fig.~\ref{fig2}, their integration ranges which
contribute significantly to $\Gamma^s_2({\bf k},\omega)$ are, for
$k_{_{\rm B}}T < \sim$$\omega^{\lambda}({\bf Q})$, within the cone-like
areas around those ${\bf Q}_m$ points where ${\bf k} - {\bf Q}_m$ is
within the LE areas; the shape of the integration ranges is modified when
$k_{_{\rm B}}T$ exceeds $\sim$$\omega^{\lambda}({\bf Q})$. 

As was mentioned above, at a point ${\bf k}^{\prime}$ within the LE
areas, $A^s({\bf k}^{\prime}, \omega^{\prime})$ is not small in the
range $\omega^{\prime} \simeq k_{_{\rm B}}T$, due the effect of
$\Gamma^s_q$ at $T > T^q_{\rm p}$; by Eq.~(\ref{eq13a}), its $T$
dependence could then be approximated (within the range where it is
significant) in terms of a scaling factor $\alpha^s$ between
$\omega^{\prime}$ and $(T - T_0)$ which yields, taking into account
normalization: 
\begin{eqnarray} 
&\ & (T^{\prime} - T_0) A^s({\bf k}^{\prime}, \omega^{\prime} =
\pm\alpha^s (T^{\prime}-T_0)) \;@\; T=T^{\prime} \nonumber \\ &\cong&
(T^{\prime\prime} - T_0) A^s({\bf k}^{\prime}, \omega^{\prime} =
\pm\alpha^s (T^{\prime\prime}-T_0)) \;@\; T=T^{\prime\prime}, \nonumber
\\ &\ & \text{for }{\bf k}^{\prime} \in \text{the LE areas.}
\label{eq17a} 
\end{eqnarray} 
Since the $\cosh{^2}$ and $\sinh{^2}$ factors in the expression for
$\Gamma^s_{\tau {\rm s}}$, in Eq.~(\ref{eq17}), approximately scale in
the LE areas with $1/(T-T_0)$, they introduce a $1/(T-T_0)$ factor to the
integrals within these areas. Furthermore, since the size of the LE areas
scales with $(T-T_0)^2$, and $(T - T_0) A^s({\bf k}^{\prime},
\omega^{\lambda})$ has a scaling dependence on
$\omega^{\lambda}/(T-T_0)$ (see Eq.~(\ref{eq17a})), one could,
approximately, replace the variable $\omega^{\lambda}$ in these
integrals by $\omega^{\lambda}/(T-T_0)$, in the low $|\omega|/k_{_{\rm
B}}T$ limit at least for $k_{_{\rm B}}T < \sim$$\omega^{\lambda}({\bf
Q})$. 

This results in $\Gamma^s_2({\bf k},\omega)$ which approximately scales,
in the low $|\omega|/k_{_{\rm B}}T$ limit, with
$\omega/[(T-T_0)^2(1-T_0/T)]$, if ${\bf k}$ is within the LE areas, and
with $\omega/[(T-T_0)(1-T_0/T)]$, if it is within the part of ME areas
where there exists a ${\bf Q}_m$ point for which ${\bf k} - {\bf Q}_m$
is within the LE areas (which applies when $k_{_{\rm B}}T <
\sim$$\omega^{\lambda}({\bf Q})$). Thus, the $\omega$ dependence of
$\Gamma^s_2({\bf k},\omega)$ in the low $|\omega|/k_{_{\rm B}}T$ limit
is similar to that of $\Gamma^s_q({\bf k},\omega)$ (see above) in the
$k_{_{\rm B}}T \ll \Gamma^q_{0{\rm a}}$ regime, but it decreases faster
with $T$ there; however, this behavior of $\Gamma^s_2$ persists in the
$k_{_{\rm B}}T \gg \Gamma^q_{0{\rm a}}$ regime, where $\Gamma^s_q$
decreases faster with $T$, and it is somewhat modified when $k_{_{\rm
B}}T$ exceeds $\sim$$\omega^{\lambda}({\bf Q})$. 

Similarly to $\Gamma^s_q$ above, an integration on the
low-$(|\omega|/k_{_{\rm B}}T)$ $\Gamma^s_2$, over a part of the BZ,
results in a contribution of the LE areas which, approximately, scales
with $\omega /(1-T_0/T)$; for $k_{_{\rm B}}T >
\sim$$\omega^{\lambda}({\bf Q})$ this contribution persists as the
dominant part of the ${\bf k}$-integrated low-$(|\omega|/k_{_{\rm B}}T)$
$\Gamma^s$, and it approximately scales with $\omega$. 

In the high-$|\omega|/k_{_{\rm B}}T$ limit, $\Gamma^s_2$ is determined
by the $\Gamma^s_{\omega {\rm s}}$ terms in Eq.~(\ref{eq17}). The $A^s$
functions in the integrals there can then be approximated by
$\delta$-functions around the corresponding $\pm {\bar \epsilon}^s({\bf
k}^{\prime})$, resulting in factors $|\omega| - {\bar \epsilon}^s({\bf
k}^{\prime}) - \omega^{\lambda s}_{0{\rm s}}$ (thus depending
linearly on $\omega$), for $|\omega| > {\bar \epsilon}^s({\bf
k}^{\prime})$; however, there are also multiplicative factors of
$\cosh{^2(\xi_{{\bf k}^{\prime}}^s({\bar \epsilon}^s({\bf
k}^{\prime}))}$ or $\sinh{^2(\xi_{{\bf k}^{\prime}}^s({\bar
\epsilon}^s({\bf k}^{\prime}))}$ and of $\cosh{(2\xi_{\bf
k}^s(\omega))}$ which has an opposite $\omega$-dependence for $|\omega|
< \sim$$J$. 

\section{Evaluation of the Electron Spectral Functions}

By Eq.~(\ref{eq18}), $A^d_0({\bf k}, \omega)$ is a convolution of QE and
svivon spectral functions, weighted by the $b_{_T}(\omega^{\prime}) +
f_{_T}(\omega^{\prime} \mp\omega)$ factors. In order to study the
effect of these factors, let us approximate them through their low- and
high-$|\omega^{\prime}| / k_{_{\rm B}}T$ asymptotic behaviors; thus they
are expressed as: (a) $b_{_T}(\omega^{\prime}) + f_{_T}(\omega^{\prime}
\mp\omega) \cong k_{_{\rm B}}T / \omega^{\prime}$, for
$|\omega^{\prime}| < k_{_{\rm B}}T$; (b) $b_{_T}(\omega^{\prime}) +
f_{_T}(\omega^{\prime} \mp\omega) \cong 0$, for $|\omega^{\prime}| \;
\&\; |\omega^{\prime} \mp\omega| > k_{_{\rm B}}T$ and
$\text{sign}(\omega^{\prime}) = \text{sign}(\omega^{\prime} \mp\omega)$;
(c) $b_{_T}(\omega^{\prime}) + f_{_T}(\omega^{\prime} \mp\omega) \cong
\omega^{\prime} / |\omega^{\prime}|$, for $|\omega^{\prime}| \; \&\;
|\omega^{\prime} \mp\omega| > k_{_{\rm B}}T$ and
$\text{sign}(\omega^{\prime}) \ne \text{sign}(\omega^{\prime}
\mp\omega)$; (d) $b_{_T}(\omega^{\prime}) + f_{_T}(\omega^{\prime}
\mp\omega) \cong \omega^{\prime} / 2|\omega^{\prime}|$, for
$|\omega^{\prime}| > k_{_{\rm B}}T$ and $|\omega^{\prime} \mp\omega| <
k_{_{\rm B}}T$. Eq.~(\ref{eq18}) is then approximated as: 
\begin{eqnarray}
A^d_0({\bf k}, \omega) &=& A^d_{0{\rm b}}({\bf k}, \omega) + A^d_{0{\rm
c}}({\bf k}, \omega), \ \ \ \text{where} \nonumber \\ 
A^d_{0{\rm b}}({\bf k}, \omega) &\cong& {k_{_{\rm B}}T \over N}
\sum_{{\bf k}^{\prime}} \int_{-k_{_{\rm B}}T}^{k_{_{\rm B}}T}
{d\omega^{\prime} \over \omega^{\prime}} A^s({\bf k}^{\prime},
\omega^{\prime}) \nonumber \\ &\ & \times \big[ A^q({\bf k} - {\bf
k}^{\prime}, \omega - \omega^{\prime}) \cosh{^2(\xi_{{\bf
k}^{\prime}}^s(\omega^{\prime}))} \nonumber \\ &\ & + A^q({\bf k} - {\bf
k}^{\prime}, \omega + \omega^{\prime}) \sinh{^2(\xi_{{\bf
k}^{\prime}}^s(\omega^{\prime}))} \big], \nonumber \\ 
A^d_{0{\rm c}}({\bf k}, \omega) &\cong& {1 \over N} \sum_{{\bf
k}^{\prime}} \bigg\{ \int_{k_{_{\rm B}}T}^{|\omega| + k_{_{\rm B}}T}
d\omega^{\prime} A^s({\bf k}^{\prime}, \omega^{\prime})\nonumber \\ &\ &
\times \big[1 - \half \theta(k_{_{\rm B}}T - |\omega| + \omega^{\prime})
\big] \nonumber \\ &\ & \times \big[ \theta(\omega) A^q({\bf k} - {\bf
k}^{\prime}, \omega - \omega^{\prime}) \cosh{^2(\xi_{{\bf
k}^{\prime}}^s(\omega^{\prime}))} \nonumber \\ &\ & + \theta(-\omega)
A^q({\bf k} - {\bf k}^{\prime}, \omega + \omega^{\prime})
\sinh{^2(\xi_{{\bf k}^{\prime}}^s(\omega^{\prime}))} \big] \nonumber \\
&\ & -\int_{-|\omega| - k_{_{\rm B}}T}^{-k_{_{\rm B}}T} d\omega^{\prime}
A^s({\bf k}^{\prime}, \omega^{\prime}) \label{eq19} \\ &\ & \times
\big[1 - \half \theta(k_{_{\rm B}}T - |\omega| - \omega^{\prime}) \big]
\nonumber \\ &\ & \times \big[ \theta(-\omega) A^q({\bf k} - {\bf
k}^{\prime}, \omega - \omega^{\prime}) \cosh{^2(\xi_{{\bf
k}^{\prime}}^s(\omega^{\prime}))} \nonumber \\ &\ & + \theta(\omega)
A^q({\bf k} - {\bf k}^{\prime}, \omega + \omega^{\prime})
\sinh{^2(\xi_{{\bf k}^{\prime}}^s(\omega^{\prime}))} \big] \bigg\}.
\nonumber 
\end{eqnarray}
 
$A^d_{0{\rm c}}({\bf k}, \omega)$ in Eq.~(\ref{eq19}) represents a
convolution of QE and modified svivon spectral functions, where the
low-energy ($|\omega^{\prime}| < k_{_{\rm B}}T$) tails of $A^s$ have
been truncated (see Eq.~(\ref{eq14}) and Fig.~\ref{fig6}(b)), while
$A^d_{0{\rm b}}({\bf k}, \omega)$ is generated by these $A^s$ tails and
QE spectral functions. Since the QE spectrum (see Fig.~\ref{fig3}(a-c))
includes almost dispersionless low-energy peaks extending over ranges of
the BZ, the effect of the $A^q({\bf k} - {\bf k}^{\prime}, \omega \mp
\omega^{\prime})$ terms in the expression for $A^d_{0{\rm b}}$ is to
introduce their low-energy peaks which are modified (due to the ${\bf
k}^{\prime}$ summation and $\omega^{\prime}$ integration) to increase
the peak widths by $\sim$$2 k_{_{\rm B}} T$; this results in an
effective low-energy (LE) electron band of a linewidth which depends
linearly on $T$, for $T > T^q_{\rm p}$. 

The contribution of svivons to the ${\bf k}^{\prime}$ summation
(including averaging over the svivon condensates) in the expression for
$A^d_{0{\rm b}}$ in Eq.~(\ref{eq19}) is determined by that of the LE
svivons, plus some contribution of ME svivons. For LE svivons, the
$\cosh{^2}$ and $\sinh{^2}$ factors in this expression approximately
scale with $1/(T-T_0)$ (see Eq.~(\ref{eq13b})), and the $T$ dependence
of the $A^s({\bf k}^{\prime}, \omega^{\prime})$ terms there is
approximated through the scaling relation specified in
Eq.~(\ref{eq17a}), introducing a factor of $1 / (T-T_0)$, and replacing
the integration variable $\omega^{\prime}$ by $x = \omega^{\prime} /
k_{_{\rm B}} (T-T_0)$ (with integration limits $\pm T/(T-T_0)$). For ME
svivons the $\cosh{^2}$, $\sinh{^2}$ and $A^s$ factors have a weak $T$
dependence. 

As was discussed above, the size of the BZ areas of LE svivons scales
with $(T-T_0)^2$ (on the expense of areas of the ME svivons). The
expression for $A^d_{0{\rm b}}({\bf k}, \omega)$ in Eq.~(\ref{eq19})
introduces, for ${\bf k}^{\prime}$ points corresponding to LE svivons, a
BZ section (of the same size) of ${\bf k} - {\bf k}^{\prime}$ points of
QEs, out of which only its subsection of low-energy QEs contributes to
the LE $A^d_{0{\rm b}}$ effective band. The size of this subsection
approximately scales with $(T-T_0)^{\alpha^q({\bf k})}$, where the
$T$-dependent exponent $\alpha^q({\bf k}) \ge 0$ is generally $\le 2$.
Consequently, the ${\bf k}^{\prime}$ summation over LE svivons in
Eq.~(\ref{eq19}) introduces a factor of $(T-T_0)^{\alpha^q({\bf k})}$ to
their contribution to the LE $A^d_{0{\rm b}}({\bf k}, \omega)$. 

At low $T > T^q_{\rm p}$, these low-QE-energy ${\bf k} - {\bf
k}^{\prime}$ points are shifted from the electron ${\bf k}$ point
(choosing the QE BZ as in Fig.~\ref{fig3}(a-c)) by $\sim(\delta {\bf
q}_m/2)$ and $\sim({\bf Q}_m - \delta {\bf q}_m/2)$, for the four values
of $m$ (see Eq.~(\ref{eq5})). As can be viewed in Fig.~\ref{fig3}(c),
$\alpha^q({\bf k})$ is close to 2 for antinodal ${\bf k}$ points, within
this temperature regime, and one can then also assume that $k_{_{\rm
B}}T \ll \Gamma^q_0({\bf k})$, where the $T$ dependence of the stripon
$A^q$ is weak (see Eq.~(\ref{eq10a}) and discussion in Appendix A).
Thus, one gets by Eq.~(\ref{eq19}) and the above discussion that the
antinodal LE $A^d_{0{\rm b}}({\bf k}, \omega)$ approximately scales with
$T$ around its maximum; since the width of the effective electron band,
it represents, is characterized by a constant plus a linear $T$ term,
this implies that for antinodal electrons at low $T > T^q_{\rm p}$, the
spectral weight within the LE effective band anomalously increases with
$T$. There is some increase with $T$, in this regime, also in the
spectral weight within the nodal LE effective band, mainly due to the
contribution of ME svivons. 

On the other hand, for antinodal and nodal ${\bf k}$ points in the
$k_{_{\rm B}}T \gg \Gamma^q_0({\bf k})$ regime (in the nodal case this
corresponds to $T > T^q_{\rm p}$), $\alpha^q({\bf k})$ is close to 1
(see Fig.~\ref{fig3}(c)), and $A^q$ in Eq.~(\ref{eq19}) scales with
$1/T$ (see discussion in Appendix A). This results in LE $A^d_{0{\rm
b}}({\bf k}, \omega)$ which approximately scales with $1/T$ around its
maximum; thus since the linewidth of the effective band it represents
scales with $T$, one gets that the spectral weight within this band is
approximately $T$ independent for $k_{_{\rm B}}T \gg \Gamma^q_0({\bf
k})$. Note, however, that when $k_{_{\rm B}}T$ approaches $\omega_{_{\rm
H}}$, the humpons merge with the stripons in the antinodal BZ areas,
resulting in a further increase in the spectral weight within the
effective LE electron band there, saturating at $k_{_{\rm B}}T \gta
\omega_{_{\rm H}}$.


\begin{thebibliography}{99}
\bibitem{pairing}For simplicity, one QE pairing temperature $T^q_{\rm
p}$ is mentioned, though as is discussed in the text, pairing of
``antinodal'' QEs occurs in the PG phase, and it is extended to
``nodal'' QEs (except for those on the lines of nodes) in the SC state.
\end{thebibliography}
\end{document}